\documentstyle{cupconf}

\ifoldfss
\else
  \ifnfssone
    \newmathalphabet{\mathit}
      \addtoversion{normal}{\mathit}{cmr}{m}{it}
      \addtoversion{bold}{\mathit}{cmr}{bx}{it}
    \newmathalphabet{\mathcal}
      \addtoversion{normal}{\mathcal}{cmsy}{m}{n}
    \else
    \ifnfsstwo
    \fi
  \fi
\fi

\let\sec=\section
\let\ssec=\subsection

\def\key{}
\def\idx#1{}
\def\boxit{}
\newcount\japfignum
\newcount\tmpfig
\def\nextfig{\tmpfig=\japfignum \advance\tmpfig by 1 \the\tmpfig}
\def\lastfig{\tmpfig=\japfignum \the\tmpfig}
\def\del{\nabla}
\def\upto{$\,$--$\,$}
\def\half{\frac{1}{2}}
\def\sssec#1{\vskip 1em\rm\smallskip\noindent{\hbox{\bf #1}}\quad}	

\def\japfig#1#2#3#4#5#6#7{
\begin{figure}
\global\advance\japfignum by 1
\strut\vglue 1em
\centering\leavevmode\epsfxsize=#5\hsize\epsfbox[#1 #2 #3 #4]{jap_#6.eps}
\caption{#7}
\label{fig:#6}
\end{figure}
}

\def\japfigtwo#1#2#3{
\begin{figure}
\global\advance\japfignum by 1
\strut\vglue 1em
\centering\leavevmode
\epsfxsize=0.49\textwidth 
\epsfbox[58 208 510 595]{jap_#1.eps}
\hglue 1em
\epsfxsize=0.49\textwidth
\epsfbox[58 208 515 595]{jap_#2.eps}
\caption{#3}
\label{fig:#1}
\end{figure}
}

\def\japfigbasic#1#2#3{
\begin{figure}[ht]
\global\advance\japfignum by 1
\strut\vglue 1em
\centering\leavevmode
\epsfxsize=#2\textwidth \epsfbox{jap_#1.eps}
\caption{#3}
\label{fig:#1}
\end{figure}
}

\def\japref{\parskip=0pt\par\noindent\hangindent\parindent
    \parskip =2ex plus .5ex minus .1ex}

\def\gs{\mathrel{\lower0.6ex\hbox{$\buildrel {\textstyle >}
 \over {\scriptstyle \sim}$}}}
\def\ls{\mathrel{\lower0.6ex\hbox{$\buildrel {\textstyle <}
 \over {\scriptstyle \sim}$}}}
\newcount\japequationnum
\global\japequationnum=0
\def\bookdisp#1$${\leftline{\hfill{$\displaystyle#1$}
    \global\advance\japequationnum by 1
    \hfill (\the\japequationnum )}$$}
\everydisplay{\bookdisp}
\def\japsub{\rm\scriptscriptstyle}

\def\kms{{\,\rm km\,s^{-1}}}

\def\hompc{{\,h\,\rm Mpc^{-1}}}
\def\mpcoh{{\,h^{-1}\,\rm Mpc}}
\def\japitem#1{\medskip\noindent\rlap{#1}\hglue 3em\hangindent 3em}
\def\endjapitem{\medskip\noindent}
\def\japkey#1{#1}

\input epsf

\def\m@th{\mathsurround=0pt }
\def\eqalign#1{\null\,\vcenter{\openup1\jot \m@th
 \ialign{\strut\hfil$\displaystyle{##}$&$\displaystyle{{}##}$\hfil
 \crcr#1\crcr}}\,}

\def\hexnumber#1{\ifcase#1 0\or1\or2\or3\or4\or5\or6\or7\or8\or9\or
 A\or B\or C\or D\or E\or F\fi }

\makeatletter
\ifx\CUP@mtlplain@loaded\undefined
\else
\fi
\makeatother
 \makeatletter
 \ifx\CUP@mtlplain@loaded\undefined
   \font\tenbmi=cmmib10 at 10pt
   \font\sevenbmi=cmmib10 at 7pt
   \font\fivebmi=cmmib10 at 5pt

   \newfam\bmifam
   \textfont\bmifam=\tenbmi
   \scriptfont\bmifam=\sevenbmi
   \scriptscriptfont\bmifam=\fivebmi
   
 \fi
 \makeatother
\ifnfsstwo

\fi
\ifnfssone

\fi
\ifoldfss

\fi

\mathchardef\varLambda="0103

\makeatletter
\ifx\CUP@mtlplain@loaded\undefined
\else
\fi
\makeatother
\makeatletter
\ifx\CUP@mtlplain@loaded\undefined
  \font\tenbms=cmbsy10
  \font\sevenbms=cmbsy10 at 7pt
  \font\fivebms=cmbsy10 at 5pt
  \newfam\bmsfam
  \textfont\bmsfam=\tenbms
  \scriptfont\bmsfam=\sevenbms
  \scriptscriptfont\bmsfam=\fivebms

  \edef\bsy@{\hexnumber\bmsfam}
  \mathchardef\bnabla="0\bsy@72
\fi
\makeatother
\begin{document}
\ifnfssone
\else
  \ifnfsstwo
  \else
    \ifoldfss
      \let\mathcal\cal
      \let\mathrm\rm
      \let\mathsf\sf
    \fi
  \fi
\fi

  \title[Surveys and cosmic structure]{Large-scale surveys and\\
cosmic structure}

  \author[J.A. Peacock]{%
J.A. PEACOCK}
  \affiliation{%
Institute for Astronomy, University of Edinburgh,\\
Royal Observatory, Edinburgh EH9 3HJ, UK}

  \maketitle

\begin{abstract}
These lectures deal with our current knowledge of the matter 
distribution in the universe, focusing
on how this is studied via the large-scale
structure seen in galaxy surveys.
We first assemble the necessary basics needed to
understand the development of density fluctuations in an
expanding universe, and discuss how galaxies are located within
the dark-matter density field. Results from the 2dF
Galaxy Redshift Survey are presented and contrasted with
theoretical models. We show that the combination of large-scale 
structure and data on microwave-background anisotropies can
eliminate almost all degeneracies, and yield a completely
specified cosmological model. This is the `concordance' universe:
a geometrically flat combination of vacuum energy and cold dark matter.
The study of cosmic structure is able to establish this in
a manner independent of external information, such as the
Hubble diagram; this extra information can however be used
to limit non-standard alternatives, such as a variable
equation of state for the vacuum.
\end{abstract}

\firstsection % if your document starts with a section,
              % remove some space above using this command.

\sec{Preamble}

\ssec{The perturbed universe}

It has been clear since the 1930s that galaxies are not
distributed at random in the universe (Hubble 1934). For decades, our
understanding of this fact was limited by the lack
of a three-dimensional picture, although some
impressive progress was made: the dedication of
pioneers such as Shane \& Wirtanen in compiling
galaxy catalogues by eye is humbling to consider.
However, studies of the galaxy distribution came of age
in the 1980s, via redshift surveys, in which Hubble's
$v=H d$ law is used to turn spectroscopic redshifts
into estimates of distance
(e.g. Davis \& Peebles 1983; de Lapparant, Geller \& Huchra 1986; 
Saunders et al. 1991).  We were then able to
see clearly (e.g. figure~\ref{fig:cfa}) a wealth of
large-scale structures  of size exceeding 100~Mpc.
The existence of these cosmological
structures must be telling us something important about the
initial conditions of the big bang, and about the physical
processes that have operated subsequently. 
These lectures cover some of what we have learned in this regard.

\japfig{55}{190}{485}{605}{0.65}{cfa}
{One of the iconic pictures of the large-scale
structure in the galaxy distribution is this slice made
from John Huchra's ZCAT compilation of galaxy redshifts,
reflecting the state of our knowledge in the mid-1980s.
The survey coverage is
not quite complete; as well as the holes due to the galactic
plane around right ascensions $6^{\rm h}$ and $19^{\rm h}$, the
rich clusters are  somewhat over-represented with respect
to a true random sampling of the galaxy population.
Nevertheless, this plot emphasizes nicely both the large-scale
features such as the `great wall' on the left, the totally
empty void regions, and the radial `fingers of God' caused
by virialized motions in the clusters. One of the principal challenges
in cosmology is to explain this pattern.}

Throughout, it will be convenient to adopt a notation
in which the density (of mass, light, or any property) is
expressed in terms of a dimensionless density
perturbation $\delta$:
$$
\boxit{
1+\delta({\bf x}) \equiv \rho({\bf x}) / \langle\rho\rangle, 
}
$$
where $\langle\rho\rangle$ is the global mean density.
The quantity $\delta$ need not be small, but writing things
in this form naturally suggests an approach via perturbation
theory in the important linear case where $\delta \ll 1$.
As we will see, this was a good approximation at early times.
The existence of this field in the universe raises two
questions: what generated it, and how does it evolve?
A popular answer for the first question is inflation, in which quantum
fluctuations are able to seed density fluctuations. So far,
despite some claims, this theory is not tested, and we consider
later some ways in which this might be accomplished.
Mainly, however, we will be concerned here with the
question of evolution.

\ssec{Relativistic viewpoint and gauge issues}

Many of the key aspects of the evolution of structure
in the universe can be dealt with via a deceptively simple 
Newtonian approach, but honesty requires a brief
overview of some of the difficult issues that will
be evaded by taking this route.

Because relativistic physics equations are written in
a covariant form in which all quantities are independent
of coordinates, relativity does not distinguish between
{\it active\/} changes of coordinate (e.g. a Lorentz boost)
or {\it passive\/} changes (a mathematical change of variable,
normally termed a gauge transformation).
This generality is a problem, since it is not trivial
to know which coordinates should be used.
To see how the problems arise,
ask how tensors of different order change under a gauge
transformation $x^\mu\rightarrow x'{}^\mu=x^\mu+\epsilon^\mu$.
Consider first a scalar quantity $S$ (which might be
density, temperature etc.). 
A scalar quantity in relativity is normally taken to be
independent of coordinate frame, but this is only for the
case of Lorentz transformations, which do not involve a
change of the spacetime origin. A gauge transformation
therefore not only induces the usual transformation
coefficients $dx'{}^\mu/dx^\nu$, but also involves
a translation that relabels spacetime points.
We therefore have to deal with 
$S'(x^\mu+\epsilon^\mu)=S(x^\mu)$, so the rule
for the gauge transformation of scalars is
$$
S'(x^\mu)=S(x^\mu)-\epsilon^\alpha \partial S / \partial x^{\alpha}.
$$
Similar reasoning yields
the gauge transformation laws for higher tensors, although
we need to account not only for the translation
of the origin, but also for the usual effect of the coordinate
transformation on the tensor. 

Consider applying this to the case of a
uniform universe; here $\rho$ only depends on time, so that
$$
\rho'=\rho-\epsilon^0 \dot\rho.
$$
An effective density perturbation is thus produced by a local
alteration in the time coordinate: when we look at
a universe with a fluctuating density, should we really think
of a uniform model in which time is wrinkled?
This ambiguity may seem absurd, and in the laboratory it
could be resolved empirically by constructing 
the coordinate system directly -- in principle by using light
signals. This shows that the cosmological horizon plays an
important role in this topic: perturbations with wavelength
$\lambda\ls ct$ inhabit a regime in which gauge ambiguities
can be resolved directly via common sense. The real difficulties
lie in the super-horizon modes with $\lambda\gs ct$. However,
at least within inflationary models, these difficulties can be overcome.
According to inflation, perturbations on scales greater than the
horizon were originally generated via quantum fluctuations on
small scales within the horizon of a nearly de Sitter exponential
expansion. There is thus no problem in
understanding how the initial density field is described, since
the simplest coordinate system can once again be constructed
directly.

The most direct way of solving these difficulties 
is to construct perturbation
variables that are explicitly independent of gauge.
Comprehensive technical discussions of this method are
given by Bardeen (1980), Kodama \& Sasaki (1984),
Mukhanov, Feldman \& Brandenberger (1992).
The starting point for a discussion of metric perturbations is
to devise a notation that will classify the possible
perturbations. Since the metric is symmetric, there are
10 independent degrees of freedom in $g^{\mu\nu}$; a
convenient scheme that captures these possibilities is to write
the cosmological metric as
$$
d\tau^2 = a^2(\eta)\left\{
(1+2\phi)d\eta^2 + 2 w_i d\eta\, dx^i -
\left[(1-2\psi)\gamma_{ij}+2h_{ij}\right]dx^i\,dx^j \right\}.
$$
In this equation, $\eta$ is conformal time, and
$\gamma_{ij}$ is the comoving spatial part of the
Robertson-Walker metric.

The total number of degrees of freedom here is
apparently 2 (scalar fields $\phi$ and $\psi$)
+ 3 (3-vector field $\bf w$) + 6 (symmetric 3-tensor $h_{ij}$)
$= 11$. To get the right number, the tensor $h_{ij}$ is
required to be traceless: $\gamma^{ij} h_{ij}=0$.
The perturbations can be split into three
classes: \key{scalar perturbations}, which are described
by scalar functions of spacetime coordinate, and which
correspond to the growing density perturbations studied
above; \key{vector perturbations}, which correspond to
vorticity perturbations, and \key{tensor perturbations},
which correspond to gravitational waves. 
Here, we shall concentrate mainly on scalar perturbations.
Since vectors and tensors can be generated from derivatives
of a scalar function, the most general scalar perturbation
actually makes contributions to all the $g_{\mu\nu}$
components in the above expansion:
$$
\delta g_{\mu\nu} = a^2 \, \left(\matrix{ 2\phi & - B_{,i} \cr
-B_{,i} & 2[\psi \delta_{ij} - E_{,ij}]\cr}\right),
$$
where four scalar functions $\phi$, $\psi$, $E$ and $B$
are involved. It turns out that this situation can be
simplified by defining variables that are
unchanged by a gauge transformation:
$$
\boxit{
\eqalign{
\Phi &\equiv \phi + {1\over a}\; [(B-E')a]' \cr
\Psi &\equiv \psi - {a'\over a}\ (B-E'), \cr
}
}
$$
where primes denote derivatives with respect to conformal time.

These gauge-invariant `potentials' have a fairly direct
physical interpretation, since they are closely related
to the Newtonian potential. The easiest way to evaluate
the gauge-invariant  fields is to make a specific gauge
choice and work with the \key{longitudinal gauge} in
which $E$ and $B$ vanish, so that $\Phi=\phi$ and
$\Psi=\psi$. A second key result is that inserting the longitudinal
metric into the Einstein equations shows that
$\phi$ and $\psi$ are identical in the case of fluid-like
perturbations where off-diagonal elements of the
energy--momentum tensor vanish.  
In this case, the
longitudinal gauge becomes identical to the Newtonian
gauge, in which perturbations are described by a single scalar
field, which is the gravitational potential. 
The conclusion is thus that the gravitational potential 
can for many purposes give an effectively gauge-invariant measure of
cosmological perturbations, and this provides a sounder
justification for the Newtonian approach that we now adopt.

\sec{Newtonian equations of motion}

\ssec{Matter-dominated universe}

In the Newtonian approach, we treat dynamics of cosmological matter exactly 
as we would in the laboratory, by finding the equations of motion induced by
either pressure or gravity.
In what follows, it should be remembered that we probably need
to deal in practice with two rather different kinds of material:
dark matter that is collisionless and interacts only via
gravity, and baryonic material which is a collisional
fluid, coupled to dark matter only via gravity
(and to photons via Thomson scattering, so that the dominant part
of the pressure derives from the radiation).

Also, the problem of cosmological dynamics has to deal with the characteristic feature
of the Hubble expansion. This means that it is convenient to introduce
comoving length units, and to consider primarily peculiar
velocities -- i.e. deviations from the Hubble flow.
The standard notation that includes these aspects is
$$
\boxit{
\eqalign{
{\bf x}(t) &= a(t) {\bf r}(t) \cr
{\bf \delta v}(t) &= a(t) {\bf u}(t),\cr
}
}
$$
so that ${\bf x}$ has units of proper length, i.e. it is 
an \key{Eulerian coordinate}.
First note that the comoving peculiar velocity $\bf u$ is
just the time derivative of the comoving coordinate $\bf r$:
$$
\dot {\bf x} = \dot a {\bf r} + a\dot{\bf r} = H{\bf x} + a\dot{\bf r},
$$
where the rhs must be equal to the Hubble flow $H{\bf x}$, plus the
peculiar velocity $\delta{\bf v}=a{\bf u}$. 
In this equation, dots stand for exact convective time
derivatives -- i.e. time derivatives measured by
an observer who follows a particle's trajectory -- rather
than partial time derivatives $\partial/\partial t$.

The equation of motion follows from writing the Eulerian
equation of motion as $\ddot{\bf x} = {\bf g_{\rm 0} + g}$, where
${\bf g}=-\del \Phi/a$ is the peculiar acceleration,
and $\bf g_{\rm 0}$ is the acceleration that acts on
a particle in a homogeneous universe (neglecting
pressure forces to start with, for simplicity). Differentiating 
${\bf x}=a{\bf r}$ twice gives
$$
\ddot {\bf x}=a\dot{\bf u}+2\dot a {\bf u} + {\ddot a\over a}\, {\bf x} = {\bf g_{\rm 0} + g}.
$$
The unperturbed equation corresponds to zero peculiar 
velocity and zero peculiar acceleration:
$(\ddot a/a)\, {\bf x} = {\bf g_{\rm 0}}$; subtracting this gives the
perturbed equation of motion 
$$
\dot{\bf u}+2(\dot a/a) {\bf u} = {\bf g}/a.
$$
The only point that needs a little more thought is
the nature of the unperturbed equation of motion. This cannot be derived
from Newtonian gravity alone,
since general relativity is really needed for
a proper derivation of the homogeneous equation of motion.
However, as long as we are happy to accept that $\bf g_{\rm 0}$ is given, then
it is a well-defined procedure to add a peculiar acceleration that is the
gradient of the potential derived from the density perturbations.

The equation of motion for the peculiar velocity shows that
$\bf u$ is affected by gravitational acceleration and by
the \key{Hubble drag} term, $2(\dot a/a) {\bf u}$. This
arises because the peculiar velocity falls with time
as a particle attempts to catch up with successively more distant
(and therefore more rapidly receding) neighbours.
If the proper peculiar velocity is $v$, then after time $dt$ 
the galaxy will have moved a proper distance $x=v\, dt$ from its original location.
Its near neighbours will now be galaxies with recessional velocities
$H\, x=H\, v\, dt$, relative to which the peculiar velocity
will have fallen to $v-Hx$.
The equation of motion is therefore just
$$
\dot v=-H\, v=-{\dot a\over a}\, v,
$$
with the solution $v\propto a^{-1}$: peculiar
velocities of nonrelativistic objects suffer redshifting by exactly the same
factor as photon momenta. This becomes
$\dot u = -2H\, u$ when rewritten in comoving units.

The peculiar velocity is directly related to the evolution of
the density field, through conservation of mass.
This is expressed via the continuity equation, which takes the form
$$
{d\over dt}\rho_0(1+\delta) = -\rho_0(1+\delta)\, {\bf \del\cdot u}.
$$
Here, spatial derivatives are 
with respect to comoving coordinates:
$$
   \del \equiv a\,\del_{\rm proper},
$$
which we will assume hereafter, and
the time derivative is once more a convective one:
$$
{d\over dt} = {\partial\over \partial t} + {\bf u \cdot \del}.
$$
Finally, when using comoving length units, the
background density $\rho_0$ independent of time, and so
the full continuity equation can be written as
$$
{d\over dt} \delta = - (1+\delta) {\bf \del\cdot u}.
$$
Unlike the equation of motion for $\bf u$, this is not linear in the
perturbations $\delta$ and $\bf u$. To cure this, we restrict ourselves
to the case $\delta \ll 1$ and linearize the equation,
neglecting second-order terms like $\delta \times {\bf u}$,
which removes the distinction between convective and partial
time derivatives. The linearized equations for conservation of momentum and matter 
as experienced by fundamental observers moving with the
Hubble flow are then:
$$
\boxit{
\eqalign{
&\dot{\bf u}+ 2{\dot a\over a}\,{\bf u} = {{\bf g}\over a}\cr
&\dot\delta = -{\bf \del\cdot u},\cr
}}
$$
where the peculiar gravitational acceleration $-{\bf\del}\Phi/a$
is denoted by $\bf g$. 

The solutions of these equations
can be decomposed into modes either parallel to $\bf g$ or
independent of $\bf g$ (these are the homogeneous and inhomogeneous
solutions to the equation of motion).
The homogeneous case corresponds to no peculiar gravity -- i.e.
zero density perturbation. This is consistent with the
linearized continuity equation, ${\bf\del\cdot u} = -\dot\delta$,
which says that it is possible to have \key{vorticity modes}
with ${\bf\del\cdot u} = 0$ for which $\dot\delta$ vanishes, so there
is no growth of structure in this case.
The proper velocities of these vorticity modes 
decay as $v=au\propto a^{-1}$,
as with the kinematic analysis for a single particle.

\sssec{Growing mode}
For the growing mode, it is most convenient to eliminate
$\bf u$ by taking the divergence of the equation of motion
for $\bf u$, and the time derivative of the continuity
equation. This requires a knowledge of $\bf \del\cdot g$,
which comes via Poisson's equation: ${\bf \del\cdot g}=4\pi Ga\rho_0\delta$.
The resulting 2nd-order equation for $\delta$ is
$$
\ddot\delta + 2 {\dot a\over a}\dot\delta =
4\pi G\rho_0 \,   \delta.
$$
This is easily solved for the $\Omega_m=1$ case, where
$4\pi G\rho_0 = 3H^2/2 = 2/3t^2$, and a power-law solution works:
$$
\delta(t) \propto t^{2/3} \quad{\rm or} \quad t^{-1}.
$$
The first solution, with $\delta(t)\propto a(t)$ is the growing mode, corresponding
to the gravitational instability of density perturbations.
Given some small initial seed fluctuations, this is the
simplest way of creating a universe with any desired
degree of inhomogeneity.

An alternative way of looking at the growing mode
is that we want to try looking for
a homogeneous solution ${\bf u}=F(t)\bf g$. Then using continuity
plus ${\bf \del\cdot g}=4\pi Ga\rho_0\delta$,
gives us
$$
{\bf\delta v} = {2f(\Omega_m)\over 3 H\Omega_m}{\bf g},
$$
where the function $f(\Omega_m)\equiv d\ln\delta/d\ln a$.
A very good approximation to this (Peebles 1980) is
$f\simeq\Omega^{0.6}$ (a result that is almost independent of $\Lambda$;
Lahav et al. 1991).

\sssec{Jeans scale}
So far, we have mainly considered the collisionless component.
For the photon-baryon gas, all that changes is that the peculiar
acceleration gains a term from the pressure gradients:
$$
{\bf g} = -{\bf\del}\Phi/a - {\bf \del} p / (a \rho_0).
$$
The pressure fluctuations are related to the
density perturbations via the sound speed
$$
c_s^2\equiv {\partial p\over \partial \rho}.
$$
Now think of a plane-wave disturbance $\delta\propto
e^{-i{\bf k\cdot r}}$, where $\bf k$ is a comoving wavevector;
in other words, suppose that the wavelength of a single Fourier mode stretches
with the universe.
All time dependence is carried by the amplitude of the
wave, and so the spatial dependence can be factored out
of time derivatives in the above equations
(which would not be true with a constant comoving wavenumber $k/a$).
The equation of motion for $\delta$ then gains an extra term
on the rhs from the pressure gradient:
$$
\boxit{
\ddot\delta + 2 {\dot a\over a}\dot\delta =
  \delta \bigl( 4\pi G\rho_0 - c_s^2 k^2/a^2\bigr).
}
$$
This shows that there is a critical proper wavelength,
the \key{Jeans length}, at which we switch from
the possibility of gravity-driven growth for long-wavelength
modes to standing sound waves at short wavelengths.
This critical length is
$$
\boxit{
\lambda_{\japsub J} = {2\pi \over k_{\japsub J} a} =  c_s \sqrt{{\pi\over G\rho}}.
}
$$
Qualitatively, we expect to have no growth when
the `driving term' on the rhs is negative.
However, owing to the expansion, $\lambda_{\japsub J}$ will
change with time, and so a given perturbation
may switch between periods of growth and stasis.
These effects help to govern the form of the
perturbation spectrum that propagates to the
present universe from early times.

\sssec{The general case}
How does the matter-dominated growth $\delta(a) \propto a$
change at late times when $\Omega_m \ne 1$? The differential equation
for $\delta$ is as before, but $a(t)$ is altered.
Provided the vacuum
equation of state is exactly $p=-\rho c^2$, or
if the vacuum energy is negligible, the solutions to the
growth equations can be written as
$$
\boxit{
\delta\propto \cases{
({\dot a / a})\, \int_0^a (\dot a)^{-3}\, da \quad\; &(growing mode) \cr
({\dot a / a}) \quad\; &(decaying mode). \cr
}
}
$$
(Heath 1977; see also section 10 of Peebles 1980).
For the most general case, e.g. a vacuum with time-varying
density, these do not apply, and
the differential equation for $\delta$ must be
integrated directly.

In any case, the equation for the growing mode requires numerical
integration unless the vacuum energy vanishes.
A very good approximation to the answer is given by Carroll et al. (1992):
$$
\boxit{
{\delta(z=0,\Omega)\over\delta(z=0,\Omega=1)} \simeq 
\frac{5}{2}\Omega_m\left[\Omega_m^{4/7}-\Omega_v+
 (1+\half\Omega_m)(1+\frac{1}{70}\Omega_v)\right]^{-1}.
}
$$
This fitting formula for the growth suppression in
low-density universes is an invaluable practical tool.
For flat models with $\Omega_m+\Omega_v=1$, it says that the
growth suppression is less marked than for an open
universe -- approximately $\Omega^{0.23}$ as against
$\Omega^{0.65}$ in the $\Lambda=0$ case.
This reflects the more rapid
variation of $\Omega_v$ with redshift; if the cosmological
constant is important dynamically, this only became
so very recently, and the universe spent more of its
history in a nearly Einstein--de Sitter state by comparison
with an open universe of the same $\Omega_m$.

\ssec{Radiation-dominated universe}

At early enough times, the universe was radiation dominated
($c_s=c/\sqrt{3}$) and the analysis so far does not apply.
It is common to resort to general relativity perturbation theory
at this point. However, the fields are still weak, and
so it is possible to generate the results we need by
using special relativity fluid mechanics and Newtonian gravity
with a relativistic source term:
$$
\nabla^2\Phi = 4\pi G(\rho +3p/c^2),
$$
in Eulerian units.
The main change from the previous treatment come from factors
of 2 and $4/3$ due to this $(\rho +3p/c^2)$ term, and other contributions
of the pressure to the relativistic equation of motion.
The resulting evolution equation for $\delta$ is
$$
\boxit{
\ddot\delta + 2 {\dot a\over a}\dot\delta =
  {32\pi \over 3} G\rho_0  \delta, 
}
$$
so the net result of all the relativistic corrections is
a driving term on the rhs that is a factor $8/3$ higher
than in the matter-dominated case
(see e.g. Section 15.2 of Peacock 1999 for the details).

In both matter- and radiation-dominated universes with $\Omega=1$, we 
have $\rho_0\propto 1/t^2$:
$$
\eqalign{
{\rm matter\  domination}\ (a\propto t^{2/3}): \quad &4\pi G\rho_0 = {2\over 3t^2}\cr
{\rm radiation\ domination}\ (a\propto t^{1/2}): \quad &32\pi G\rho_0/3 = {1\over t^2}.\cr
}
$$
Every term in the equation for $\delta$ is thus the product of
derivatives of $\delta$ and powers of $t$, and  a power-law
solution is obviously possible. If we try $\delta\propto t^n$, then
the result is $n=2/3$ or $-1$ for matter domination;
for radiation domination, this becomes $n=\pm 1$.
For the growing mode, these can be combined rather conveniently
using the \key{conformal time} $\eta\equiv\int dt/a$:
$$
\boxit{
\delta \propto \eta^2.
}
$$
The quantity $\eta$ is proportional to the comoving size of the
cosmological particle horizon.

One further way of stating this result is that gravitational
potential perturbations are independent of time (at least while
$\Omega=1$). Poisson's equation tells us that $-k^2\Phi/a^2\propto \rho\,\delta$;
since $\rho\propto a^{-3}$ for matter domination or $a^{-4}$
for radiation, that gives $\Phi\propto \delta/a$ or $\delta/a^2$
respectively, so that $\Phi$ is independent of $a$ in either case. In other words,
the metric fluctuations resulting from potential perturbations are
frozen, at least for perturbations with wavelengths greater than the horizon size.

\ssec{M\'esz\'aros effect}

What about the case of collisionless matter in a radiation
background? The fluid treatment is not appropriate here,
since the two species of particles can interpenetrate.
A particularly interesting limit is for perturbations
well inside the horizon: the radiation can then be treated as
a smooth, unclustered background that affects only the overall
expansion rate. This is analogous to the effect of $\Lambda$,
but an analytical solution does exist in this case.
The perturbation equation is as before
$$
\ddot\delta + 2 {\dot a\over a}\dot\delta = 4\pi G\rho_m  \delta, 
$$
but now $H^2=8\pi G(\rho_m+\rho_r)/3$. If we change variable to
$y\equiv \rho_m/\rho_r=a/a_{\rm eq}$, and use the Friedmann
equation, then the growth equation becomes
$$
\delta'' + {2+3y\over 2y(1+y)}\, \delta' -{3\over 2y(1+y)}\, \delta = 0
$$
(for $k=0$, as appropriate for early times).
It may be seen by inspection that a growing solution exists with $\delta''=0$:
$$
\boxit{
\delta\propto y + 2/3.
}
$$
It is also possible to derive the decaying mode. This is simple in
the radiation-dominated case ($y\ll 1$): $\delta\propto -\ln y$ is
easily seen to be an approximate solution in this limit.

What this says is that, at early times, the dominant energy of
radiation drives the universe to expand so fast that the matter
has no time to respond, and $\delta$ is frozen at a constant value.
At late times, the radiation becomes negligible, and the growth
increases smoothly to the Einstein--de Sitter $\delta\propto a$ behaviour
(M\'esz\'aros 1974).
The overall behaviour is therefore similar to the effects of
pressure on a coupled fluid: for scales greater than the horizon,
perturbations in matter and radiation can grow together, but this
growth ceases once the perturbations enter the horizon.
However, the explanations of these two phenomena are completely different. In the
fluid case, the radiation pressure prevents the perturbations
from collapsing further; in the collisionless case, the photons
have free-streamed away, and the matter perturbation fails to
collapse only because radiation domination ensures that the
universe expands too quickly for the matter to have time to
self-gravitate. Because matter perturbations enter the horizon (at $y=y_{\rm entry}$)
with $\dot\delta>0$, $\delta$ is not frozen quite at the horizon-entry
value, and continues to grow until this initial `velocity'
is redshifted away, giving a total boost factor of roughly
$\ln y_{\rm entry}$. This log factor may be seen below in the
fitting formulae for the CDM power spectrum.

\ssec{Coupled perturbations}

We will often be concerned with the evolution of perturbations
in a universe that contains several distinct components
(radiation, baryons, dark matter). It is easy to treat such
a mixture if only gravity is important (i.e. for large
wavelengths). Look at the perturbation equation in the form
$$
L\,\delta = {\rm driving\ term},\quad\quad
L\equiv{\partial^2\over
   \partial t^2} + {2\dot a\over a}{\partial\over\partial t}.
$$
The rhs represents the effects of gravity, and particles
will respond to gravity whatever its source. The coupled
equations for several species are thus given by summing
the driving terms for all species.

\sssec{Matter plus radiation}
The only subtlety is that we must take into account the
peculiarity that radiation and pressureless matter respond
to gravity in different ways, as seen in the equations
of fluid mechanics.
The coupled equations for perturbation growth are thus
$$
L\left({\delta_m\atop \delta_r}\right) = 4\pi G
  \left(\matrix{\rho_m & 2\rho_r\cr 4\rho_m/3 & 8\rho_r/3\cr}
  \right)\left({\delta_m\atop \delta_r}\right).
$$
Solutions to this will be simple if the matrix
has time-independent eigenvectors. Only one of
these is in fact time independent: $(1,  4/3)$.
This is the \key{adiabatic mode} in which
$\delta_r=4\delta_m/3$ at all times. This corresponds
to some initial disturbance in which matter particles
and photons are compressed together. The entropy per
baryon is unchanged, $\delta(T^3)/(T^3)=\delta_m$, hence
the name `adiabatic'. In this case, the perturbation amplitude 
for both species obeys $L\delta=4\pi G(\rho_m+8\rho_r/3)\delta$.
We also expect the baryons and photons to obey this
adiabatic relation very closely even on small scales:
the \key{tight coupling approximation} says that Thomson
scattering is very effective at suppressing motion of
the photon and baryon fluids relative to each other.

\sssec{Isocurvature modes}
The other perturbation mode is harder to see until we realize
that, whatever initial conditions we choose for $\delta_r$ and
$\delta_m$, any subsequent changes to matter and radiation on large scales
must be adiabatic (only gravity is acting). 
Suppose that the radiation field is initially chosen to be uniform;
we then have
$$
\boxit{
\delta_r = {4\over 3} (\delta_m-\delta_i),
}
$$
where $\delta_i$ is some initial value of $\delta_m$. The
equation for $\delta_m$ becomes
$$
L\delta_m={32\pi G\over 3}\left[\left(\rho_r+{3\over 8}\rho_m
\right)\delta_m -\rho_r\delta_i\right],
$$
which is as before if $\delta_i=0$. The other solution is therefore
a particular integral with $\delta\propto\delta_i$.
For $\Omega=1$, the answer can
be expressed most neatly in terms of $y\equiv\rho_m/\rho_r$
(Peebles 1987):
$$
\eqalign{
\delta_m/\delta_i &= {4\over y}- {8\over y^2}(\sqrt{1+y }-1)
    \simeq 1 - y/2 + \cdots\cr
\delta_r/\delta_i &= 4(\delta_m/\delta_i -1)/3\simeq -2y/3+\cdots\cr
}
$$
At late times, $\delta_m\rightarrow 0$, while $\delta_r
\rightarrow -4\delta_i/3$.
This mode is called the \key{isocurvature mode}, since it
corresponds to a total density perturbation
$\delta\rho/\rho\rightarrow 0$ as $t_i\rightarrow 0$.
Subsequent evolution attempts to preserve constant density
by making the matter perturbations decrease while the amplitude
of $\delta_r$ increases.
An alternative name for this mode is an \key{entropy  perturbation}.
This reflects the fact that one only perturbs the initial
ratio of photon and matter number densities.
The late-time evolution is then easily understood:
causality requires that, on large scales, the initial
entropy perturbation is not altered. Hence, as the universe
becomes strongly matter dominated, the entropy perturbation
becomes carried entirely by the photons. This
leads to an increased amplitude of microwave-background
anisotropies in isocurvature models (Efstathiou \& Bond 1986), which is one
reason why such models are not popular.
Of course, a small admixture of isocurvature perturbations
is always going to be hard to rule out
(e.g. Bucher, Moodley \& Turok 2002), so neglect of this mode
is primarily justified by the fact that the simplest model for
the generation of cosmological perturbations (single-field
inflation) produces pure adiabatic modes.
Models with multiple fields, such as the decaying
\key{curvaton} of Lyth \& Wands (2002) tend to generate
order-unity isocurvature contributions, which are impossible
to reconcile with CMB data (e.g. Gordon \& Lewis 2002).

\sssec{Baryons and dark matter}
This case is simpler, because both components have the same equation of state:
$$
L\left({\delta_b\atop \delta_d}\right) = {4\pi G\rho\over\Omega}
  \left(\matrix{\Omega_b& \Omega_d\cr \Omega_b & \Omega_d\cr}
  \right)\left({\delta_b\atop \delta_d}\right).
$$
Both eigenvectors are time independent: $(1, 1)$ and
$(\Omega_d, -\Omega_b)$. The time dependence of these modes is
easy to see for an $\Omega=1$ matter-dominated universe:
if we try $\delta\propto t^n$, then we obtain respectively
$n=2/3$ or $-1$ and $n=0$ or $-1/3$ for the two modes.
Hence, if we set up a perturbation with $\delta_b=0$, this
mixture of the eigenstates will quickly evolve to be dominated 
by the fastest-growing mode with $\delta_b=\delta_d$:
the baryonic matter falls into the dark potential wells.
This is one process that allows universes containing
dark matter to produce smaller anisotropies in the
microwave background: radiation drag allows the dark matter
to undergo growth between matter--radiation equality and
recombination, while the baryons are held back.

This is the solution on large scales, where
pressure effects are negligible. On small scales,
the effect of pressure will prevent the baryons
from continuing to follow the dark matter. We can
analyse this by writing down the
coupled equation for the baryons, but now adding
in the pressure term (sticking to the matter-dominated era,
to keep things simple):
$$
L\, \delta_b = L\, \delta_d -k^2 c_{\japsub S}^2 \delta_b/a^2.
$$
In the limit that dark matter dominates the gravity,
the first term on the rhs can be taken as an imposed
driving term, of order $\delta_d/t^{2}$.
In the absence of pressure, we saw that $\delta_b$
and $\delta_d$ grow together, in which case the second term on the
rhs is smaller than the first if $kc_{\japsub S}t/a \ll 1$.
Conversely, for large wavenumbers ($kc_{\japsub S}t/a \gg 1$),
baryon pressure causes the growth rates in the baryons and
dark matter to differ; the main behaviour of the
baryons will then be slowly declining sound waves, 
and we can write the WKB solution.
$$
\delta_b\propto (a c_{\japsub S})^{-1/2} \exp 
\left[ \pm i \int k c_{\japsub S}\; d\eta\right],
$$
where $\eta$ is conformal time.
An alternative way to see that the baryons are damped
is to write the coupled equations as
$$
L\left({\delta_b\atop \delta_d}\right) = {4\pi G\rho\over\Omega}
  \left(\matrix{\Omega_b -\kappa^2 & \Omega_d\cr \Omega_b & \Omega_d\cr}
  \right)\left({\delta_b\atop \delta_d}\right),
$$
where $\kappa\equiv k/k_{\japsub J}$. In the special case
$\Omega_b \rightarrow 0$ and $\kappa = {\rm constant}$, a solution is
clearly
$$
\delta_b = {\delta_d \over 1+ \kappa^2},
$$
and this is found to be the asymptotic solution in more
general cases (Nusser 2000).

This oscillatory behaviour holds so long as pressure forces
continue to be important. However, the sound speed drops by a large
factor at recombination, and we would then expect the oscillatory
mode to match on to a mixture of the pressure-free growing and
decaying modes. This behaviour can be illustrated in a simple
model where the sound speed is constant until recombination at conformal time $\eta _r$
and then instantly drops to zero. The behaviour of the density
field before and after $t_r$ may be written as
$$
\delta = \cases{
\delta_0\, \sin(\omega \eta)/(\omega \eta) \quad\; &\hbox{($\eta<\eta _r$)} \cr
A\eta^2 + B \eta^{-3} \quad\; &\hbox{($\eta>\eta _r$)}, \cr
}
$$
where $\omega\equiv k c_{\japsub S}$.
Matching $\delta$ and its time derivative on either side of the
transition allows the decaying component to be eliminated, giving
the following relation between the growing-mode amplitude after
the transition to the amplitude of the initial oscillation:
$$
A\eta_r^2 = {\delta_0 \over 3}\, \cos \omega \eta_r.
$$
The amplitude of the growing mode after recombination depends
on the phase of the oscillation at the time of recombination.
The output is maximised when the input density perturbation is zero
and the wave consists of a pure velocity perturbation; this
effect is known as \key{velocity overshoot}.
The post-recombination transfer function will thus 
display oscillatory features, peaking for wavenumbers that
had particularly small amplitudes prior to recombination.
Such effects can be seen at work in determining the relative
positions of small-scale features in the power spectra of
matter fluctuations and microwave-background fluctuations.

\ssec{Transfer functions and characteristic scales}

The transfer function for models with the full above list
of ingredients was first computed accurately by Bond \& Szalay (1983),
and is today routinely available via public-domain codes
such as {\sc cmbfast} (Seljak \& Zaldarriaga 1996).
These calculations are a technical challenge
because we have a mixture of
matter (both collisionless dark particles and
baryonic plasma) and relativistic particles
(collisionless neutrinos and collisional photons),
which does not behave as a simple fluid.
Particular problems are caused by the change in
the photon component from being a fluid tightly coupled
to the baryons by Thomson scattering, to being
collisionless after recombination. Accurate
results require a solution of the Boltzmann
equation to follow the evolution in detail.

Some illustrative results are shown in figure~\nextfig. Leaving aside
the isocurvature models, all adiabatic cases have $T\rightarrow 1$
on large scales -- i.e. there is growth at the universal
rate (which is such that the amplitude of potential perturbations
is constant until the vacuum starts to be important at $z\ls 1$).
The different shapes of the functions can be understood intuitively
in terms of a few special length scales, as follows:

\japfigbasic{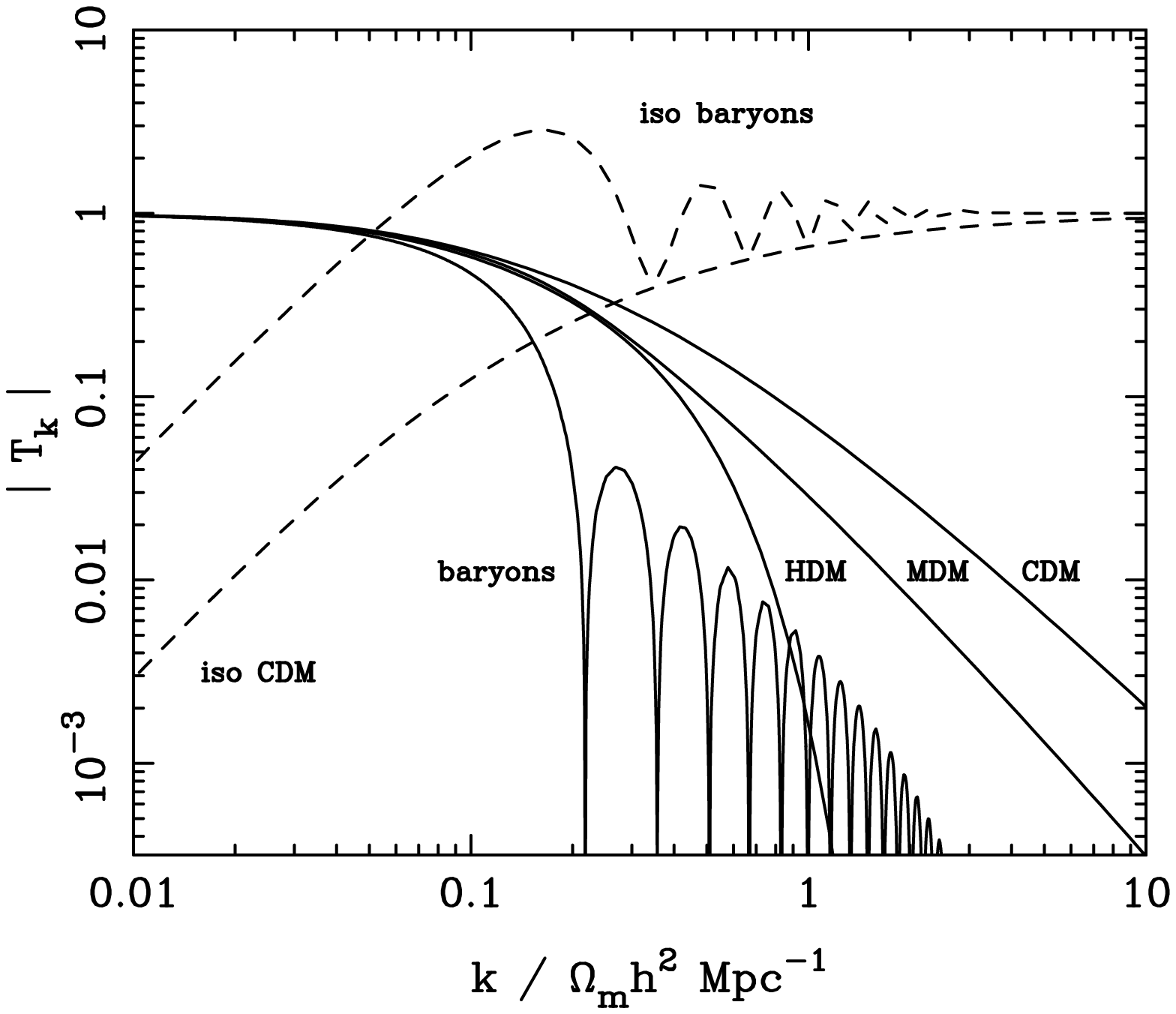}{0.7}
{A plot of transfer functions for various
adiabatic models, in which $T_k\rightarrow 1$ at small $k$.
A number of possible matter contents are illustrated:
pure baryons; pure CDM; pure HDM.
For dark-matter models, the characteristic wavenumber scales
proportional to $\Omega_m h^2$, marking the break 
scale corresponding to the horizon length at 
matter-radiation equality. The scaling for baryonic
models does not obey this exactly; the plotted case corresponds
to $\Omega_m=1$, $h=0.5$.
}

{\bf (1) Horizon length at matter-radiation equality.}
The main bend visible in all 
transfer functions is due to the M\'esz\'aros effect, which 
arises because the universe is radiation
dominated at early times. 
Fluctuations in the matter can only grow if dark
matter and radiation fall together. This does not
happen for perturbations of small wavelength, because
photons and matter can separate.
Growth only occurs for perturbations of
wavelength larger than the horizon distance, where there has been no time for the
matter and radiation to separate.
The relative diminution in fluctuations at high $k$ is the amount of growth 
missed out on between horizon entry and $z_{\rm eq}$, which would be
$\delta\propto D_{\japsub H}^2$ in the absence of the M\'esz\'aros effect.
Perturbations with larger $k$ enter the horizon 
when $D_{\japsub H}\simeq 1/k$; they are then frozen
until $z_{\rm eq}$, at which point they can grow again.
The missing growth factor is just the square of
the change in $D_{\japsub H}$ during this period, which is $\propto k^2$.
The approximate limits of the CDM
transfer function are therefore
$$
T_k \simeq \cases{
1 \quad\quad &\hbox{$kD_{\japsub  H}(z_{\rm eq})\ll 1$} \cr
[kD_{\japsub  H}(z_{\rm eq})]^{-2} \quad\quad &\hbox{$kD_{\japsub  H}(z_{\rm eq})\gg 1$}. \cr
}
$$

This process continues, until $z_{\rm eq}=23\,900\,\Omega_m h^2$,
where the universe becomes matter dominated.
We therefore expect a characteristic `break' in the
fluctuation spectrum around the comoving horizon length at this time.
Using a distance--redshift relation that ignores vacuum energy at high $z$,
$$
R_0 dr={c\over H_0} {dz\over (1+z)\sqrt{1+\Omega_m z+(1+z)^2\Omega_r}},
$$
we obtain
$$
D_{\japsub  H}(z_{\rm eq})  =
(\sqrt{2}-1)
\frac{2c}{H_0}(\Omega_m z_{\rm eq})^{-1/2} =
16\, (\Omega_m h^2)^{-1}{\rm Mpc}.
$$
Since distances in cosmology always scale as $h^{-1}$, this means
that $\Omega_m h$ should be observable.

{\bf (2) Free-streaming length.}
This relatively gentle filtering away of the initial
fluctuations is all that applies to a universe dominated
by Cold Dark Matter, in which random velocities are
negligible. 
A CDM universe thus contains fluctuations
in the dark matter on all scales, and
structure formation proceeds via 
hierarchical process in which nonlinear structures
grow via mergers.

Examples of CDM would be thermal relic WIMPs with masses
of order 100~GeV. Relic particles that were never in
equilibrium, such as axions, also come under
this heading, as do more exotic possibilities
such as primordial black holes.
A more interesting case arises when thermal relics
have lower masses.
For collisionless dark matter, perturbations can be
erased simply by free streaming: random
particle velocities cause blobs to disperse.
At early times ($kT>mc^2$), the particles will
travel at $c$, and so any perturbation that has
entered the horizon will be damped.
This process switches off when the particles
become non-relativistic, so that perturbations
are erased up to proper lengthscales of 
$ \simeq ct(kT = mc^2)$.
This translates to a comoving horizon scale
($2ct/a$ during the radiation era) at $kT = mc^2$ of
$$
L_{\rm free-stream} = 112\, (m/{\rm eV})^{-1}\, {\rm Mpc}
$$
(in detail, the appropriate figure for neutrinos will
be smaller by $(4/11)^{1/3}$ since they have a smaller
temperature than the photons).
A light neutrino-like relic that decouples while
it is relativistic satisfies
$$
\Omega_\nu h^2 = m / 93.5\,{\rm eV}
$$
Thus, the damping
scale for HDM (Hot Dark Matter) is of order the bend scale.
Alternatively, if the particle decouples sufficiently
early, its relative number density is boosted by annihilations,
so that the critical particle  mass 
to make $\Omega_m=1$ can be boosted to
around 1--10~keV (Warm Dark Matter).
The existence of galaxies at $z\simeq 6$ tells us that the
coherence scale must have been below about 100~kpc, so
WDM is close to being ruled out.
A similar constraint is obtained from small-scale
structure in the Lyman-alpha forest (Narayanan et al. 2000):
$m > 0.75$~keV.

A more interesting (and probably practically relevant) case
is when the dark matter is a mixture of hot and cold
components. The free-streaming length for the hot
component can therefore be very large, but within
range of observations. The dispersal of HDM fluctuations
reduces the CDM growth rate on all scales below
$L_{\rm free-stream}$ -- or, relative to small
scales, there is an enhancement in large-scale power.

{\bf (3) Acoustic horizon length.}
The horizon at matter-radiation equality also enters in
the properties of the baryon component. Since the sound speed
is of order $c$, the largest scales that can undergo a single
acoustic oscillation are of order the horizon. 
The transfer function for a pure baryon universe shows
large modulations, reflecting the number of oscillations
that have been completed before the universe becomes
matter dominated and the pressure support drops.
The lack of such large modulations in real data is
one of the most generic reasons for believing in 
collisionless dark matter. Acoustic oscillations
persist even when baryons are subdominant, however, and
can be detectable as lower-level modulations in the transfer
function (e.g. Goldberg \& Strauss 1998; Meiksin et al. 1999).

{\bf (4) Silk damping length.}
Acoustic oscillations are also damped on small scales, where
the process is called Silk damping: the
mean free path of photons due to scattering by the plasma
is non-zero, and so radiation can diffuse out of a
perturbation, convecting the plasma with it. 
The typical distance of a random walk in terms
of the diffusion coefficient $D$ is $x\simeq\sqrt{Dt}$,
which gives a damping length of
$$
\lambda_{\japsub S}\simeq\sqrt{\lambda D_{\japsub H}},
$$
the geometric mean of the horizon size and the
mean free path.
Since $\lambda=1/(n\sigma_{\japsub T}) = 44.3 (1+z)^{-3}(\Omega_{b}h^2)^{-1}$
proper Gpc, we obtain a comoving damping length of
$$
\lambda_{\japsub S}=16.3\, (1+z)^{-5/4} (\Omega_{b}^2\Omega_m h^6)^{-1/4}\; {\rm Gpc}.
$$
This becomes close to the horizon length by the time of last
scattering, $1+z\simeq 1100$.
The resulting damping effect
can be seen in figure~\lastfig\ at $k\sim 10 k_{\japsub H}$.

\sssec{Fitting formulae}
It is invaluable in practice to have some
accurate analytic formulae that fit the
numerical results for transfer functions.
We give below results for some common models of particular interest
(illustrated in figure~\lastfig, along with other
cases where a fitting formula is impractical).
For the models with collisionless dark matter,
$\Omega_{b}\ll\Omega_m$ is assumed, so that all lengths
scale with the horizon size at matter--radiation equality,
leading to the definition 
$$
q\equiv {k\over \Omega h^2 {\rm Mpc}^{-1}}.
$$
We consider the following cases: (1) Adiabatic CDM; (2) Adiabatic
massive neutrinos (1 massive, 2 massless); (3) Isocurvature
CDM; these expressions come
from Bardeen et al. (1986; BBKS).
Since the characteristic length-scale  in the transfer function
depends on the horizon size at matter--radiation equality,
the temperature of the CMB enters. In the above formulae,
it is assumed to be exactly 2.7 K; for other values, the
characteristic wavenumbers scale $\propto T^{-2}$.
For these purposes massless neutrinos count as radiation,
and three species of these contribute a total density
that is 0.68 that of the photons.
$$
\eqalign{
{\rm (1)}\quad T_k&={\ln(1+2.34q)\over 2.34 q}\left[1+3.89q+(16.1q)^2 +(5.46q)^3
  +(6.71q)^4\right]^{-1/4}\cr
{\rm (2)}\quad T_k&=\exp(-3.9q -2.1 q^2)\cr
{\rm (3)}\quad T_k&=(5.6 q)^2\, \left(1+ 
\left[15.0q + (0.9q)^{3/2} + (5.6q)^2\right]^{1.24}\right)^{-1/1.24}\cr
}
$$
The case of \key{mixed dark matter} (MDM:\idx{MDM} a mixture of massive
neutrinos and CDM) is more complex. Ma (1996) gives the following expression:
$$
{T_{\japsub MDM}\over T_{\japsub CDM}} = 
\left[{1+(0.081x)^{1.630} + (0.040x)^{3.259} \over 1 + (2.080x_0)^{3.259}}
\right]^{\Omega_\nu^{1.05}/2},
$$
where $x\equiv k/\Gamma_\nu$, $\Gamma_\nu\equiv a^{1/2}\Omega_\nu h^2$
and $x_0$ is the value of $x$ at $a=1$. The scale-factor dependence is such that the
damping from neutrino free-streaming is less severe at high redshift,
but the spectrum is very nearly of constant shape for $z\ls 10$.
See Pogosyan \& Starobinsky (1995) for a more complicated
fit of higher accuracy.

These expressions are useful for work at a level of 10\% precision,
but increasingly it is necessary to do better. In particular,
these expressions do not include the weak oscillatory features that are
expected if the universe has  a significant baryon content.
Eisenstein \& Hu (1998) give an accurate (but long)
fitting formula that describes these wiggles for the CDM transfer function.
This was extended to cover MDM in Eisenstein \& Hu (1999).

\sec{Nonlinear evolution of cosmic structure}

The equations of motion are nonlinear, and
we have only solved them in the limit of linear perturbations.
We now discuss evolution beyond the linear regime, first 
for a few special density models, and then considering
full numerical solution of the equations of motion.

\ssec{The Zeldovich approximation}

Zeldovich (1970) invented a {\it kinematical\/} approach to the 
formation of structure. In this method, we work out the
initial displacement of particles and assume
that they continue to move in this initial direction.
Thus, we write for the proper coordinate of a given particle
$$
\boxit{
{\bf x}(t) = a(t) {\bf q} + b(t){\bf f(q)}.
}
$$
This looks like Hubble expansion with some
perturbation, which will become negligible
as $t\rightarrow 0$. The coordinates
$\bf q$ are therefore equal to the usual
comoving coordinates at $t=0$, and $b(t)$ is a function
that scales the time-independent displacement field ${\bf f(q)}$.
In fluid-mechanical terminology,
$\bf x$ is said to be the \key{Eulerian position},
and $\bf q$ the \key{Lagrangian position}.

To get the Eulerian density, we need the Jacobian
of the transformation between $\bf x$ and $\bf q$,
in which frame $\rho$ is constant.
This strain tensor is symmetric, provided
we assume that the density perturbation originated from a
growing mode. The displacement field is then irrotational, so that
we can write it in terms of a potential
$$
{\bf f}({\bf q}) = \del \psi({\bf q}) \quad\Rightarrow\quad {\partial f_i\over \partial q_j}
= {\partial^2 \psi\over \partial q_i\partial q_j}.
$$
The strain tensor is thus characterized by its three eigenvalues, and the
density becomes infinite when the most negative eigenvalue
reaches $-1$.

If we linearize the density relation, then the relation
to density perturbations is
$$
\delta= -{b\over a}{\bf\del\cdot f}.
$$
This is first-order \key{Lagrangian perturbation theory}, in contrast to the earlier
approach, which carried out perturbation theory in
Eulerian space (higher-order Lagrangian theory is discussed
by Bouchet et al. 1995).
When the density fluctuations are
small, a first-order treatment from either point of
view should give the same result. Since the
linearized density relation is $\delta=-(b/a){\bf\del\cdot f}$,
we can tell immediately that $[b(t)/a(t)]=D(t)$, where
$D(t)$ is the linear density growth law.
Without doing any more work, we therefore know that the first-order
form of Lagrangian perturbations must be
$$
{\bf x}(t) = a(t)[ {\bf q} + D(t) {\bf f(q)}],
$$
so that $b(t)=a(t)D(t)$.
The advantage of the Zeldovich approximation is
that it normally breaks down later than 
Eulerian linear theory -- i.e. first-order
Lagrangian perturbation theory can give results
comparable in accuracy to Eulerian theory with higher-order
terms included.
This method is therefore commonly used to set up
quasi-linear initial conditions for $N$-body
simulations, as discussed below.
The same arguments that we used earlier in discussing peculiar
velocities show that the growing-mode
comoving displacement field $\bf f$ is parallel to $\bf k$
for a given Fourier mode, so that
$$
{\bf f}_k = -i\, {\delta_k\over k^2} \; {\bf k}.
$$
Given the desired linear density mode amplitudes,
the corresponding displacement field can then be constructed.

\ssec{The spherical model}

An overdense sphere is a very useful nonlinear
model, as it behaves in exactly the
same way as a closed sub-universe.
The density perturbation need not be a uniform
sphere: any spherically symmetric perturbation
will  clearly evolve at a given radius in the same way
as a uniform sphere containing the same
amount of mass. In what follows, therefore, density
refers to the {\it mean\/} density inside a given sphere.
The equations of motion are the same as for the scale
factor, and we can therefore write down the cycloid
solution immediately. For a matter-dominated universe,
the relation between the proper radius of the sphere and time is
$$
\eqalign{
r&= A(1-\cos\theta)\cr
t&= B(\theta-\sin\theta),\cr
}
$$
and $A^3=GMB^2$, just from $\ddot r = -GM/r^2$.
Expanding these relations up to order $\theta^5$
gives $r(t)$ for small $t$:
$$
r\simeq{A\over 2} \left({6t\over B}\right)^{2/3}
  \left[ 1- {1\over 20} \left({6t\over B}\right)^{2/3} \right],
$$
and we can identify the density perturbation within the sphere:
$$
\delta\simeq {3\over 20}\left({6t\over B}\right)^{2/3}.
$$
This all agrees with what we knew already: at early times
the sphere expands with the $a\propto t^{2/3}$ Hubble flow
and density perturbations grow proportional to $a$.

We can now see how linear theory breaks down as the
perturbation evolves. There are three interesting epochs
in the final stages of its development, which we can
read directly from the above solutions. 
Here, to keep things simple, we compare only with linear theory for
an $\Omega=1$ background.

\japitem{(1)} \key{Turnround}. The sphere breaks away from the
general expansion and reaches a maximum radius
at $\theta=\pi$,  $t=\pi B$. At this point, the true density enhancement
with respect to the background is just $[A(6t/B)^{2/3}/2]^3/r^3
=9\pi^2/16\simeq 5.55$. By comparison, 
extrapolation of linear $\delta\propto t^{2/3}$ theory predicts
$\delta_{\rm lin}=(3/20)(6\pi)^{2/3}\simeq 1.06$.

\japitem{(2)} \key{Collapse}. If only gravity operates, then the
sphere will collapse to a singularity at $\theta=2\pi$.
This occurs when $\delta_{\rm lin}=(3/20)(12\pi)^{2/3}\simeq 1.69$.

\japitem{(3)} \key{Virialization}. 
Clearly, collapse will never occur in practice; dissipative
physics will eventually intervene and
convert the kinetic energy of collapse into random
motions. How dense will the resulting body be?
Consider the time at which the sphere has collapsed by a
factor 2 from maximum expansion. At this point, it has
kinetic energy  $K$ related to potential energy $V$
by $V=-2K$. This is the condition for equilibrium, according
to the \key{virial theorem}.
For this reason, many workers take this epoch as
indicating the sort of density contrast to be expected
as the endpoint of gravitational collapse.
This occurs at $\theta=3\pi/2$, and the corresponding
density enhancement is $(9\pi+6)^2/8\simeq 147$,
with $\delta_{\rm lin}\simeq 1.58$.
Some authors prefer to assume that this virialized size is
eventually achieved only at collapse, in which case the contrast
becomes $(6\pi)^2/2\simeq 178$.

\endjapitem
These calculations are the basis for a common `rule of thumb',
whereby one assumes that linear theory applies until
$\delta_{\rm lin}$ is equal to some $\delta_c$ a little greater
than unity, at which point virialization is deemed to
have occurred. 
Although the above only applies for $\Omega=1$,
analogous results can be worked out from the
full $\delta_{\rm lin}(z,\Omega)$ and $t(z,\Omega)$
relations; $\delta_{\rm lin}\simeq 1$ is a good
criterion for collapse for any value of
$\Omega$ likely to be of practical relevance.
The full density contrast at virialization may be approximated by
$$
1+\delta_{\rm vir}\simeq 178\, \Omega_m^{-0.7}(t_{\rm vir})
$$
(although open models show a slightly stronger
dependence on $\Omega_m$ than
flat $\Lambda$-dominated models;
Eke et al. 1996).
The faster expansion of low-density universes
means that, by the time a perturbation has turned
round and collapsed to its final radius, a larger
density contrast has been produced.
For real non-spherical systems, it is not clear that
this distinction is meaningful, and in practice
a density contrast of around 200 is used to define
the \key{virial radius} that marks the boundary of an object.

\ssec{N-body models}

The exact evolution of the density field is usually
performed by means of an \key{N-body simulation}, in which
the density field is represented by the sum of a
set of fictitious discrete particles. The equations of motion
for each particle depend on solving for the gravitational
field due to all the other particles, finding the
change in particle positions and velocities over
some small time step, moving and accelerating
the particles, and finally re-calculating the
gravitational field to start a new iteration.
Using comoving units for length and velocity
(${\bf v}=a{\bf u}$), we have previously seen the
equation of motion
$$
{d\over dt}{\bf u}=-2{\dot a\over a}{\bf u} - {1\over a^2}{\bf\del}\Phi,
$$
where $\Phi$ is the Newtonian gravitational potential
due to density perturbations.
The time derivative is already in the required form of
the convective time derivative observed by a particle,
rather than the partial $\partial /\partial t$.
If we change time variable from $t$ to $a$, this becomes
$$
{d\over d\, \ln a}(a^2{\bf u}) ={a\over H}\,{\bf g}
={G\over aH}\; \sum_i m_i\; {{\bf x}_i-{\bf x}\over |{\bf x}_i-{\bf x}|^3}.
$$
Here, the gravitational acceleration has been written exactly
by summing over all particles, but this becomes prohibitive
for very large numbers of particles. Since the problem is
to solve Poisson's equation, a faster approach is to use
Fourier methods, since this allows the use of the
fast Fourier transform (\key{FFT}) algorithm
(see chapter 13 of Press et al. 1992). 
If the density perturbation field
(not assumed small) is expressed as $\delta=\sum \delta_k\exp(-i{\bf k\cdot x})$,
then Poisson's equation becomes $-k^2\Phi_k=4\pi G a^2\bar\rho\, \delta_k$, and
the required $k$-space components of ${\bf\del}\Phi$ are just
$$
({\bf\del}\Phi)_k=-i\Phi_k{\bf k}={-i 4\pi G a^2 \bar\rho\over k^2}\;\delta_k\, {\bf k}.
$$
If we finally eliminate matter
density in terms of $\Omega_m$, the equation of motion for a
given particle is
$$
{d\over d\, \ln a}(a^2{\bf u}) =\sum {\bf F}_k\exp(-i{\bf k\cdot x}),
\quad\quad
{\bf F}_k =-i{\bf k}\;{3\Omega_m H a^2\over 2 k^2}\; \delta_k.
$$

This can be expressed more neatly by defining dimensionless
units that incorporate the
the side of the box, $L$:
$$
\eqalign{
{\bf X}&= {\bf x}/L \cr
{\bf U}&= \delta {\bf v}/(HLa)= {\bf u}/HL.\cr
}
$$
For $N$ particles, the density is $\rho=Nm/(aL)^3$,
so the mass of the particles and the gravitational
constant can be eliminated and the
equation of motion can be cast in an attractively
dimensionless form:
$$
\boxit{
{d\over d\, \ln a}[f(a){\bf U}]
={3\over 8\pi}\,\Omega_m(a) f(a)\; {1\over N} \sum_i  {{\bf X_i-X}\over |{\bf X_i-X}|^3}.
}
$$
The function $f(a)$ is proportional to $a^2H(a)$, and has an
arbitrary normalization -- e.g. unity at the initial epoch.

Particles are now moved according to $d{\bf x}={\bf u}\; dt$,
which becomes
$$
d{\bf X}={\bf U}\; d\ln a.
$$
It only remains to set up the initial conditions;
this is easy to do if the  initial epoch is
at high enough redshift that $\Omega_m=1$, since then
${\bf U}\propto a$ 
and the earlier discussion of Lagrangian perturbations
shows that velocities and the initial displacements are related by
$$
\Delta{\bf X}=\bf U.
$$

The simplest $N$-body algorithm for solving the
equations of motion is the \key{particle--mesh (PM) code}.
This averages the density field onto a grid and
uses the FFT algorithm both to perform the transformation
of density and to perform the (three) inverse transforms to
obtain the real-space force components from their $k$-space
counterparts
(see Hockney \& Eastwood 1988; Efstathiou et al. 1985).
The resolution of
a PM code is clearly limited to about the size of
the mesh. To do better, one can use
a \key{particle--particle--particle--mesh (P$\bf ^3$M) code},
also discussed by the above authors. Here, the
direct forces are evaluated between particles in the
nearby cells, with the grid estimate being used
only for particles in more distant cells. 
An alternative approach is to use \key{adaptive mesh codes},
in which high-density regions are re-gridded to use a finer
mesh (e.g. Kravtsov, Klypin \& Khokhlov 1997).
A similar effect, although without the use of a mesh, is achieved
by \key{tree codes} (e.g. Hernquist, Bouchet \& Suto 1991).

In practice, however, the increase in resolution gained from
these methods is limited to a factor of $\ls 10$. This is
because each particle in a cosmological $N$-body
simulation in fact stands for a large number of less
massive particles. Close encounters of these spuriously
large particles can lead to wide-angle scattering, whereas
the true physical systems are completely collisionless.
To prevent collisions, the forces must be \key{softened}, i.e. set
to a constant below some critical separation, rather than
rising as $1/r^2$. If there are already few particles
per PM cell, the softening must be some significant
fraction of the cell size, so there is a limit to the
gain over pure PM.

\japfig{0}{78}{396}{255}{0.8}{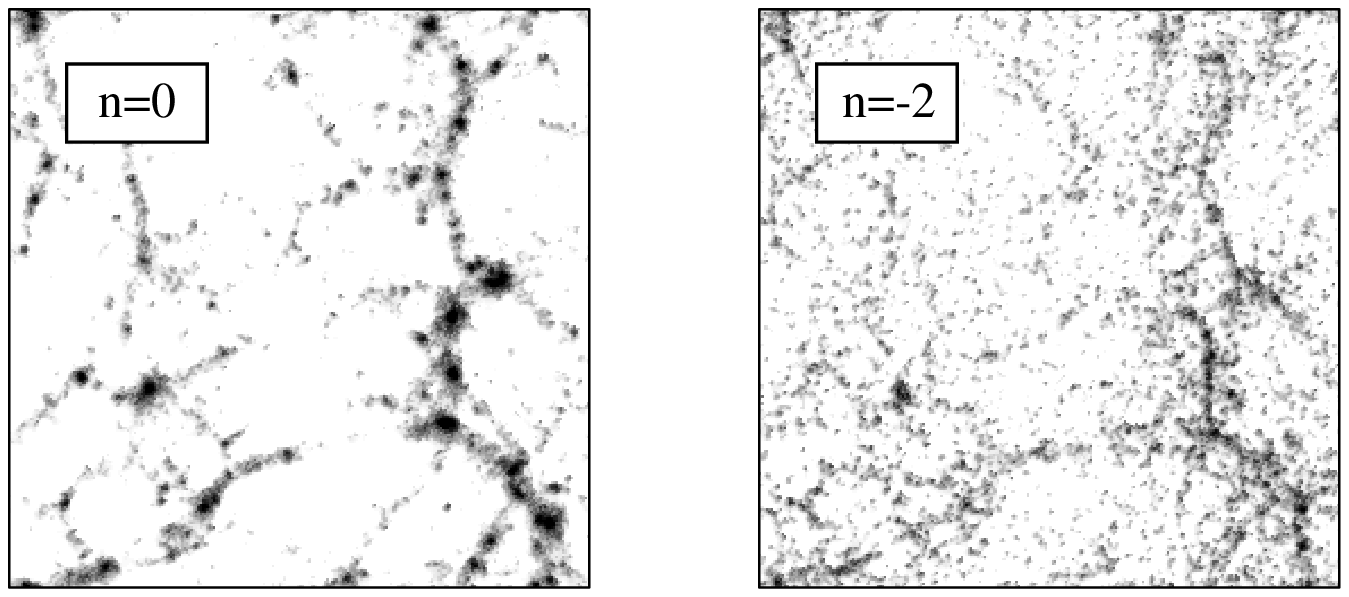}
{Slices through $N$-body simulations of different
power spectra, using the same set of random phases
for the modes in both cases.
The slices are $1/15$ of the box in
thickness, and density from 0.5 to 50 is log encoded.
The box-scale power is the same in both cases, and produces
much the same large-scale filamentary structure. However, the
$n=0$ spectrum has much more small-scale power, and this
manifests itself as far stronger clumping within the overall skeleton of the structure.}

Despite these caveats,  the results of $N$-body dynamics
paint an impressive picture of the large-scale mass 
distribution. Consider figure \lastfig, which shows slices
through the computational density field for two
particular sets of initial conditions, with different relative
amplitudes of long and short wavelengths, but with the
same amplitude for the modes with wavelengths equal to the side
of the computational box. Although the
small-scale `lumpiness' is different, both display
a similar large-scale network of filaments and voids -- bearing
a striking resemblance to the features seen in reality.

The state of the art in these calculations now routinely involves
$10^8$ to $10^9$ particles, covering box sizes from
the minimum necessary so that the box-scale modes
do not saturate ($\sim 100\mpcoh$) to effectively
the entire observable universe (e.g. Evrard et al. 2002).
The resolution available in the smaller boxes is
sufficient that the nonlinear evolution of collisionless
mass distributions is now effectively a solved
problem, and nonlinear clustering statistics for model
universes of practical interest can be measured to
a few \% precision (e.g. Jenkins et al. 1998).
Further improvements in these sort of calculations are unlikely
to be of practical importance, because of the need to include
the evolution of the baryonic component, which makes up
around 20\% of the total matter density. The history of
the gas is immensely complex, since it is strongly influenced
by \key{feedback} of energy from the stars that form within it.
The limitation of our modelling of such processes lies not
so much in simple numerical aspects such as resolution, but
in the simplifying assumptions used to treat processes that occur on scales
very far below the resolution of any simulation.
See e.g. Katz, Weinberg \& Hernquist (1996); Pearce et al. (2001).

\sec{Statistics of cosmological density fields}

Having discussed the main elements of the theory of cosmological
structure formation, we now turn to the statistical
treatment of data -- which is how theory and observation will
be confronted.
The density perturbation field, $\delta$,
inhabits a universe that is isotropic and homogeneous
in its large-scale properties, suggesting that
the statistical properties of $\delta$
should also be homogeneous.
This statement sounds contradictory, and yet it makes
perfect sense if there exists an \key{ensemble of universes}.
The concept of an ensemble is used every time we apply probability
theory to an event such as tossing a coin: we imagine an
infinite sequence of repeated trials, half of which
result in heads, half in tails. 
The analogy of coin tossing in cosmology is that the density at a given
point in space will have different values in each member
of the ensemble, with some overall variance $\langle \delta^2 \rangle$
between members of the ensemble. Statistical homogeneity
of the $\delta$ field then means that this variance must be independent of position.
The actual field found
in a given member of the ensemble is a \key{realization} of the
statistical process.

There are two problems with this line of argument:
(i) we have no evidence that the ensemble exists;
(ii) in any case, we only get to observe one realization,
so how is the variance $\langle \delta^2 \rangle$ to be measured?
The first objection applies to coin tossing, and
may be evaded if we understand the physics that
generates the statistical process -- we only
need to {\it imagine\/} tossing the coin many times,
and we do not actually need to perform the exercise.
The best that can be done in answering the second
objection is to look at widely separated parts of
space, since the $\delta$ fields there should be
causally unconnected; this is therefore as good as taking
measurements from two different member of the ensemble.
In other words, if we measure the variance $\langle \delta^2 \rangle$
by averaging over a sufficiently large volume, the results
would be expected to approach the true ensemble variance,
and the averaging operator $\langle \cdots \rangle$ is often
used without being specific about which kind of average is intended.
Fields that satisfy this property, whereby
$$
\rm{volume\ average} \quad\leftrightarrow\quad {\rm ensemble\ average}
$$
are termed \key{ergodic}.
Giving a formal proof of ergodicity for a random process
is not always easy (Adler 1981); in cosmology it is perhaps best
regarded as a common-sense axiom.

\ssec{Fourier analysis of density fluctuations}

It is often convenient to consider building up a general
field by the superposition of many modes.
For a flat comoving geometry, the natural tool for
achieving this is via Fourier analysis.
For other models, plane waves are not a complete
set and one should use instead the eigenfunctions
of the wave equation in a curved space. Normally
this complication is neglected: even in an open
universe, the difference only matters on scales
of order the present-day horizon.

How do we make a Fourier expansion of the
density field in an infinite universe? If the field were periodic
within some box of side $L$, then we would just
have a sum over wave modes:
$$
F({\bf x}) = \sum F_{\bf k} e^{-i{\bf k\cdot x}}.
$$
The requirement of periodicity restricts the allowed wavenumbers
to \key{harmonic boundary conditions}
$$
k_x=n\;{2\pi\over L},\;\quad n=1,2\cdots,
$$
with similar expressions for $k_y$ and $k_z$.
Now, if we let the box become arbitrarily large, then the
sum will go over to an integral that incorporates the
density of states in $k$-space, exactly as in
statistical mechanics; this is how the general idea of the 
Fourier transform is derived.
The Fourier relations in $n$ dimensions are thus
$$
\boxit{
\eqalign{
F(x)&=\left({ L \over 2\pi}\right)^n \int F_k(k) \exp \bigl(-i{\bf
k\cdot x}\bigr)\; d^n k\cr
F_k(k)&=\left({1\over  L }\right)^n \int F(x) \exp
\bigl(i{\bf k\cdot x}\bigr)\; d^n x.\cr }
}
$$
One advantage of this particular Fourier convention
is that the definition of convolution involves just a simple
volume average, with no gratuitous factors of $(2\pi)^{-1/2}$:
$$f*g\equiv {1\over L^n}\int f({\bf x-y}) g({\bf y}) d^ny.$$ 
Although one can make all the manipulations on density fields that follow
using either the integral or sum formulations, it is
usually easier to use the sum. This saves having to introduce
$\delta$-functions in $k$-space.
For example, if we have $f=\sum f_k \exp(-ikx)$, the obvious
way to extract $f_k$ is via $f_k=(1/ L )\int f\exp(ikx)\; dx$:
because of the harmonic boundary conditions, all oscillatory
terms in the sum integrate to zero, leaving only $f_k$ to be
integrated from 0 to $ L $. There is less chance of
committing errors of factors of $2\pi$ in this way than
considering $f=( L /2\pi)\int f_k \exp(-ikx)\; dk$ and then
using $\int\exp[i(k-K)x]\; dx=2\pi\delta_{\japsub D}(k-K)$.

\sssec{Correlation functions and power spectra}
As an immediate example of the Fourier
machinery in action, consider the important quantity
$$
\boxit{
\xi({\bf r}) \equiv \left\langle \delta({\bf x})\delta({\bf x+ r})\right\rangle,
}
$$
which is the autocorrelation function of the density field --
usually referred to simply as the \key{correlation function}.
The angle brackets indicate an averaging over the normalization
volume $V$. Now express $\delta$ as a sum and note that $\delta$ is real,
so that we can replace one of the two $\delta$'s by
its complex conjugate, obtaining
$$
\xi = \left\langle \sum_{\bf k}\sum_{\bf k'}\delta_{\bf k}
  \delta^*_{\bf k'} e^{i{\bf  (k'-k)\cdot x}} e^{-i{\bf k\cdot r}}\right\rangle.
$$
Alternatively, this sum can be obtained 
without replacing $\langle \delta \delta \rangle$
by $\langle \delta \delta^* \rangle$, from the 
relation between modes with opposite wavevectors
that holds for any real field:
$\delta_{\bf k}(-{\bf k})=\delta^*_{\bf k}({\bf k})$.
Now, by the periodic boundary conditions, all the cross terms
with $\bf k'\ne k$ average to zero. Expressing the
remaining sum as an integral, we have
$$
\boxit{
\xi({\bf  r})=
{V \over (2\pi)^3}\int|\delta_{\bf k}|^2 e^{-i{\bf  k\cdot r}} d^3 k.
}
$$
In short, the correlation function is the Fourier transform of
the \key{power spectrum}. 
This relation has been obtained by volume averaging, so it applies
to the specific mode amplitudes and correlation function measured in any
given realization of the density field. Taking ensemble averages
of each side, the relation clearly also holds for the ensemble
average power and correlations -- which are really the quantities
that cosmological studies aim to measure.
We shall hereafter often use the 
alternative notation 
$$
\boxit{
P(k)\equiv\langle|\delta_k|^2\rangle
}
$$ 
for the ensemble-average power
(although this only applies for a Fourier series with discrete
modes).
The distinction between the ensemble average and the actual power
measured in a realization is clarified below in the section on Gaussian
fields.

In an isotropic universe, the density perturbation
spectrum cannot contain a preferred direction,
and so we must have an \key{isotropic power spectrum}:
$\langle |\delta_{\bf k}|^2({\bf k})\rangle = |\delta_k|^2(k)$.
The angular part of the $k$-space integral can therefore
be performed immediately: introduce spherical polars
with the polar axis along $\bf k$, and use the reality
of $\xi$ so that $e^{-i{\bf k\cdot x}}\rightarrow
\cos(kr\cos\theta)$.
In three dimensions, this yields
$$
\xi(r) = {V\over (2\pi)^3}\int P(k)\, {\sin kr\over kr}\, 4\pi k^2\; dk.
$$
The 2D analogue of this formula is
$$
\xi(r) = {A\over (2\pi)^2}\int P(k)\, J_0(kr)\, 2\pi k\; dk.
$$

We shall usually express
the power spectrum in dimensionless form, as the variance per $\ln k$
($\Delta^2(k) =d \langle \delta^2 \rangle/d\ln k \propto k^3 P[k]$):
$$
\boxit{
\Delta^2(k)\equiv {V\over (2\pi)^3} \, 4\pi k^3\, P(k)
={2\over \pi}k^3\int_0^\infty\xi(r)\,
{\sin kr\over kr}\, r^2\, dr.
}
$$
This gives a more easily visualizable meaning to the power
spectrum than does the quantity $V P(k)$, which has
dimensions of volume: $\Delta^2(k)=1$ means that there
are order-unity density fluctuations from modes
in the logarithmic bin around wavenumber $k$.
$\Delta^2(k)$ is therefore the natural choice for
a Fourier-space counterpart to the dimensionless quantity $\xi(r)$.

\sssec{Power-law spectra}
The above shows that the power spectrum is a central quantity
in cosmology, but how can we predict its functional form?
For decades, this was thought to be impossible, and so
a minimal set of assumptions was investigated.
In the absence of a physical theory, we should not assume
that the spectrum contains any preferred length scale,
otherwise we should then be compelled to explain this feature.
Consequently, the spectrum must be a featureless power law:
$$
\boxit{
\left\langle|\delta_k|^2\right\rangle \propto k^n
}
$$
The index $n$ governs the balance between large-
and small-scale power.
The meaning of different values of $n$ can be seen by
imagining the results of filtering the density
field by passing over it a box of some characteristic
comoving size $x$ and averaging the density over the box. 
This will filter out waves with $k\gs 1/x$,
leaving a variance 
\smash{$\langle\delta^2\rangle \propto \int_0^{1/x}
k^n 4\pi k^2dk\propto x^{-(n+3)}$}. 
Hence, in terms of a mass $M\propto x^3$, we have
$$
\delta_{\rm rms} \propto M^{-(n+3)/6}.
$$

Similarly, a power-law spectrum implies a power-law
correlation function.
If $\xi(r)=(r/r_0)^{-\gamma}$, with $\gamma=n+3$,
the corresponding 3D power spectrum is
$$
\Delta^2(k)={2\over\pi}\,(kr_0)^{\gamma}\, \Gamma(2-\gamma) \,
   \sin {(2-\gamma)\pi\over 2}
\equiv \beta (kr_0)^\gamma
$$
($=0.903 (kr_0)^{1.8}$ if $\gamma=1.8$).
This expression is only valid for $n<0$ ($\gamma<3$);
for larger values of $n$, $\xi$ must become
negative at large $r$ (because $P(0)$ must vanish,
implying $\int_0^\infty \xi(r)\, r^2\, dr=0$).
A cutoff in the spectrum at large $k$ is needed
to obtain physically sensible results.

What general constraints can we set on the value of $n$?
Asymptotic homogeneity clearly requires $n>-3$.
An upper limit of $n<4$ comes from an argument due to
Zeldovich.
Suppose we begin with a totally
uniform matter distribution and then group it
into discrete chunks as uniformly as possible.
It can be shown  that conservation of momentum
in this process means that we cannot create
a power spectrum that goes to zero at small
wavelengths more rapidly than $\delta_k\propto k^2$.
Thus, discreteness of matter produces
the \key{minimal spectrum}, $n=4$.
More plausible alternatives lie between these extremes.
The value $n=0$ corresponds to \key{white noise},
the same power at all wavelengths.
This is also known as the Poisson
power spectrum, because it corresponds to fluctuations
between different cells that scale as $1/\sqrt{M_{\rm cell}}$.
A density field created by throwing
down a large number of point masses at random would
therefore consist of white noise.
Particles placed at random within cells, one per cell,
create an $n=2$ spectrum on large scales.

Practical spectra in cosmology, conversely, 
have negative effective values of $n$ over a large
range of wavenumber. 
For many years, the data on the galaxy correlation
function were consistent with a single power law:
$$
\boxit{
\xi_g(r) \simeq \left({r\over 5 \mpcoh}\right)^{-1.8}\quad\quad
\left(1 \ls \xi \ls 10^4\right);
}
$$
see Peebles (1980), Davis \& Peebles (1983). This corresponds to $n\simeq -1.2$.
By contrast with the above examples, large-scale
structure is `real', rather than reflecting the low-$k$
Fourier coefficients of some small-scale process.

\sssec{The Zeldovich spectrum}
Most important of all is the \key{scale-invariant spectrum}, 
which corresponds to the value $n=1$, i.e.
$\Delta^2\propto k^4$. To see how the name arises,
consider a perturbation $\Phi$ in the gravitational potential:
$$
\nabla^2\Phi= 4\pi G\rho_0\delta
\quad\Rightarrow\quad \Phi_k = -4\pi G\rho_0\delta_k/k^2.
$$
The two powers of $k$ pulled down by $\nabla^2$ mean
that, if $\Delta^2\propto k^4$ for the power spectrum of 
density fluctuations, then $\Delta^2_\Phi$ is a constant.
Since potential perturbations govern the flatness
of spacetime, this says that the scale-invariant
spectrum corresponds to a metric that is
a \key{fractal}: spacetime has the same degree of
`wrinkliness' on each resolution scale.
The total curvature fluctuations diverge, but only
logarithmically at either extreme of wavelength.

Another way of looking at this spectrum is in terms
of perturbation growth balancing the
scale dependence of $\delta$: $\delta\propto x^{-(n+3)/2}$.
We know that $\delta$ viewed on a given
comoving scale will increase with 
the size of the horizon:
$\delta\propto r_{\japsub H}^2$. 
At an arbitrary time, though, the only natural length
provided by the universe (in the absence of non-gravitational
effects) is the horizon itself:
$$
\delta(r_{\japsub H})\propto r_{\japsub H}^2 r_{\japsub H}^{-(n+3)/2} =
 r_{\japsub H}^{-(n-1)/2}.
$$
Thus, if $n=1$, the growth of both $r_{\japsub H}$ and $\delta$ with time
cancels out so that the universe always looks the same
when viewed on the scale of the horizon; such a universe
is self-similar in the sense of always appearing the same
under the magnification of cosmological expansion.
This spectrum is often known as the \key{Zeldovich spectrum}
(sometimes hyphenated with Harrison and Peebles, who invented it independently).

The generic nature of the scale-invariant spectrum makes it
difficult to use as a test, since many theories may be expected
to have a chance of yielding something like a fractal
spacetime. The interesting aspect to focus on is therefore where theory
predicts deviations from this rule. Inflation is an interesting
case, since the horizon-scale amplitude is expected to change
logarithmically with scale in simple models (Hawking 1982):
$$
\delta_{\japsub H}\propto[-\ln(k r_{\japsub H}^{\rm infl})]^\alpha,
$$
where $\alpha$ is a constant of order unity that depends on the
inflationary potential ($\alpha=2$ for $V(\phi)=m^2\phi^2/2$, for example).
Since the proper horizon scale at the end of inflation cannot be
infinitely small ($a^{\rm infl} r_{\japsub H}^{\rm infl} > \ell_{\rm Planck}$),
we see that $\delta_{\japsub H}$ should vary by a small
but definite amount over the range of scales that can be
probed by the CMB and large-scale structure (a change by a 
factor 1.07 between $k=0.1 \hompc$ and $10^{-3}\hompc$,
taking $\alpha=1$,  $r_{\japsub H}^{\rm infl} = \ell_{\rm Planck}/a^{\rm infl}$
and $a^{\rm infl}\simeq 2.73/T_{\rm Planck}$, so that $r_{\japsub H}^{\rm infl} =
10^{-3.08}{\rm m}$).
This is pretty close to scale-invariance, but shows that small
amounts of \key{tilt} are potentially observable if sufficiently
accurate observations can be made.

\ssec{CDM models for structure formation}

The elements discussed so far assemble into the $\Lambda$CDM cosmological
model, which is the simplest possibility that is
consistent with current evidence.
The overall matter power spectrum is written dimensionlessly
as the logarithmic contribution to the fractional
density variance, $\sigma^2$:
$$
\Delta^2(k)={d\sigma^2\over d\ln k}  \propto k^3 |\delta_k|^2 \propto k^{3+n},
$$
which undergoes linear growth
$$
\delta_k(a) = \delta_k(a_0)\; \left[{D(a)\over D(a_0)}\right] \; T_k,
$$
where the linear growth law is $D(a)=a\, g[\Omega(a)]$
in the matter era,
and the growth suppression factor for a
density parameter $\Omega_m\ne 1$ is
$$
\boxit{
g(\Omega_m)  \simeq 
\frac{5}{2}\Omega_m\left[\Omega_m^{4/7}-\Omega_v+
 (1+\half\Omega_m)(1+\frac{1}{70}\Omega_v)\right]^{-1}.
}
$$
The transfer function $T_k$ depends on the dark-matter
content as discussed earlier, in particular displaying a horizon-scale break
at $k \simeq 0.06 (\Omega_m h)^{-1} \hompc$.
Weak oscillatory features are also expected if the universe has 
a significant baryon content.
Eisenstein \& Hu (1998) give an accurate
fitting formula that describes these wiggles.
This detailed fit of the CDM spectrum is to be preferred to
the older notation in which the spectrum was described by
the zero-baryon form, but with an effective value of
$\Omega_m h$ that allowed for the main effects of the baryon content:
$$
(\Omega_m h)_{\rm eff} \equiv \Gamma = \Omega_m h\, \exp\bigl[-\Omega_{b}(1+\sqrt{2h}/\Omega)\bigr]
$$
(Sugiyama 1995).

\sec{Comparison with 2dFGRS data}

\ssec{Survey overview}

The largest dataset for which a thorough comparison with the
above picture has been made is the
2dF Galaxy Redshift Survey (2dFGRS).
This survey was designed around the 2dF multi-fibre spectrograph on the
Anglo-Australian Telescope, which is capable of observing up to 400
objects simultaneously over a 2~degree diameter field of view. 
For details of
the instrument and its performance 
see {\tt http://www.aao.gov.au/2df/}, and also
Lewis et~al.\ (2002).
The source catalogue for the survey is a revised and extended version of
the APM galaxy catalogue (Maddox et~al.\ 1990a,b,c); this
includes over 5~million galaxies down to $b_{\japsub J}=20.5$ in both
north and south Galactic hemispheres over a region of almost
$10^4\, {\rm deg}^2$.
The $b_{\japsub J}$
magnitude system 
is related to the Johnson--Cousins system by $b_{\japsub J} = B -0.304(B-V)$,
where the colour term is estimated from comparison with the SDSS Early
Data Release (Stoughton et al. 2002).

\japfigbasic{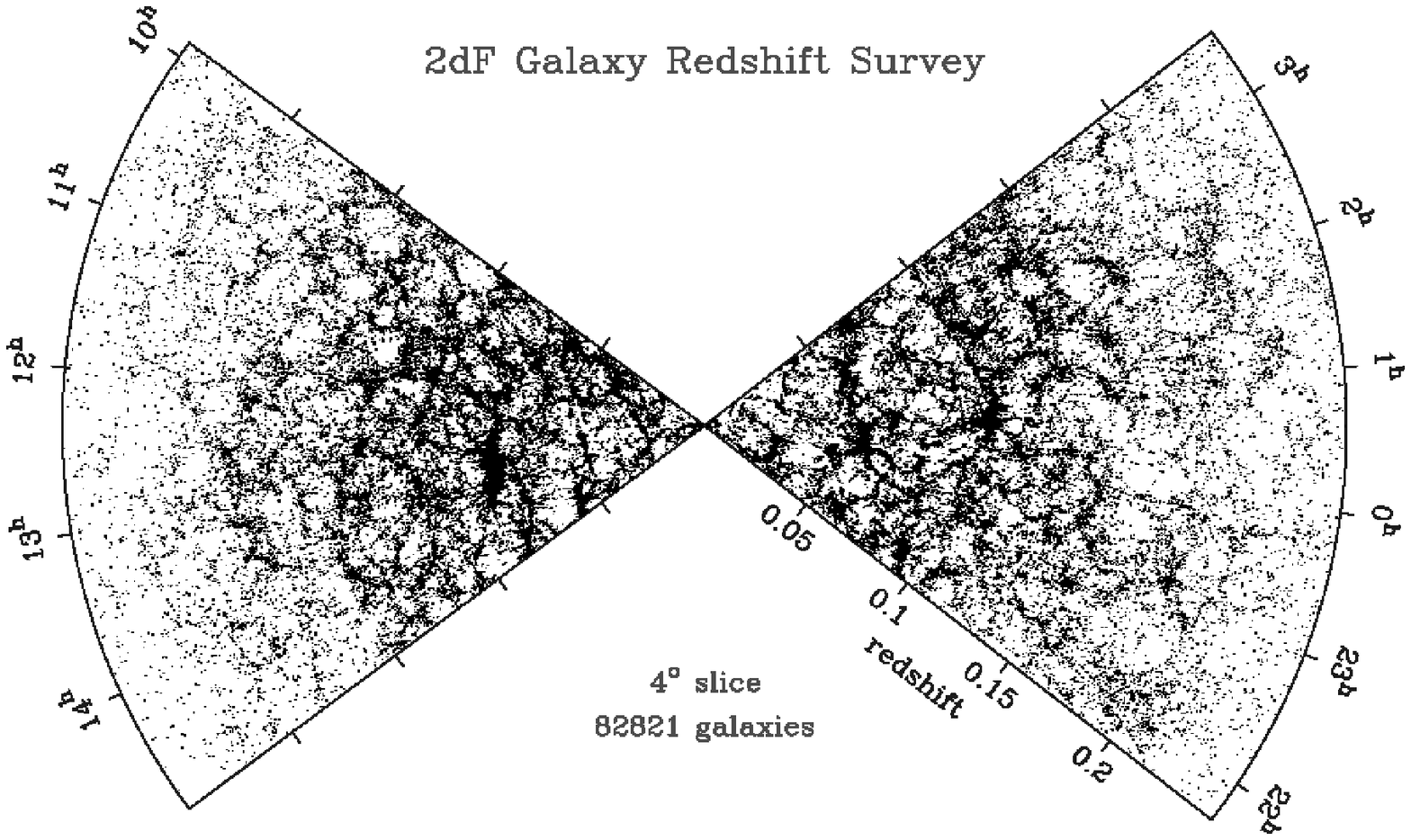}{1.0}
{The distribution of galaxies in part of the 2dFGRS:
slices $4^\circ$ thick, centred at declination
$-2.5^\circ$ in the NGP and $-27.5^\circ$ in the SGP.
This magnificently detailed image of large-scale structure
provides the basis for measuring the shape of the
primordial fluctuation spectrum and hence constraining the
matter content of the universe.}

The 2dFGRS geometry consists of two contiguous
declination strips, plus 100 random 2-degree fields. One strip is in the
southern Galactic hemisphere and covers approximately
75$^\circ$$\times$15$^\circ$ centred close to the SGP at
($\alpha, \delta$)=($01^h$,$-30^\circ$); the other strip is in the northern
Galactic hemisphere and covers $75^\circ \times 7.5^\circ$ centred at
($\alpha, \delta$)=($12.5^h$,$+0^\circ$). The 100 random fields are spread
uniformly over the 7000~deg$^2$ region of the APM catalogue in the
southern Galactic hemisphere. 
The sample is limited to be brighter than an extinction-corrected
magnitude of $b_{\japsub J}=19.45$ (using the extinction maps of Schlegel et~al.\
1998). This limit gives a good match between the density on the sky of
galaxies and 2dF fibres.

After an extensive period of commissioning of the 2dF instrument,
2dFGRS observing began in earnest in May 1997, and terminated
in April 2002. 
In total, observations were made of 899 fields,
yielding redshifts and identifications for 232,529 galaxies, 13976 stars
and 172 QSOs, at an overall completeness of 93\%. 
The galaxy redshifts are assigned a quality flag from 1 to 5,
where the probability of error is highest at low $Q$. Most analyses
are restricted to $Q\ge 3$ galaxies, of which there are currently
221,496.
An interim data release took place in July 2001,
consisting of approximately 100,000 galaxies (see Colless et al. 2001
for details). A public release of the full photometric and spectroscopic
database is scheduled for July 2003.
The completed 2dFGRS yields a striking
view of the galaxy distribution over large cosmological volumes.
This is illustrated in
figure~\nextfig, which shows the projection of a subset of
the galaxies in the northern and southern strips onto $(\alpha,z)$
slices. This picture is the culmination of decades of effort in 
the investigation of large-scale structure, and we are
fortunate to have this detailed view for the first time.

\japfigbasic{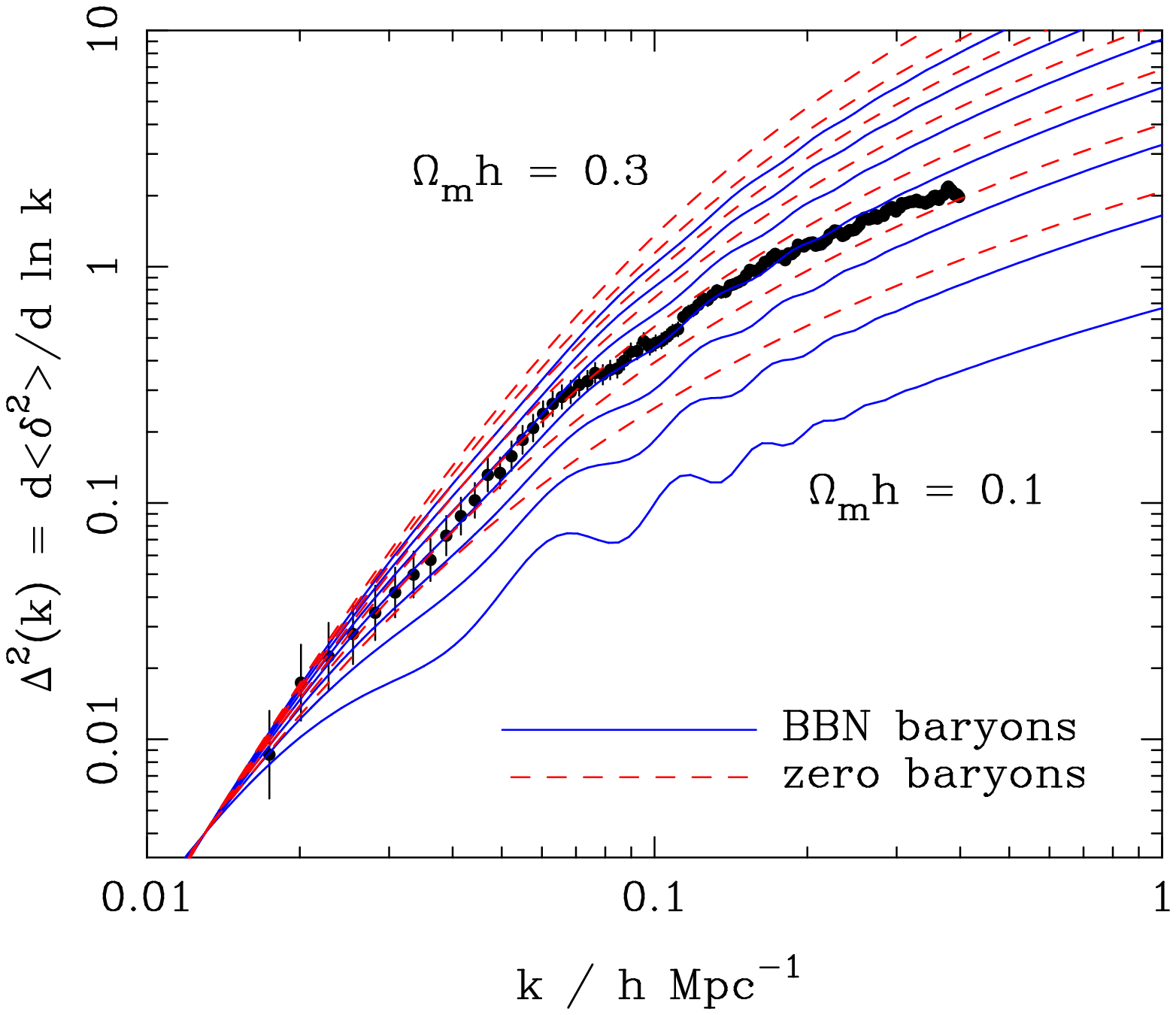}{0.7}
{The 2dFGRS redshift-space dimensionless power spectrum, 
$\Delta^2(k)$,
estimated according to the FKP procedure. The solid points
with error bars show the power estimate. The window
function correlates the results at different $k$ values,
and also distorts the large-scale shape of the power spectrum
An approximate correction for the latter effect has been applied.
The solid and dashed lines show various CDM models, all assuming
$n=1$. For the case with non-negligible baryon content,
a big-bang nucleosynthesis value of $\Omega_b h^2=0.02$ is
assumed, together with $h=0.7$. A good fit is clearly obtained
for $\Omega_m h \simeq 0.2$. Note that the observed power at
large $k$ will be boosted by nonlinear effects, but damped by 
small-scale random peculiar velocities. It appears that these
two effects very nearly cancel, but model fitting is generally
performed only at $k<0.15 \hompc$ in order to avoid these complications.}

\ssec{The 2dFGRS power spectrum and CDM models}

Perhaps the key aim of the 2dFGRS was to perform an accurate
measurement of the 3D clustering power spectrum, in order
to improve on the APM result,
which was deduced by deprojection of angular
clustering (Baugh \& Efstathiou 1993, 1994). 
The results of this direct estimation of the 3D power
spectrum are shown in figure~\lastfig\ (Percival et al. 2001).
This power-spectrum estimate uses the FFT-based approach
of Feldman, Kaiser \& Peacock (1994), and needs to be interpreted
with care. Firstly, it is a raw redshift-space estimate, so
that the power beyond $k\simeq 0.2 \hompc$ is severely damped
by smearing due to peculiar velocities, as well as being
affected by nonlinear evolution.
Finally, the FKP estimator yields the
true power convolved with the window function. This
modifies the power significantly on large scales (roughly
a 20\% correction). An approximate correction for
this has been made in figure~\lastfig. 

\japfigtwo{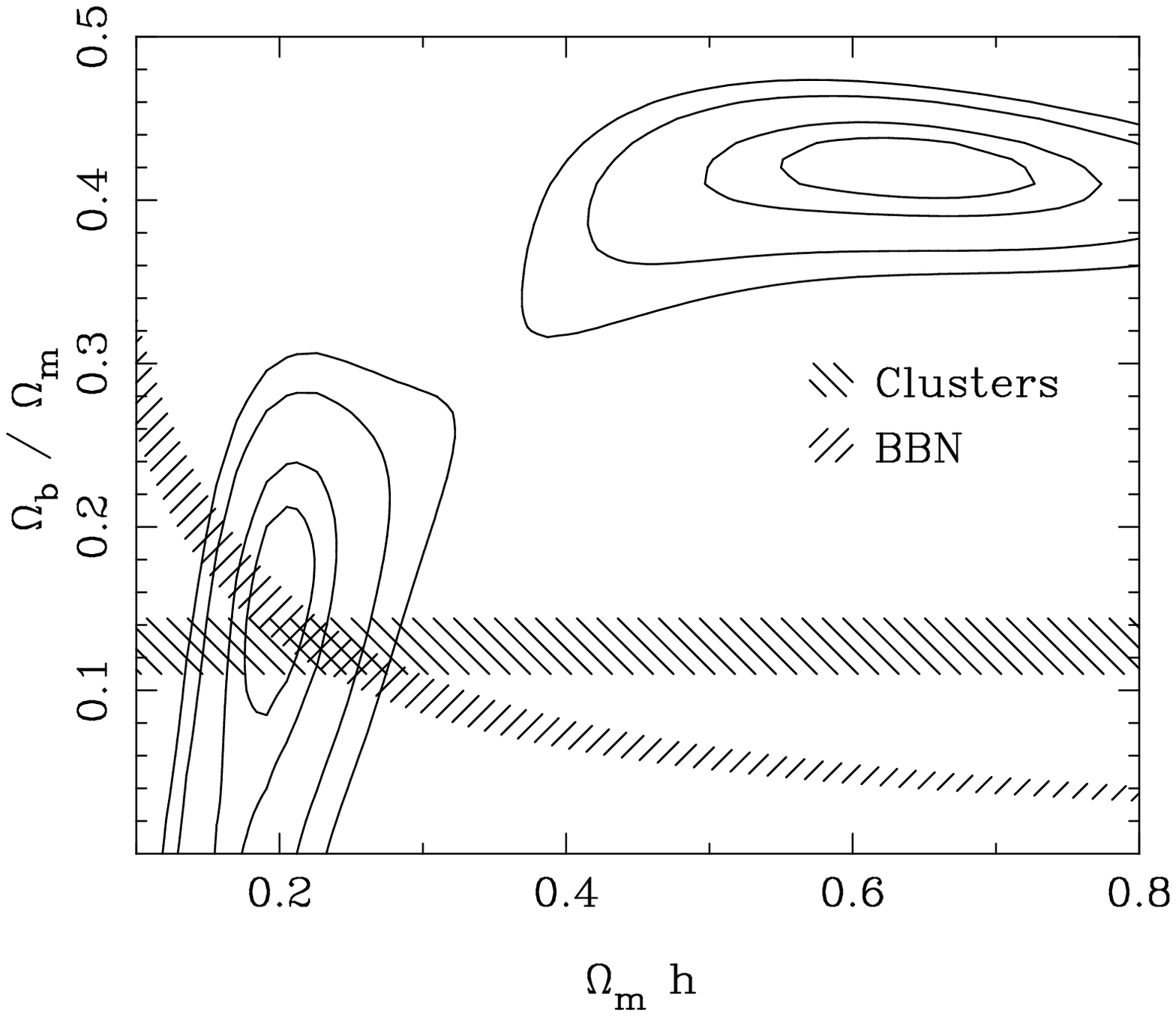}{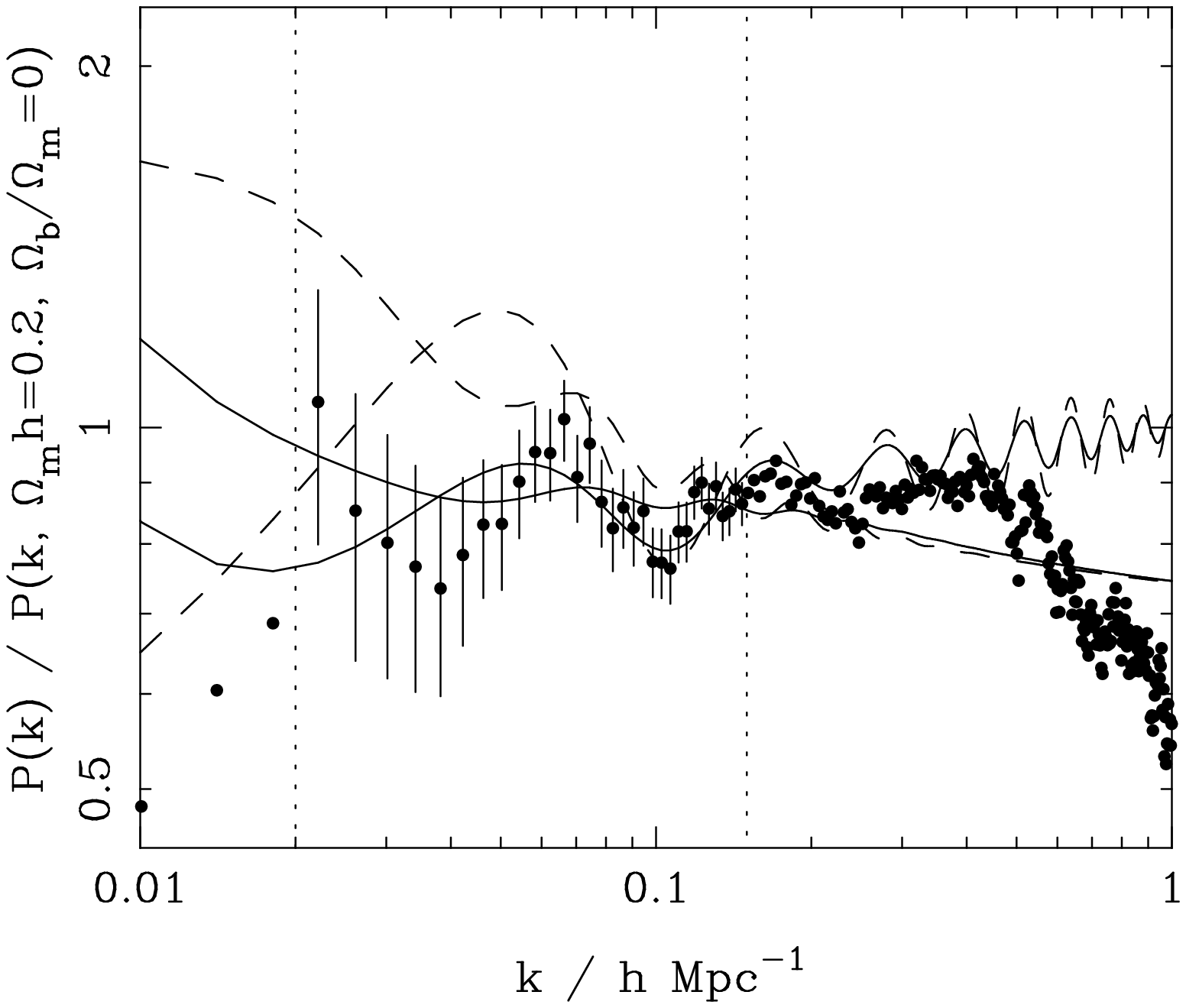}
{Likelihood contours for the best-fit linear
  CDM fit to the 2dFGRS power spectrum
  over the region $0.02<k<0.15$. Contours are plotted
  at the usual positions for one-parameter confidence of 68\%, and
  two-parameter confidence of 68\%, 95\% and 99\% (i.e. $-2\ln({\cal
  L}/{\cal L_{\rm max}}) = 1, 2.3, 6.0, 9.2$). We have marginalized
  over the missing free parameters ($h$ and the power spectrum
  amplitude).
  A prior on $h$ of $h=0.7\pm 10\%$ was assumed. 
  This result is compared to estimates from X-ray cluster
  analysis (Evrard 1997) and big-bang nucleosynthesis (Burles et al.  2001).
The second panel shows the 2dFGRS data compared with the two preferred models from
  the Maximum Likelihood fits convolved with the window function
  (solid lines). The unconvolved models are also shown (dashed
  lines). The $\Omega_m h \simeq 0.6$, $\Omega_b/\Omega_m=0.42$,
  $h=0.7$ model has the higher bump at $k\simeq 0.05\hompc$. The
  smoother $\Omega_m h \simeq 0.20$, $\Omega_b/\Omega_m=0.15$, $h=0.7$
  model is a better fit to the data because of the overall shape.
A preliminary analysis of the complete final 2dFGRS sample yields
a slightly smoother spectrum than the results shown
here (from Percival et al. 2001), so that the high-baryon solution becomes
disfavoured.
}

The fundamental assumption is that, on large scales, linear biasing
applies, so that the nonlinear galaxy power spectrum in redshift space has a shape
identical to that of linear theory in real space.
This assumption is valid for $k<0.15\hompc$;
the detailed justification comes from analyzing realistic 
mock data derived from $N$-body simulations (Cole et al. 1998).
The free parameters in fitting CDM models are thus the primordial spectral
index, $n$, the Hubble parameter, $h$, the total matter
density, $\Omega_m$, and the baryon fraction, $\Omega_b/\Omega_m$.
Note that the vacuum energy does not enter. Initially, we
show results assuming $n=1$; this assumption is relaxed later.

An accurate model comparison requires
the full covariance matrix of the data, because
the convolving effect of the window function 
causes the power at adjacent $k$ values to be correlated.
This covariance matrix was estimated by applying the survey window to a
library of Gaussian realisations of linear density fields, and
checked against a set of mock catalogues.
It is now possible to explore the space of CDM models, and
likelihood contours in $\Omega_b/\Omega_m$ versus $\Omega_mh$
are shown in figure~\lastfig. At each point in this
surface we have marginalized by integrating the likelihood surface
over the two free parameters, $h$ and the power spectrum
amplitude. 
We have added a Gaussian prior $h=0.7\pm
10\%$, representing external constraints such as the HST key project
(Freedman et al. 2001); this has only a minor effect on the results.

Figure~\lastfig\ shows that there is a degeneracy between
$\Omega_mh$ and the baryonic fraction $\Omega_b/\Omega_m$. However, there
are two local maxima in the likelihood, one with $\Omega_mh \simeq 0.2$
and $\sim 20\%$ baryons, plus a secondary solution $\Omega_mh \simeq 0.6$
and $\sim 40\%$ baryons. The high-density model can be rejected through a variety
of arguments, and the preferred solution is
$$
  \Omega_m h = 0.20 \pm 0.03; \quad\quad \Omega_b/\Omega_m = 0.15 \pm 0.07.
$$
The 2dFGRS data are compared to the best-fit linear power spectra
convolved with the window function in figure~\lastfig. The low-density
model fits the overall shape of the spectrum with relatively small
`wiggles', while the solution at $\Omega_m h \simeq 0.6$ provides a
better fit to the bump at $k\simeq 0.065\hompc$, but fits the overall
shape less well.
A preliminary analysis of $P(k)$ from the full final dataset
shows that $P(k)$ becomes smoother: 
the high-baryon solution becomes disfavoured, and
the uncertainties narrow slightly around the lower-density solution:
$\Omega_m h = 0.18 \pm 0.02$; $\Omega_b/\Omega_m = 0.17 \pm 0.06$.
The lack of large-amplitude oscillatory features in the power
spectrum is one general reason
for believing that the universe is dominated
by collisionless nonbaryonic matter.
In detail, the constraints on the collisional nature of dark matter
are weak, since all we require is that the effective sound speed
for modes of 100-Mpc wavelength is less than about $0.1c$.
Nevertheless, if a pure-baryon model is ruled out, the next simplest
alternative would arguably be to introduce a weakly-interacting
relic particle, so there is at least circumstantial evidence
in this direction from the power spectrum.

It is interesting to compare these conclusions with other
constraints. These are shown on figure~\lastfig, again assuming 
$h=0.7\pm 10\%$.
Estimates of the Deuterium to Hydrogen ratio in QSO spectra
combined big-bang nucleosynthesis theory predict $\Omega_bh^2 =
0.020\pm 0.001$ (Burles et al. 2001), which translates to the
shown locus of $f_{b}$ vs $\Omega_m h$. X-ray
cluster analysis yields a baryon fraction
$\Omega_b/\Omega_m=0.127\pm0.017$ (Evrard 1997) which is within
$1\sigma$ of our value. These loci intersect very close
to our preferred model.

Perhaps the main point to emphasise here is that the 2dFGRS results are not
greatly sensitive to the assumed tilt of the primordial spectrum. 
As discussed below, CMB data show that $n=1$ is a very
good approximation; in any case, very substantial tilts ($n\simeq 0.8$)
are required to alter the conclusions significantly.

\japfigtwo{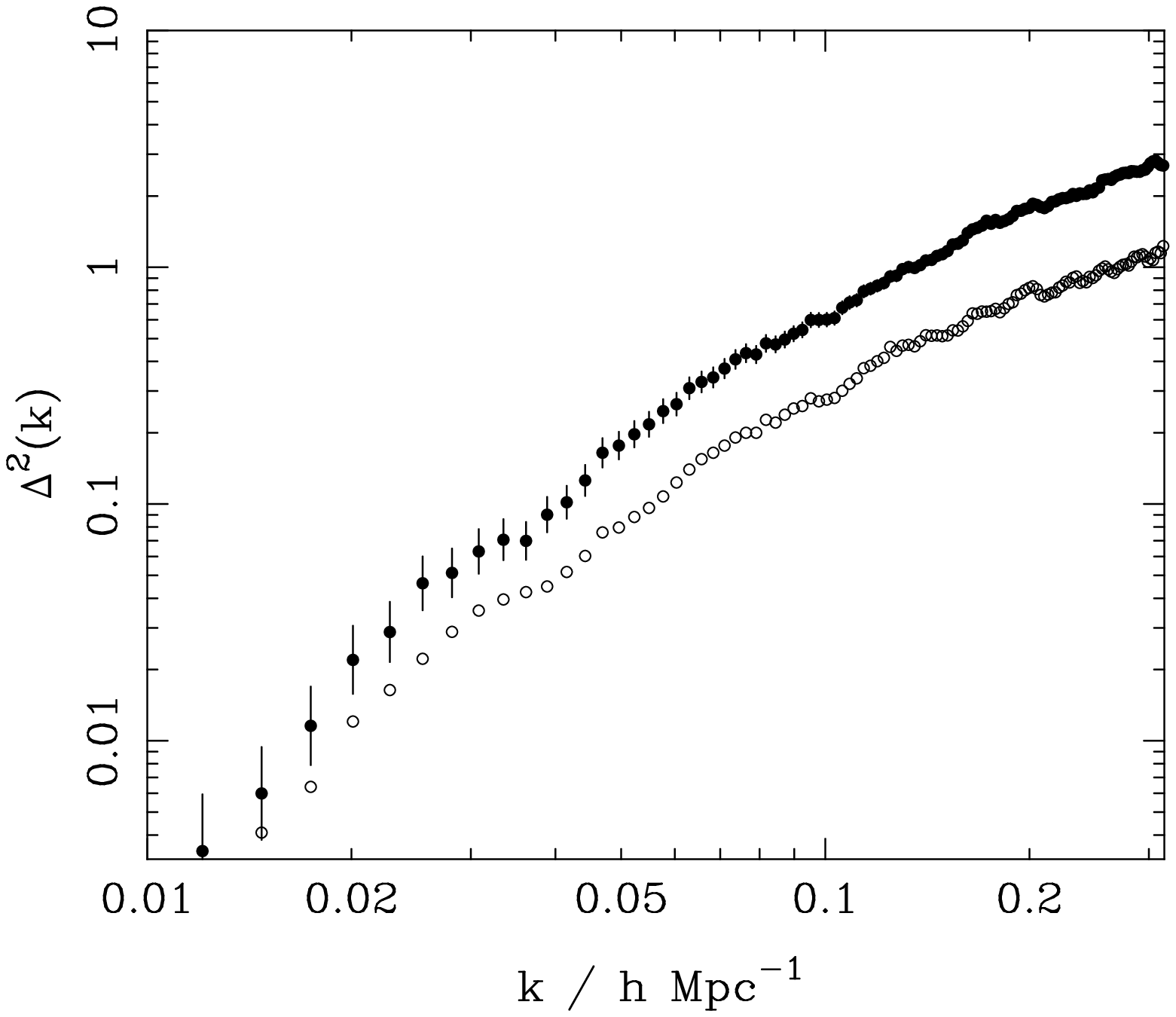}{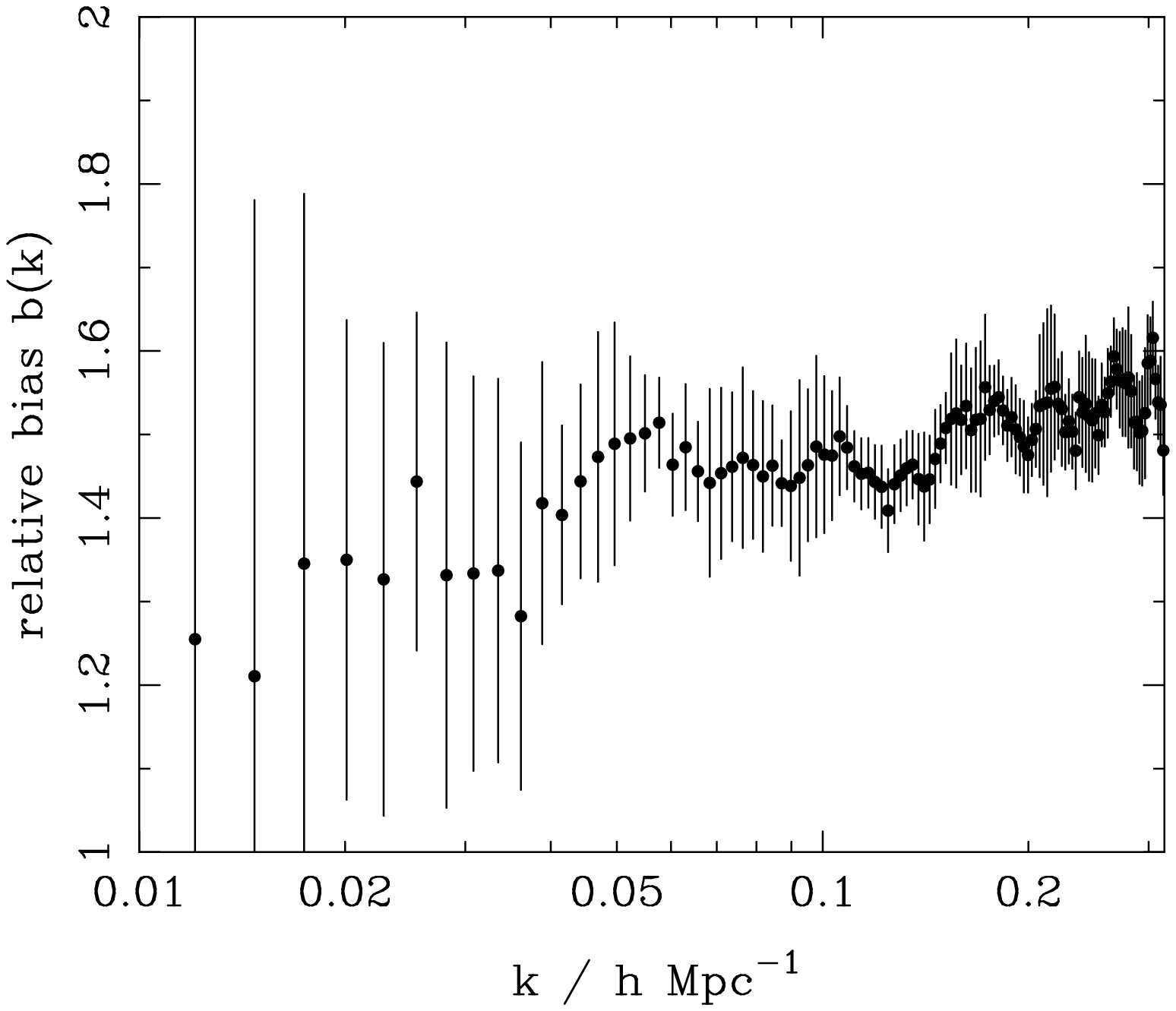}
{The power spectra of red galaxies (filled circles) and blue galaxies (open circles),
divided at photographic $B-R=0.85$.
The shapes are strikingly similar, and
the square root of the ratio yields the right-hand panel:
the relative bias in redshift space of red and blue galaxies.
The error bars are obtained by a jack-knife analysis. The relative
bias is consistent with a constant value of 1.4 over the range used for
fitting of the power-spectrum data ($0.015 < k < 0.15 \hompc$).
}

\ssec{Robustness of results}

The main residual worry about accepting the above conclusions is 
probably whether the assumption of linear bias can really be valid. 
In general, concentration towards higher-density regions both
raises the amplitude of clustering, but also steepens the correlations,
so that bias is largest on small scales, as discussed below.
We need to be clear of the regime in which the bias depends
on scale.

One way in which
this issue can be studied is to consider subsamples with very
different degrees of bias. Colour information has recently 
been added to the 2dFGRS database using SuperCosmos scans of the
UKST red plates (Hambly et al. 2001), and a division at rest-frame photographic
$B-R=0.85$ nicely separates ellipticals from spirals.
Figure~\lastfig\ shows the power spectra for the 2dFGRS divided in this
way. The shapes are almost identical (perhaps not so surprising,
since the cosmic variance effects are closely correlated in these
co-spatial samples). However, what is impressive is that the
relative bias is almost precisely independent of scale,
even though the red subset is rather strongly biased
relative to the blue subset (relative $b\simeq 1.4$). This provides some
reassurance that the large-scale $P(k)$ reflects the underlying
properties of the dark matter, 
rather than depending on the particular class
of galaxies used to measure it.

\sec{Relation of galaxies and dark matter}

\ssec{History and general aspects of bias}

In order to make full use of the cosmological information encoded
in large-scale structure, it is essential to understand the relation between the
number density of galaxies and the mass density field.
It was first appreciated during the 1980s that these two fields need
not be strictly proportional. Until this time, the general assumption was that
galaxies `trace the mass'. Since the mass density is a continuous field and galaxies
are point events, the approach is to postulate a
\key{Poisson clustering hypothesis}, in which the number of galaxies in a given volume
is a Poisson sampling from a fictitious number-density field that is
proportional to the mass. Thus within a volume $V$,
$$
\langle N_g(V) \rangle \propto M(V).
$$
With allowance for this discrete sampling, the observed numbers of
galaxies, $N_g$, would give an unbiased estimate of the mass in a given
region.

The first motivation for considering that galaxies might in fact be biased 
mass tracers came from attempts to reconcile the $\Omega_m=1$ 
Einstein--de Sitter model with observations. Although $M/L$ ratios in rich clusters
argued for dark matter, as first shown by Zwicky (1933), typical
blue values of $M/L\simeq 300h$ implied only $\Omega_m\simeq 0.2$ if
they were taken to be universal.
Those who argued that the value $\Omega_m=1$ was more natural (a greatly
increased camp after the advent of inflation)
were therefore forced to postulate that
the efficiency of galaxy formation was enhanced
in dense environments: \key{biased galaxy formation}.

We can note immediately that a consequence of this bias in
density will be to affect the velocity statistics of galaxies
relative to dark matter. Both galaxies and dark-matter
particles follow orbits in the overall gravitational potential
well of a cluster; if the galaxies are to be more strongly
concentrated towards the centre, they must clearly have
smaller velocities than the dark matter. This is the phenomenon
known as \key{velocity bias} (Carlberg, Couchman \& Thomas 1990). 

An argument for bias at the opposite extreme of density arose
through the discovery of large \key{voids} in the galaxy distribution
(Kirshner et al. 1981).
There was a reluctance to believe that such vast
regions could be truly devoid of matter --
although this was at a time before the discovery
of large-scale velocity fields.
This tendency was given further stimulus
through the work of Davis, Efstathiou, Frenk \& White (1985),
who were the first to calculate $N$-body models
of the detailed nonlinear structure arising in
CDM-dominated universes. Since the CDM spectrum
curves slowly between effective indices
of $n=-3$  and $n=1$, the correlation function
steepens with time. There is therefore
a unique epoch when $\xi$ will have the observed
slope of $-1.8$. Davis et al. identified this
epoch as the present and then noted that, for $\Omega_m=1$, it
implied a rather low {\it amplitude\/} of fluctuations:
$r_0=1.3h^{-2}$ Mpc. An independent argument for this low
amplitude came from the size of the peculiar velocities
in CDM models: if the spectrum was given an amplitude corresponding to the
$\sigma_8\simeq 1$ seen in the galaxy distribution, the
pairwise dispersion was $\sigma_p\simeq 1000$\upto$1500\kms$,
around 3 times the observed value.
What seemed to be required was
a galaxy correlation function that was an amplified
version of that for mass. This was exactly the
phenomenon analysed for Abell clusters by Kaiser (1984), and
thus was born the idea of \key{high-peak bias}: bright
galaxies form only at the sites of high peaks in the
initial density field. This was developed in some
analytical detail by Bardeen et al. (1986), and was implemented
in the simulations of Davis et al. (1985).

\japfig{0}{0}{477}{224}{0.7}{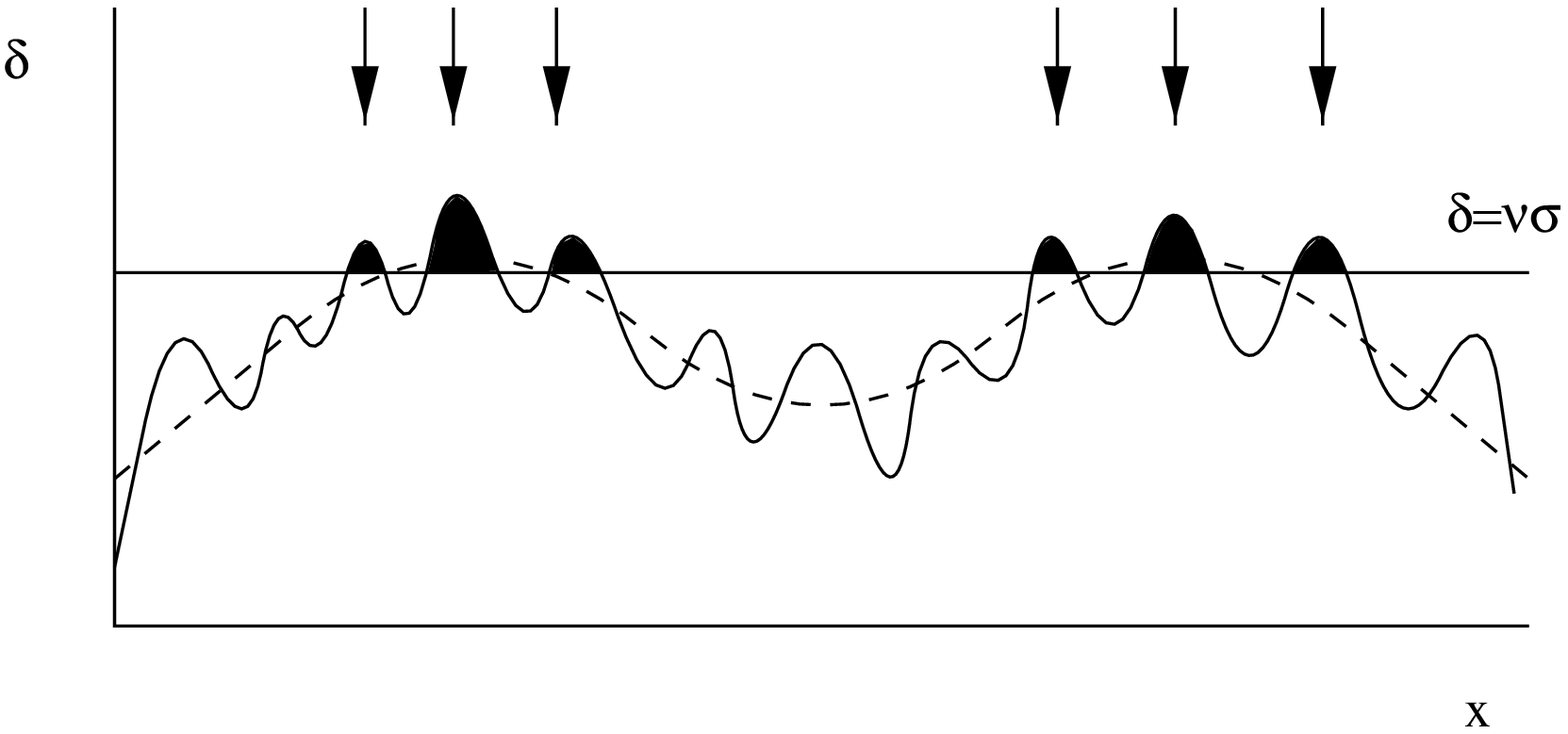}
{The high-peak bias model. If we decompose a density
field into a fluctuating component on galaxy scales,
together with a long-wavelength `swell' (shown dashed),
then those regions of density that lie above 
a threshold in density of $\nu$ times the rms will be
strongly clustered. If proto-objects are presumed to form
at the sites of these high peaks (shaded, and 
indicated by arrows), then this is a population with
Lagrangian bias -- {\it i.e.} a non-uniform spatial
distribution even prior to dynamical evolution
of the density field.  The key question is the physical
origin of the threshold; for massive objects such as clusters,
the requirement of collapse by the present imposes a threshold
of $\nu\gs 2$. For galaxies, there will be no bias without
additional mechanisms to cause star formation to favour those objects
that collapse first.}

As shown below, the high-peak model produces a linear
amplification of large-wavelength modes.
This is likely to be a general feature of other
models for bias, so it is useful to introduce the
\key{linear bias parameter}:
$$
\boxit{
\left({\delta\rho\over\rho}\right)_{\rm galaxies}
 = b\,\left({\delta\rho\over\rho}\right)_{\rm mass}.
}
$$
This seems a reasonable assumption when $\delta\rho/\rho\ll 1$,
although it leaves open the question of how the effective value of $b$
would be expected to change on nonlinear scales.
Galaxy clustering on large scales therefore allows us to determine 
mass fluctuations only if we know the value of $b$.
When we observe large-scale galaxy clustering, we are only
measuring $b^2\xi_{\rm mass}(r)$ or $b^2\Delta^2_{\rm mass}(k)$.

Later studies of bias concentrated on general models.
A fruitful assumption is that bias is \idx{local bias}{\it local\/},
so that the number density of galaxies is some nonlinear function of the mass density
$$
n_g({\bf r}) = f[\rho_m({\bf r})].
$$
Coles (1993) proved the powerful result that, whatever the function $f$
may be, the quantity
$$
b(r) \equiv \sqrt{\xi_g(r)/\xi_m(r)}
$$
had to show a monotonic dependence on scale, provided the mass density
field had Gaussian statistics.
An interesting concrete example of this is provided by the
\key{lognormal density field} (Coles \& Jones 1991); this is
generated by exponentiation of a Gaussian field:
$$
1+\delta_{\japsub LN} = \exp(\delta_{\japsub G} - \sigma^2/2),
$$
where $\sigma^2$ is the total variance in the Gaussian field.
These authors argue that this analytical form 
is a reasonable approximation
to the exact nonlinear evolution of the mass density
distribution function, preventing the unphysical values
$\delta < -1$.
This non-Gaussian model is built upon an underlying Gaussian field,
so the joint distribution of the density
at $n$ points is still known. This means that the
correlations are simple enough to calculate, the result being
$$
\xi_{\japsub LN} = \exp (\xi_{\japsub G}) - 1.
$$
This says that $\xi$ on large scales is unaltered by nonlinearities
in this model; they only add extra small-scale correlations.
Using the lognormal model as a hypothetical nonlinear density
field, we can now introduce bias. A nonlinear local transformation 
$\rho_g \propto \rho_{\japsub LN}^b$ then gives a correlation function
$1+\xi_g = (1+\xi_{\japsub LN})^{b^2}$ (Mann, Peacock \& Heavens 1998).
The linear bias parameter is $b$, but the correlations steepen
on small scales, as expected for Coles' result.

In reality, bias is unlikely to be completely causal,
and this has led some workers to explore stochastic bias
models, in which
$$
n_g=f(\rho_m) + \epsilon,
$$
where $\epsilon$ is a random field that is uncorrelated with the
mass density (Pen 1998; Dekel \& Lahav 1999).
This means we need to consider not only the bias parameter
defined via the ratio of correlation functions, but also
the correlation coefficient, $r$,  between galaxies and mass:
$$
b^2 = { \langle \delta_g \delta_g' \rangle \over \langle \delta_m \delta_m' \rangle} 
\quad\quad
r^2 = { \langle \delta_g \delta_m' \rangle^2 \over 
\langle \delta_g \delta_g' \rangle \langle \delta_m \delta_m' \rangle}.
$$
Although truly stochastic effects are possible in galaxy formation,
a relation of the above form is expected when the
galaxy and mass densities are filtered on some scale
(as they always are, in practice). Just averaging a
galaxy density that is a nonlinear
function of the mass will lead to some scatter when comparing with the
averaged mass field; a scatter will also arise when the
relation between mass and light is non-local, however, and this
may be the dominant effect.

\ssec{The peak-background split}

We now consider the central mechanism of biased clustering,
in which a rare high density fluctuation, corresponding
to a massive object, collapses 
sooner if it lies in a region of large-scale overdensity.
This `helping hand' from the long-wavelength modes means that
overdense regions contain an enhanced abundance of massive objects 
with respect to the mean, so that these systems display enhanced clustering.
The basic mechanism can be immediately understood via the
diagram in figure~\lastfig; it was first clearly
analysed by Kaiser (1984) in the context of rich clusters of galaxies.
What Kaiser did not do was consider the degree of bias
that applies to more typical objects; the generalization
to consider objects of any mass was made by
Cole \& Kaiser (1989; see also Mo \& White 1996 and Sheth et al. 2001).

The key ingredient of this analysis is the mass function of
dark-matter haloes. The universe fragments into virialized
systems such that $f(M)\, dM$ is the number density of
haloes in the mass range $dM$; conservation of mass requires
that $\int M\, f(M)\, dM = \rho_0$. A convenient related
dimensionless quantity is therefore the
\key{multiplicity function}, $M^2 f(M)/\rho_0$, which
gives the fraction of the mass of the universe contained
in haloes of a unit range in $\ln M$. 
The simplest analyses of the mass function rest on the concept
of a density threshold: collapse to a virialized object
is deemed to have occurred where 
linear-theory $\delta$ averaged over a box
containing mass $M$ reaches some critical value $\delta_c$.
Generally, we shall assume the value $\delta_c=1.686$ appropriate
for spherical collapse in an Einstein--de Sitter universe.
Now imagine that this situation is perturbed, by 
adding some constant shift $\epsilon$ to
the density perturbations over some large region.
The effect of this is to perturb the threshold: 
fluctuations now only need to reach $\delta=\delta_c-\epsilon$
in order to achieve collapse. The number density is therefore
modulated:
$$
f \rightarrow f -  {df\over d\delta_c}\,  \epsilon.
$$
This gives a bias
in the number density of haloes in Lagrangian
space: $\delta f/f= b_{\japsub L}\epsilon$, where the Lagrangian bias is 
$$
b_{\japsub L}= - {d\ln f\over d\delta_c}.
$$
In addition to this modulation of the halo properties, the large-scale
disturbance will move haloes closer together where $\epsilon$
is large, giving a density contrast of $1+\epsilon$. If $\epsilon \ll1$, the overall
fractional density contrast of haloes is therefore the sum of the
dynamical and statistical effects: $\delta_{\rm halo}=\epsilon+b_{\japsub L}\epsilon$.
The overall bias in Eulerian space ($b=\delta_{\rm halo}/\epsilon$) 
is therefore
$$
b = 1 - {d\ln f\over d\delta_c}.
$$
Of course, the field $\epsilon$ can hardly be imposed by hand;
instead, we make the \key{peak-background split}, in which
$\delta$ is mentally decomposed into a small-scale
and a large-scale component -- which we identify with
$\epsilon$. The scale above which the large-scale
component is defined does not matter so long as it lies
between the sizes of collapsed systems and the scales
at which we wish to measure correlations.

To apply this, we need an explicit expression for the mass function.
The simplest alternative is the original expression of
Press \& Schechter (1974), which can be written
in terms of the parameter $\nu=\delta_c/\sigma(M)$:
$$
M^2 f(M)/\rho_0 = \sqrt{2\over \pi}\, \nu\, \exp\left(-{\nu^2\over 2}\right).
$$
We now use $d/d\delta_c=\sigma(M)^{-1}(d/d\nu)=(\nu/\delta_c)(d/d\nu)$,
since $M$ is not affected by the threshold change, which yields
$$
\boxit{
b(\nu)= 1 + {\nu^2-1\over \delta_c}.
}
$$
This says that $M^*$ haloes are unbiased, low-mass haloes are
antibiased\idx{antibias} and high-mass haloes are positively biased, eventually
reaching the $b=\nu/\sigma$ value expected for high peaks.
The corresponding expression can readily be deduced for more
accurate fitting formulae for the mass function, such as
that of Sheth \& Tormen (1999):
$$
M^2 f(M)/\rho_0 = 0.21617[ 1 + (\sqrt{2}/\nu^2)^{0.3} ] 
\, \nu\, \exp[-\nu^2/(2\sqrt{2})].
$$

We can now understand the observation that Abell clusters are much
more strongly clustered than galaxies in general:
regions of large-scale overdensity contain systematically
more high-mass haloes than expected if the haloes traced
the mass. This phenomenon was dubbed \key{natural bias}
by White et al. (1987).
However, applying the idea to galaxies is not
straightforward: we have shown that enhanced clustering is only
expected for massive fluctuations with $\sigma \ls 1$, but
galaxies at $z=0$ fail this criterion.
The high-peak idea applies will at high redshift, where massive
galaxies are still assembling, but today there has been
time for galaxy-scale haloes to collapse in all environments.
The large bias that should exist at high redshifts is erased
as the mass fluctuations grow:
if the Lagrangian component to the biased density field is kept unaltered,
then the present-day bias will tend to unity as
$$
b(\nu)= 1 + {\nu^2-1\over(1+z_f) \delta_c}.
$$
(Fry 1986; Tegmark \& Peebles 1998).
Strong galaxy bias at $z=0$ therefore requires some form
of selection that locates present-day galaxies preferentially
in the rarer haloes with $M>M^*$
(Kauffmann, Nusser \& Steinmetz 1997).

This dilemma forced the introduction of
the idea of \key{high-peak bias}: bright
galaxies form only at the sites of high peaks in the
initial density field (Bardeen et al. 1986;
Davis et al. 1985). This idea is commonly, but
incorrectly, attributed to Kaiser (1984), but it needs
an extra ingredient, namely a non-gravitational threshold.
Attempts were therefore made to argue that the first generation of objects
could propagate disruptive signals, causing
neighbours in low-density regions to be `still-born'.
It is then possible to construct models ({\it e.g.} Bower et al. 1993)
in which the large-scale modulation of the galaxy density
is entirely non-gravitational in nature.
However, it turned out to be hard to make such
mechanisms operate: the energetics and required scale of the
phenomenon are very large (Rees 1985; Dekel \& Rees 1987).
These difficulties were only removed when the standard model
became a low-density universe, in which the dynamical
argument for high galaxy bias no longer applied.

\japfigbasic{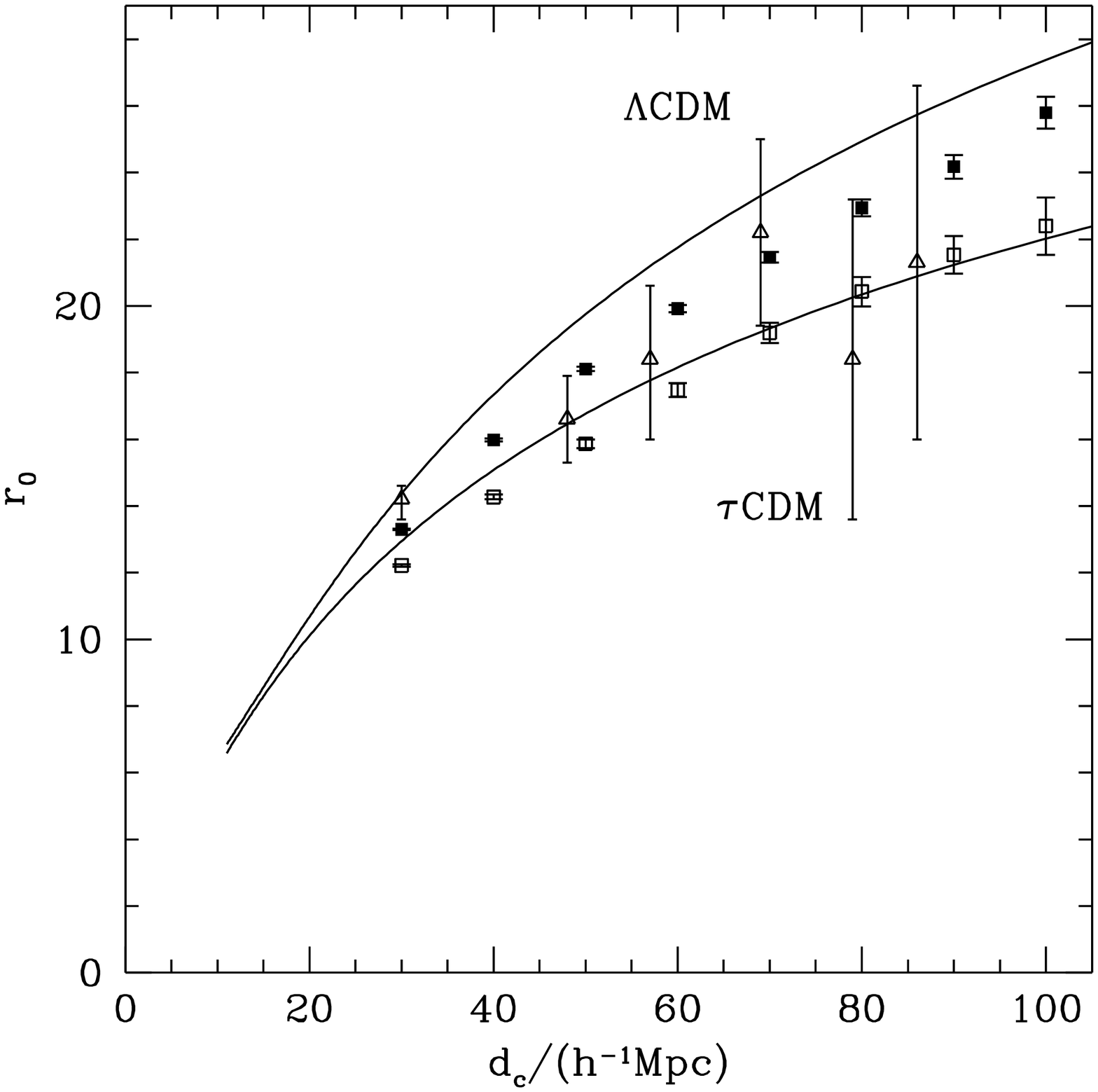}{0.6}
{The correlation length for clusters of galaxies, $r_0$, 
as a function of mean intercluster separation, $d_c$, 
taken from Colberg et al. (2000). Results 
are shown for $\tau$CDM (open squares) and
$\Lambda$CDM (filled squares) simulations. The predictions of 
Sheth et al. (2001)  are shown as solid lines. Also shown are  
data from the APM cluster catalogue (open triangles), taken from 
Croft et al. (1997).}

\ssec{Observations of biased clustering}

As indicated above, the first strong indications of biased
clustering came from measurements of the correlation function
of Abell clusters, which showed a far greater amplitude than
for galaxies in general (Klypin \& Kopylov 1983; Bahcall \& Soneira 1983).
Following Kaiser (1984), Cole \& Kaiser (1989) etc., our explanation
for this is that massive haloes show clustering that is an
increasing function of mass. This is illustrated in figure~\ref{fig:clus_b_vs_d},
which shows that the rarest and most rich clusters (as measured
by the intercluster separation) have the highest clustering,
and that the trend is in agreement with the theoretical predictions.

\japfigbasic{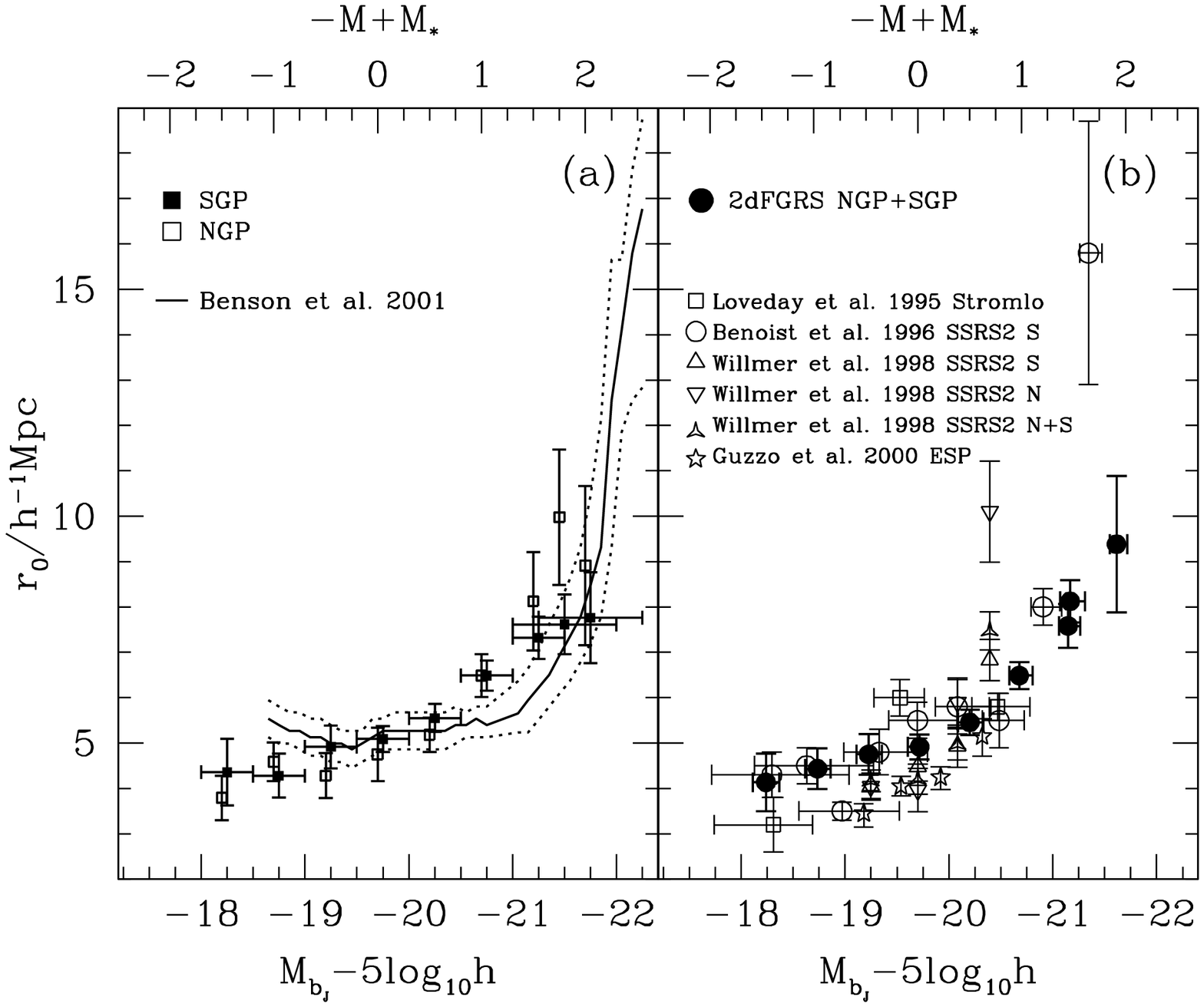}{0.7}
{(a) The correlation length in real space as a function of 
absolute magnitude. 
The solid line shows the predictions of the semi-analytic 
model of Benson et al. (2001), computed in a series of overlapping 
bins, each $0.5$ magnitudes wide. The dotted curves show an 
estimate of the errors on this prediction, including the relevant sample variance
for the survey volume.
(b) The real space correlation length estimated combining 
the NGP and SGP (filled circles). 
The open symbols show a selection of recent data from other studies.
}

Because galaxy halo masses are less extreme, it is not so
clear a priori that any trend of this sort should be expected for galaxies.
However, our empirical knowledge of luminosity functions and morphological
segregation did argue for an effect.
It has been clear for many years that elliptical
galaxies display a higher correlation amplitude than
spirals (Davis \& Geller 1976), and this makes sense
in terms of the preference of ellipticals for cluster
environments. Since ellipticals are also
more luminous on average than spirals,
some trend with luminosity is to
be expected, but the challenge is to detect it.
For a number of years, the existence of any effect was controversial
(e.g.  Loveday et al. 1995; Benoist et al. 1996), but 
Norberg et al. (2001) were able to use the 2dFGRS to demonstrate
very clearly that the effect existed,
as shown in Figure~\lastfig. The results
can be described by a linear dependence of effective bias
parameter on luminosity:
$$
b/b^* = 0.85 + 0.15\,(L/L^*),
$$
and the scale-length of the real-space correlation function for $L^*$
galaxies is approximately $r_0=4.8 \mpcoh$. 
Finally, with spectral classifications, it is possible to
measure the dependence of clustering both on luminosity and
on spectral  type, to see to what extent morphological
segregation is responsible for this result. Norberg et al. (2002)
show that, in fact, the principal effect seems to be with
luminosity: $\xi(r)$ increases with $L$ for all spectral types.

Finally, we can look at high-redshift clustering.
At high enough redshift, $M^*$ is of order a galaxy mass and
galaxies could be strongly biased relative to the mass at that time.
Indeed, there is good evidence that this is the case.
Steidel {\it et al. } (1997) have used the Lyman-limit technique
to select galaxies around redshifts $2.5 \ls z \ls 3.5$
and found their distribution to be highly inhomogeneous.
The apparent value of $\sigma_8$ for these objects is
of order unity (Adelberger et al. 1998), 
whereas the present value of $\sigma_8\simeq 0.8$
should have evolved to about 0.26 at these redshifts
(for $\Omega_m=0.3$, $k=0$). This suggests a bias parameter of $b\simeq 4$,
or $\nu\simeq 2.5$, which requires a halo mass of about $10^{12.1} h^{-1} M_\odot$
for concordance $\Lambda$CDM.
The masses of these high-redshift galaxies can be estimated
directly through their stellar masses, which are typically
$10^{10} h^{-2} M_\odot$ (Papovich, Dickinson \& Ferguson 2001), and
thus only 1\% of what is required in order to explain
the clustering. This is an unreasonably small baryon
fraction, so the correct explanation is more plausibly that
each $10^{12} h^{-1} M_\odot$ halo at $z=3$ contains a number
of Lyman-break galaxies. This theme is pursued below.

\ssec{Scale dependence of bias}

The Poisson clustering hypothesis
would propose that galaxies are simply a dilute sampling
of the mass field. If this were a correct hypothesis,
no CDM universe would be acceptable, since the correlation
functions for these models differ from the observed galaxy
correlations in a complicated scale-dependent fashion
(e.g. Klypin, Primack \& Holtzman 1996;
Peacock 1997; Jenkins et al. 1998).

\japfig{18}{144}{574}{701}{0.6}
{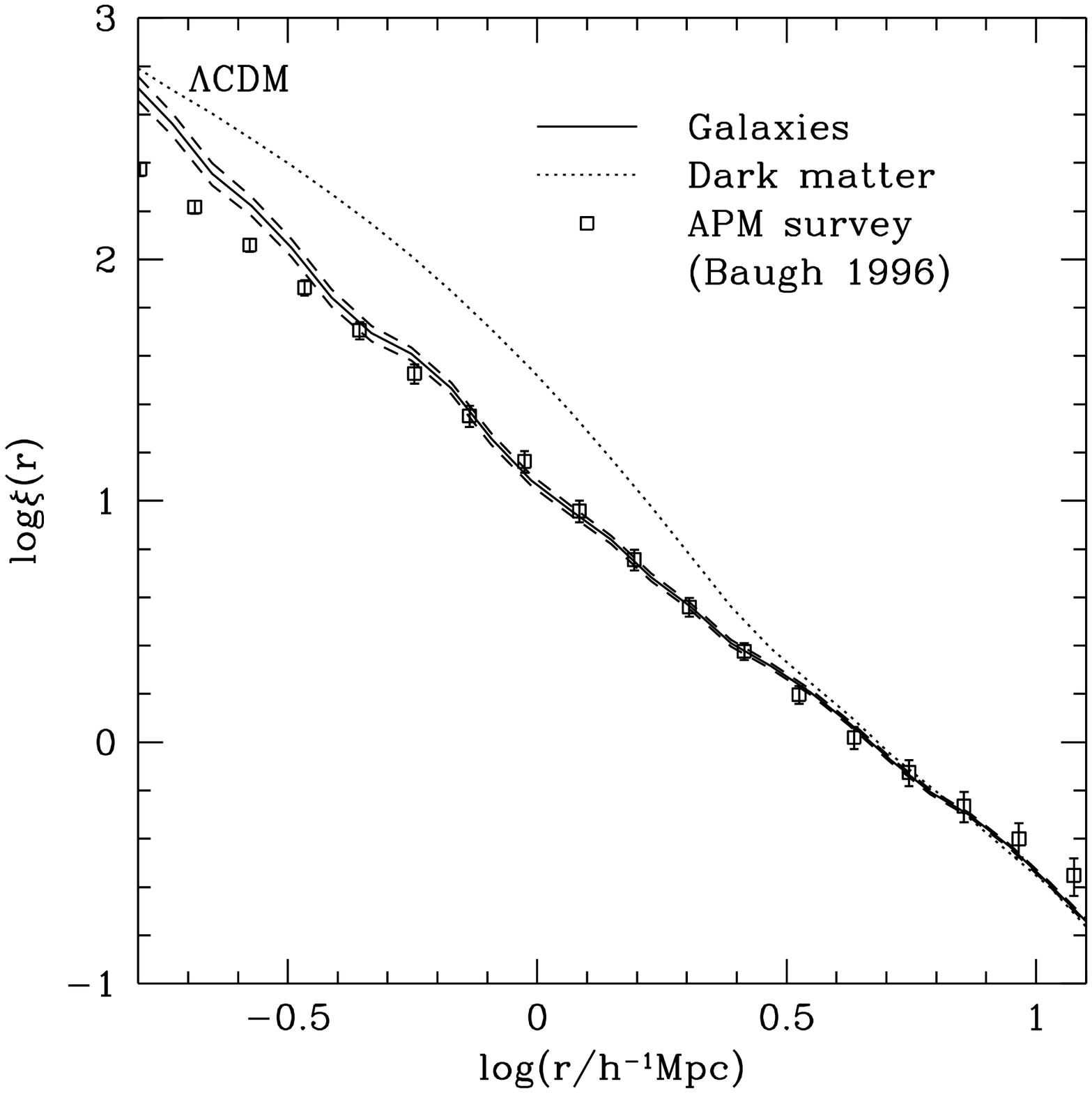}
{The correlation function of galaxies in the
semianalytical simulation of an LCDM
universe by Benson et al. (2000a).
Although the nonlinear correlations of the mass
show a characteristic convex bulge at separations
around 1~Mpc, the galaxy data follow a power law
which thus  is antibiased on these small scales.
The simulation is successful at reproducing this trend.
}

For a few years, this failure of CDM models to match the
shape of the galaxy power spectrum was seen as a serious
problem, but this was eventually resolved 
by more detailed theoretical predictions for galaxy clustering.
Two approaches  are being followed in this regard.
The brute-force method is to perform $N$-body simulations
in which the evolution of both collisionless dark matter and
dissipative gas is followed, with the physical state
of the gas ({\it i.e.} its ability to cool)
being used as a cue to insert star formation.
The stars in turn are allowed to feed energy back into the gas,
simulating the effects of mass loss and supernovae.
This determines the star formation
history of a given halo, and its appearance
can be predicted using spectral synthesis codes.
This is challenging, but starting to be feasible with
current computing power (Pearce et al. 2001). The alternative
is `semianalytic' modelling, in which the merging history
of dark-matter haloes is treated via the extended Press-Schechter
theory (Bond et al. 1991), and the location of galaxies within
haloes is estimated using dynamical-friction arguments
({\it e.g.} Kauffmann et al. 1993, 1999; Cole et al. 1994, 2000; 
Somerville \& Primack 1999;
van Kampen, Jimenez \& Peacock 1999;
Benson et al. 2000a,b). 

Both these approaches
have yielded similar conclusions, and shown how CDM models
can match the galaxy data: specifically, the low-density
flat $\Lambda$CDM model that is favoured on other
grounds can yield a correlation function that is close to a
single power law over $1000 \gs \xi \gs 1$, even though the
mass correlations show a marked curvature over this range
(Pearce et al. 1999; 2001; 
Benson et al. 2000a; see figure \ref{fig:benson}).
These results are impressive, yet it is frustrating to have a result
of such fundamental importance emerge from a complicated
calculational apparatus. 
There is thus some motivation for constructing a simpler
heuristic model that captures the main processes at work in
the full semianalytic models. The following section
describes an approach of this sort (Peacock \& Smith 2000; 
Seljak 2000; Cooray \& Sheth 2002).

\ssec{The halo model -- I: mass}

The formation of galaxies must be a non-local process to
some extent, and the modern paradigm was introduced by White \& Rees (1978):
galaxies form through the cooling of baryonic material in
virialized haloes of dark matter. The virial radii of these
systems are in excess of 0.1~Mpc, so there is the potential
for large differences in the correlation properties of
galaxies and dark matter on these scales.
The `halo model' addresses this by creating a density
field in which dark-matter haloes are superimposed.
The key feature that allows bias to be included is to encode
all the complications of  galaxy formation via
the halo occupation number: the number of galaxies
found above some luminosity threshold in a virialized halo of a given mass.

To some extent, this is a very old idea:
one of the earliest suggested models for the
galaxy correlation function was to consider a density
field composed of randomly-placed independent clumps
with some universal density profile (Neyman, Scott \& Shane 1953; Peebles 1974).
Since the clumps are placed at random (with number density $n$), the only excess neighbours
to a given mass point arise from points in the same clump, and the correlation 
function is straightforward to compute in principle.
For the case where the clumps have a power-law density profile,
$$
\rho= n B r^{-\epsilon},
$$
truncated at $r=R$, the small-$r$ behaviour
of the correlation function is $\xi\propto r^{3-2\epsilon}$,
provided $3/2 < \epsilon <3$. For smaller values of $\epsilon$, $\xi(r)$
tends to a constant as $r\rightarrow 0$.
In the isothermal $\epsilon=2$ case, the correlation function for
$r\ll R $ is
$$
\xi(r)={\pi^2 B\over 4 rR} = {\pi N \over 16 r R^2 n},
$$
where $N$ is the total number of particles per clump
(Peebles 1974).

\japfigbasic{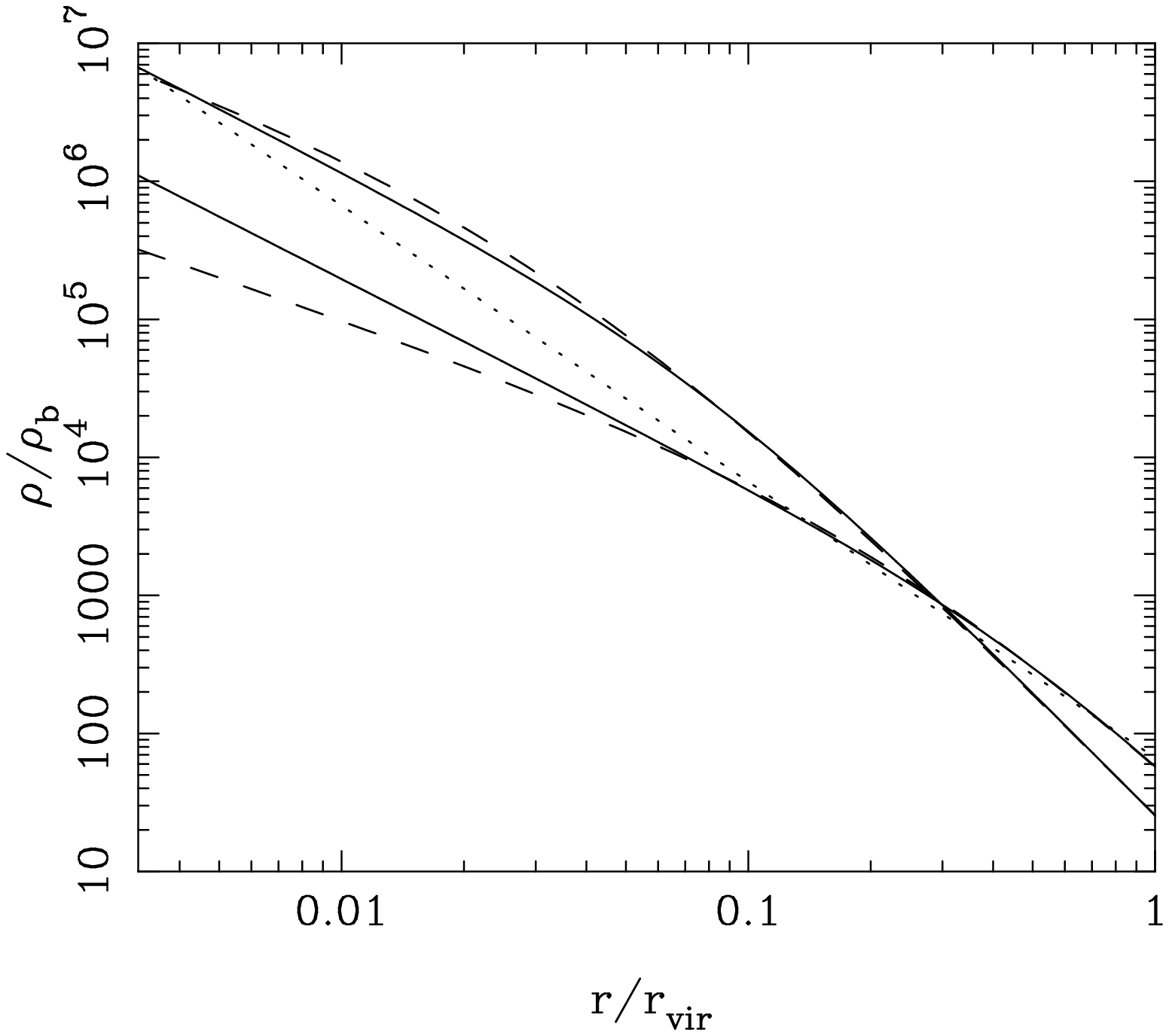}{0.7}
{A comparison of various possible density
profiles for virialized haloes. The dotted line is a
singular isothermal sphere. The solid lines show
haloes with formation redshifts of 0 and 5
according to NFW ($\Omega=1$) and M99.}

The general result is that the correlation function is
less steep at small $r$ than the clump density profile, which
is inevitable because an autocorrelation function involves convolving
the density field with itself.
A long-standing problem for this model is therefore that the
predicted correlation function is much flatter than is
observed for galaxies: $\xi \propto r^{-1.8}$ is the
canonical slope, apparently requiring 
clumps with very steep density profiles, $\epsilon=2.4$.
This is not in agreement with the profiles of dark-matter
haloes as `observed' in numerical simulations.

Traditionally, virialized systems have been found by a criterion
based on percolation (`friends-of-friends'), such that the mean
density is about 200 times the mean. Sometimes, the criterion is taken
as a density of 200 times the critical value. We shall use the
former definition:
$$
r_v = \left( {3M\over 800 \pi \rho_b} \right)^{1/3}.
$$
Thus $r_v$ is related to the Lagrangian radius containing the mass via
$r_v=R/200^{1/3}$.
Of course, the density contrast used to define the
boundary of an object is somewhat arbitrary.
Fortunately, much of the mass resides at smaller radii,
near a `core radius'. These core radii are relatively
insensitive to the exact definition of virial radius.

The simplest model for the density structure of the
virialized system is the singular isothermal sphere:
$\rho= \sigma_v^2 /(2\pi G r^2)$, or
$$
\rho/\rho_b = { 200 \over 3 y^2 }; \quad (y<1); \quad y\equiv r/r_v.
$$
A more realistic alternative is the profile proposed by
Navarro, Frenk \& White (1996; NFW):
$$
\rho/\rho_b = {\Delta_c \over y (1+y)^2 }; \quad (r<r_v); \quad y\equiv r/r_c.
$$
The parameter $\Delta_c$ is related to the core radius and the virial
radius via
$$
\Delta_c = {200 c^3/3 \over \ln(1+c) -c/(1+c)}; \quad c\equiv r_v/r_c
$$
(we change symbol from NFW's $\delta_c$ to avoid confusion with the
linear-theory density threshold for collapse, and also because our definition
of density is relative to the mean, rather than the critical density).
NFW showed that $\Delta_c$ is related to collapse redshift via
$$
\Delta_c \simeq 3000 (1+z_c)^3,
$$
An advantage of the definition of virial radius used here is that
$\Delta_c$ is independent of $\Omega$ (for given $z_c$), whereas
NFW's $\delta_c$ is $\propto \Omega$.

The above equations determine the concentration,
$c=r_v/r_c$ implicitly, hence in principle giving $r_c$ in terms of
$r_v$ once $\Delta_c$ is known. 
NFW give a procedure for determining $z_c$. A simplified argument would
suggest a typical formation era determined by $D(z_c)=1/\nu$, 
where $D$ is the linear-theory growth factor between
$z=z_c$ and the present, and $\nu$ is the dimensionless fluctuation
amplitude corresponding to the system in units of the rms: $\nu\equiv \delta_c/\sigma(M)$,
where $\delta_c\simeq 1.686$. For very massive systems with $\nu\gg 1$, only
rare fluctuations have collapsed by the present, so $z_c$ is close to zero.
This suggests the interpolation formula
$$
D(z_c) = 1 + 1/\nu;
$$
The NFW formula is actually of this form, except that the $1/\nu$ term
is multiplied by a spectrum-dependent coefficient of order unity.
It has been claimed by Moore et al. (1999; M99) that the
NFW density profile is in error at small $r$. M99 proposed the
alternative form
$$
\rho/\rho_b = {\Delta_c \over y^{3/2} (1+y^{3/2}) }; \quad (r<r_v); \quad y\equiv r/r_c.
$$
It is straightforward to use this in place of the NFW profile.

We now compute the power spectrum for the halo model.
Start by distributing point seeds throughout
the universe with number density $n$, in which case the power spectrum of the 
resulting density field is just shot noise:
$$
\Delta^2(k)={4\pi \over n}\, \left({k\over 2\pi}\right)^3.
$$
Here, we use a dimensionless notation for the power spectrum:
$\Delta^2$ is the contribution to the fractional density variance
per unit interval of $\ln k$.  In the convention of Peebles (1980), this is
$$
\Delta^2(k)\equiv{{\rm d}\sigma^2\over {\rm d}\ln k} ={V\over (2\pi)^3}
\, 4\pi \,k^3\, |\delta_k|^2
$$
($V$ being a normalization volume), and the relation to the correlation function is
$$
\xi(r)=\int \Delta^2(k)\; {dk\over k}\; {\sin kr\over kr}.
$$
The density field for a distribution of clumps is produced by
convolution of the initial field of delta-functions, so the
power spectrum is simply modified by the squared Fourier
transform of the clump density profile:
$$
\Delta^2(k)={4\pi \over n}\, \left({k\over 2\pi}\right)^3\; |W_k|^2,
$$
where
$$
W_k= {
\int \rho(r) \, {\displaystyle \sin k r \over \displaystyle k r}\, 4\pi \, r^2\; dr
\over
\int \rho(r) \, 4\pi \, r^2\; dr
}
.
$$

For a practical calculation,
we should also use the fact that hierarchical models are expected to
contain a distribution of masses of clumps.
If we use the notation $f(M)\, dM$ to denote the number density
of haloes in the mass
range $dM$, the effective number density in the shot noise
formula becomes
$$
{1\over n_{\rm eff}} = { \int M^2\, f(M)\, dM \over \left[\,\int M\, f(M)\, dM\,\right]^2 }.
$$
The window function also depends on mass, so the overall power spectrum is
$$
\Delta^2_{\rm halo}(k)=4\pi\, \left({k\over 2\pi}\right)^3\; { \int M^2\, |W_k(M)|^2
\, f(M)\, dM \over \left[\,\int M\, f(M)\, dM\,\right]^2 }.
$$
The normalization term $\int M\, f(M)\, dM$ just gives the total background
density, $\rho_b$, so there is only a single numerical integral to perform.
Using this model,
it is then possible to calculate the correlations
of the nonlinear density field, neglecting only the
large-scale correlations in halo positions. The
power spectrum determined in this way is shown in figure \ref{fig:halomodel_split},
and turns out to agree very well with the
exact nonlinear result on small and intermediate scales.
The lesson here is that a good deal of the
nonlinear correlations of the dark matter field
can be understood as a distribution of random clumps,
provided these are given the correct distribution of
masses and mass-dependent density profiles.

\japfigbasic{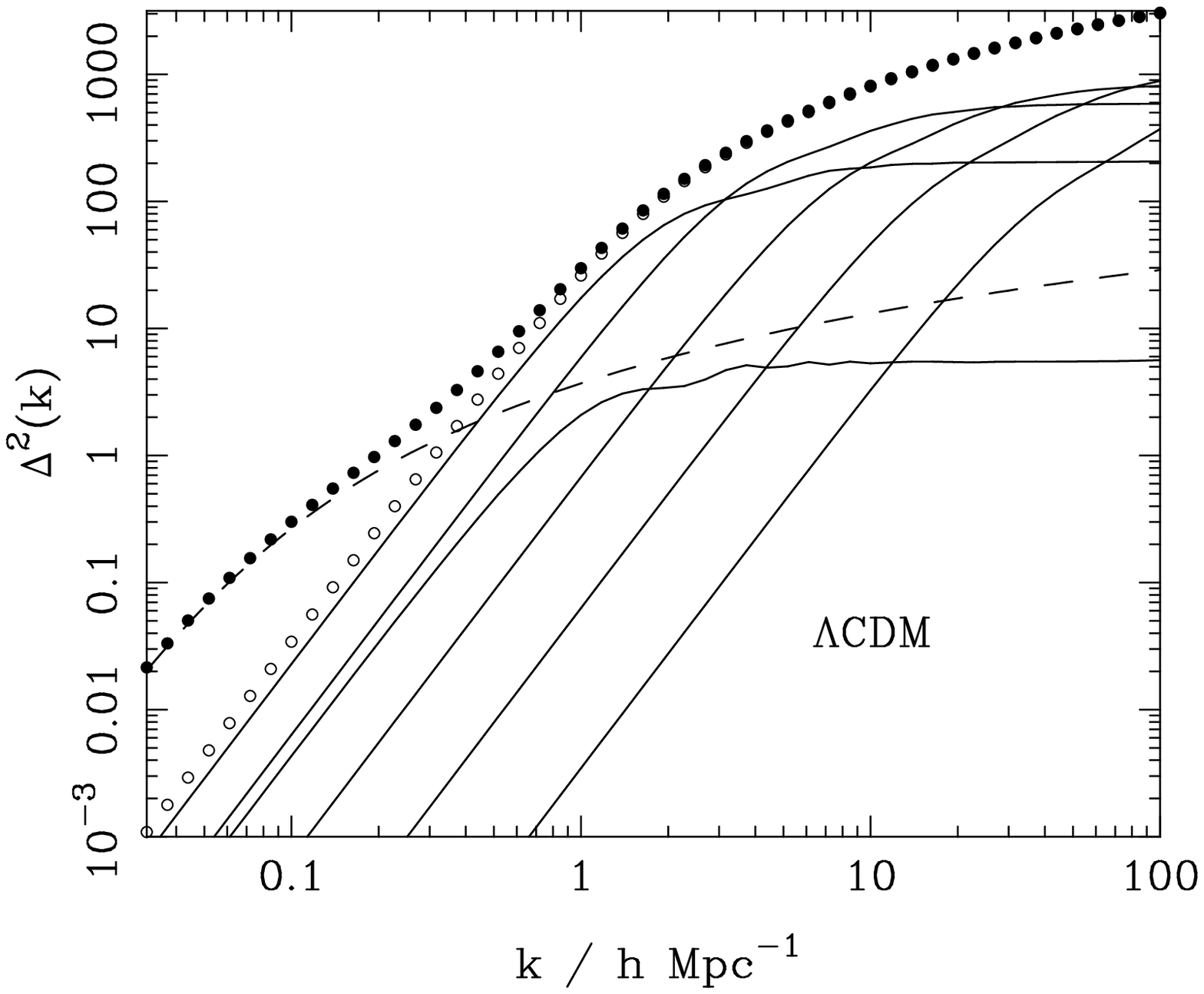}{0.7}
{The decomposition of the mass power spectrum according to the
halo model, for the flat $\Omega_m=0.3$, $\Gamma=0.2$,
$\sigma_8=0.8$ case. The dashed line shows linear theory,
and the open circles show the predicted 1-halo contribution.
Adding in linear theory to produce the correct large-scale
clustering yields the solid points. The full lines show
the contribution of different mass ranges to the 1-halo
term: bins of width a factor 10 in width, starting at
$10^{10}-10^{11} h^{-1}M_\odot$ and ending at
$10^{15}-10^{16} h^{-1}M_\odot$. The more massive haloes have
larger virial radii and hence filter the power spectrum on
progressively larger scales. The majority of the quasilinear
power is contributed by the haloes near the peak in the mass function
at $10^{14}-10^{15} h^{-1}M_\odot$.
}

So far, we have ignored any spatial correlations in the halo positions.
A simple guess for amending this is to add  the linear power spectrum
to the power generated by the halo structure:
$$
\Delta^2_{\rm tot} = \Delta^2_{\rm halo} + \Delta^2_{\rm linear}.
$$
The justification for this is that the extra small-scale power introduced
by nonlinear evolution is associated with the internal structure of the
haloes. In practice, this model
works extremely well, giving an almost perfect description of the
power spectrum on all scales.
This is a novel way of looking at the features in the
nonlinear spectrum, particularly the steep rise between
$k\simeq 0.5\hompc$ and $k\simeq 5\hompc$, and the flattening on
smaller scales. According to the ideas presented here, the
flat small-scale spectrum arises because haloes have central
density profiles rising as $r^{-1.5}$, but not much faster.
The sharp fall in power at smaller $k$ reflects the cutoff at the virial
radii of the haloes that dominate the correlation signal.

It might be objected that this model is still not completely
realistic, since we have treated haloes as smooth objects
and ignored any substructure.
At one time, it was generally believed that 
collisionless evolution would lead to the destruction
of galaxy-scale haloes when they are absorbed into the
creation of a larger-scale nonlinear system such as a group
or cluster. However, it turns out that this
`overmerging problem' was only an artefact of
inadequate resolution (see e.g. van Kampen 2000). 
When a simulation is carried out
with $\sim 10^6$ particles in a rich cluster, the cores of
galaxy-scale haloes can still be identified after many crossing
times (Ghigna et al. 1998). This substructure must have
some effect on the correlations of the density field, and indeed
Valageas (1999) has argued that the
high-order correlations of the density field seen in
$N$-body simulations are inconsistent with a model where the
density file is composed of smooth virialized haloes.
Nevertheless, substructure seems to be unimportant at the 
level of two-point correlations.

The existence of substructure is important for the obvious next step
of this work, which is to try to understand galaxy correlations
within the current framework. It is clear that the 
galaxy-scale substructure in large dark-matter haloes defines
directly where luminous  galaxies will be found, giving hope that the 
main features of galaxy formation can be understood principally in
terms of the dark-matter distribution.
Indeed, if catalogues of these `sub-haloes' are created within a cosmological-sized
simulation, their correlation function is known to differ from that
of the mass (e.g. Klypin et al. 1999; Ma 1999).
The model of a density field consisting of smooth haloes
may therefore be a useful description of the galaxy
field, and this is explored in the following section.

\ssec{The Halo model -- II: biased galaxy populations}

In relating the distribution of galaxies to that of the mass,
there are two distinct ways in which a degree of bias is inevitable:

\japitem{(1)} Halo occupation numbers. For low-mass haloes, the
probability of obtaining an $L^*$ galaxy must fall to zero.
For haloes with mass above this lower limit, the number of
galaxies will in general not scale linearly with halo mass.

\japitem{(2)} Nonlocality. Galaxies can orbit within their
host haloes, so the probability of forming a galaxy depends
on the overall halo properties, not just the density at a point.
Also, the galaxies can occupy special places within
the haloes: for a halo containing only one galaxy, the
galaxy will clearly mark the halo centre. In general,
we will {\it assume\/} one central galaxy and a number of satellites.

\endjapitem
The first mechanism leads to large-scale bias, because
large-scale halo correlations depend on mass, and are some
biased multiple of the mass power spectrum:
$\Delta^2_h = b^2(M) \Delta^2$.
As discussed earlier,
the linear bias parameter for a given class of 
haloes, $b(M)$, depends on the rareness of
the fluctuation and the rms of the underlying field:
$$
b=1+{\nu^2-1\over \nu\sigma}= 1+ {\nu^2-1\over \delta_c}
$$
(Kaiser 1984; Cole \& Kaiser 1989; Mo \& White 1996),
where $\nu = \delta_c/\sigma$, and $\sigma^2$ is the
fractional mass variance at the redshift of interest.
This formula is not perfectly accurate, but the 
deviations may be traced to the fact that the Press-Schechter
formula for the number density of haloes (which is assumed
in deriving the bias) is itself systematically in
error; see Sheth \& Tormen (1999).

If we do not wish to assume that the number of galaxies in
a halo of mass $M$ is strictly proportional to $M$, we
are in effect giving haloes a mass-dependent weight, as
was first considered by Jing, Mo \& B\"orner (1998).
A simple but instructive model for this is
$$
w(M) = \cases{
0\quad &($M<M_c$)\cr
(M/M_c)^{\alpha-1}\quad &($M>M_c$)\cr
}
$$
A model in which mass traces light would have
$M_c \rightarrow 0$ and $\alpha=1$. We will show
below that, empirically, we should choose $\alpha <1$.

The bias formula applies to haloes of a given $\nu$, i.e. of
a given mass, so the effect of mass-dependent weights is
$$
b_{\rm tot} = 1 + {
\int_\nu^\infty b(\nu)\, w(\nu)\, {dF\over d\nu}\; d\nu
\over
\int_\nu^\infty w(\nu)\, {dF\over d\nu}\; d\nu,
}
$$
Where $F(>\nu)$ is the fraction of the mass in haloes exceeding a given $\nu$;
$dF/d\nu \propto \exp(-\nu^2/2)$ according to Press-Schechter theory.
The total model for the galaxy power spectrum is then
$$
\Delta^2_g =\left\langle \Delta^2_{\rm halo} \right\rangle
+ b_{\rm tot}^2 \Delta^2_{\rm lin}
$$
where
$$
\left\langle \Delta^2_{\rm halo}(k) \right\rangle
=4\pi\, \left({k\over 2\pi}\right)^3\; { 
\int M^2\, w^2(M)\, |W_k(M)|^2
\, f(M)\, dM \over \left[\,\int M\, w(M)\, f(M)\, dM\,\right]^2 }.
$$

The key ingredient needed to make this machinery work is
the occupation number, which in principle needs to be
calculated via a detailed  numerical model
of galaxy formation.
However, for a given assumed background cosmology, the
answer may be determined empirically.
Galaxy redshift surveys have been analyzed via grouping
algorithms similar to the `friends-of-friends' method
widely employed to find virialized clumps in $N$-body
simulations. With an appropriate correction for the
survey limiting magnitude, the observed number of galaxies in
a group can be converted to an estimate of the total
stellar luminosity in a group. This allows a
determination of the All Galaxy System (AGS)
luminosity function: the distribution of virialized
clumps of galaxies as a function of their total
luminosity, from small systems like the Local Group to
rich Abell clusters.

\japfig{57}{211}{515}{588}{0.7}
{lumm}
{The empirical luminosity--mass relation
required to reconcile the observed AGS luminosity function
with two variants of CDM. $L^*$ is the characteristic
luminosity in the AGS luminosity function
($L^* = 7.6\times 10^{10}h^{-2} L_\odot$).
Note the rather flat slope around
$M=10^{13}$ to $10^{14}h^{-1}M_\odot$,
especially for $\Lambda$CDM.}

\japfigbasic{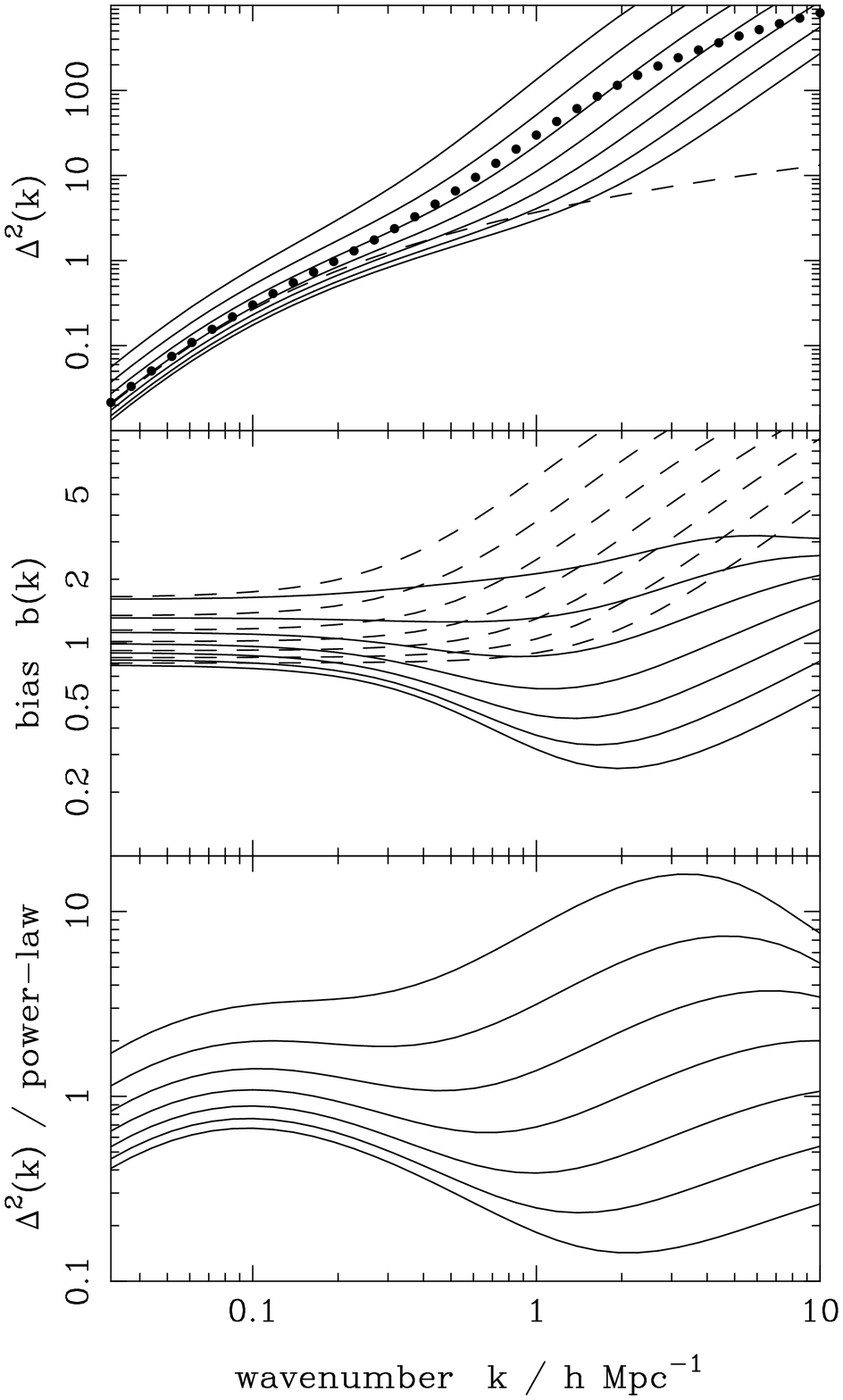}{0.6}
{The power spectrum predicted for the halo model,
for the flat $\Omega_m=0.3$, $\Gamma=0.2$,
$\sigma_8=0.8$ case. 
The halo occupation numbers are predicted according to a simple
two-parameter model, in which haloes are only included above
some minimum value, $M_{\rm min}$, and receive a weight
$\propto M^{\alpha}$. A model with `mass traces light'
would have $M_{\rm min}=0$ and $\alpha=1$. Motivated
by the results on abundances of groups, we choose
$\alpha=0.5$. Results are shown for $M_{\rm min}$ in the
range $10^{10}$ (lowest) to $10^{12.5}h^{-1} M_\odot$ (highest).
The mass power spectrum is shown as a dashed line
(linear theory) and as points (nonlinear).
The second panel shows the scale-dependent relative bias,
i.e. the square root of the ratio between galaxy and mass
power spectra. The dashed lines show the bias with respect
to linear theory. The final panel shows the ratio to the
canonical power-law spectrum; for $M_{\rm min}$ in the
region of $10^{12}h^{-1} M_\odot$, the result is within
a factor 2 of a perfect power law over a factor 300 in scale.
}

The AGS function for the CfA survey was investigated by
Moore, Frenk \& White (1993), who found that the
result in blue light was well described by
$$
d\phi = \phi^*\, \left[ (L/L^*)^\beta + (L/L^*)^\gamma \right]^{-1}\;
dL/L^*,
$$
where $\phi^*=0.00126h^3\rm Mpc^{-3}$, $\beta=1.34$, $\gamma=2.89$;
the characteristic luminosity is 
$L^*  = 7.6\times 10^{10}h^{-2} L_\odot$.
One notable feature of this function is that it is
rather flat at low luminosities, in contrast to the
mass function of dark-matter haloes (see Sheth \& Tormen 1999).
It is therefore clear that any fictitious galaxy catalogue
generated by randomly sampling the mass is unlikely to be a
good match to observation.
The simplest cure for this deficiency is to assume that the
stellar luminosity per virialized halo is a monotonic, but nonlinear,
function of halo mass. The required luminosity--mass
relation is then easily deduced by finding the luminosity
at which the integrated AGS density $\Phi(>L)$ matches the
integrated number density of haloes with mass $>M$.
The result is shown in figure \ref{fig:lumm}.

We can now calculate the halo-based galaxy power spectrum
and use semi-realistic occupation numbers, $N$, as a function of 
mass. This is needs a little care at small numbers,
however, since the number of haloes with occupation number unity
affects the correlation properties. These
haloes contribute no correlated pairs, so they simply
dilute the signal from the haloes with $N\ge 2$.
This means that we need in principle to use different
weights for the large-scale bias and the halo term:
$$
w_i = {\left\langle N_i \right\rangle \over M} \quad\quad
w_i = {\left\langle N_i(N_i-1) \right\rangle^{1/2} \over M} 
$$
respectively (Seljak 2000). In practice, this correction has
a rather small effect, provided the relation between $N$ and $M$
has no scatter. If, in contrast, the distribution of $N$
for given $M$ is assumed to obey a Poisson distribution,
the small-scale clustering properties are strongly affected, and do
not match the data well (Benson et al. 2000a).
Finally, we need to put the
galaxies in the correct location, as discussed above.
If one galaxy always occupies the halo centre, with others
acting as satellites, the small-scale correlations automatically
follow the slope of the halo density profile, which keeps them
steep. The results of this exercise are shown in figure~\ref{fig:halomodel_bias}.
This shows that, depending on the range of halo masses chosen,
the galaxies can be positively or negatively biased with
respect to the mass, as expected. What is particularly
interesting is that the shape of the galaxy spectrum is
expected to differ from that of the mass. For an appropriate
mass range, the galaxy power spectrum can be very close to
a power law, which has been a long-standing puzzle to explain.
Interestingly, the power-law should not be perfect; small
deviations have long been suspected, and were confirmed by
Hawkins et al. (2002) and Zehavi et al. (2003). The inflection
is at a scale of $\sim 0.5\hompc$, as expected from the halo model.
Figure~\ref{fig:halomodel_bias} also shows that
the results of this simple model are encouragingly similar to
the scale-dependent bias found in the detailed calculations of
Benson et al. (2000a), shown in figure \ref{fig:benson}.
There are thus grounds for optimism that we may
be starting to attain a physical understanding of the
origin of galaxy bias.

\sec{Anisotropies in the CMB}

Despite the great progress in precise measurements of large-scale
structure, we cannot achieve a complete specification of the
cosmological model in this way alone. The vacuum energy is
not probed, since this affects mainly the growth rate of
structure -- which is degenerate with bias evolution.
The matter content is only constrained if we assume that $n=1$,
and even then we only measure $\Omega_m$ if a value for $h$
is supplied. A more complete picture is obtained if we
include data on clustering at much earlier times: the 
anisotropy of the microwave background, which reaches us
from $z\simeq 1100$. In addition to breaking degeneracies,
studies of this sort also test the basic gravitational
instability theory -- which will be seen to work very well indeed
over this redshift range.
This section briefly reviews the physics of CMB
anisotropies, and presents recent data. For more details,
see e.g. Hu \& Dodelson (2002), or Dodelson (2003).

\ssec{Anisotropy mechanisms}

Fluctuations in the 2D temperature perturbation field are treated
similarly to density fluctuations, except that the field
is expanded in spherical harmonics, so modes of different scales
are labelled by multipole number, $\ell$. Once again, we
can define a `power per octave' measure for the 
temperature fluctuations:
$$
{\mathcal T}^2(\ell) = \ell(\ell+1) C_\ell / 2\pi; \quad\quad
\left\langle (\delta T/T)^2\right\rangle = \sum_\ell (2\ell+1)\, C_\ell/4\pi,
$$
where the $C_\ell$ are another common way of representing the power.
Note that ${\mathcal T}^2(\ell)$ is a power per
$\ln\ell$; the modern trend is often to plot CMB fluctuations
with a linear scale for $\ell$ -- in which case one should really
use ${\mathcal T}^2(\ell)/\ell$.

We now list the mechanisms that cause \japkey{primary anisotropies} 
in the CMB (as opposed to
\japkey{secondary anisotropies}, which are
generated by scattering along the line of sight).
There are three basic primary effects, 
illustrated in figure~\ref{fig:cmb_mech}, which are
important on respectively large, intermediate
and small angular scales:

\smallskip
\noindent
(1) Gravitational (Sachs--Wolfe) perturbations.
Photons from high-density regions at last scattering
have to climb out of potential wells, and are thus redshifted:
$$
{\delta T\over T} ={\Phi\over c^2}.
$$

\noindent
(2) Intrinsic (adiabatic) perturbations.
In high-density regions, the coupling of matter and radiation
can compress the radiation also, giving a higher temperature:
$$
{\delta T\over T} = {\delta(z_{\japsub LS})\over 3},
$$

\noindent
(3) Velocity (Doppler) perturbations. The plasma
has a non-zero velocity at recombination, which leads to
Doppler shifts in frequency and hence brightness temperature:
$$
{\delta T\over T} ={{\bf \delta v\cdot\hat r}\over c}.
$$

To the above list should be added `tensor modes': anisotropies
due to a background of primordial gravitational waves,
potentially generated during an inflationary era (see below).

There are in addition effects generated along the line of sight.
One important effect is the integrated Sachs-Wolfe effect,
which arises when the potential perturbations evolve:
$$
{\delta T\over T} = 2 \int {\dot\Phi\over c^2}\; dt.
$$
This happens both at early times (because radiation is
still important) and late times (because of $\Lambda$).
Other effects are to do with the development of nonlinear
structure, and are mainly on small scales (principally
the Sunyaev--Zeldovich effect from IGM Comptonization).
The exception is the effect of reionization; to a good
approximation, this merely damps the fluctuations on
all scales:
$$
{\delta T\over T} \quad\rightarrow\quad {\delta T\over T}\, \exp-\tau,
$$
where the optical depth must exceed
$\tau \simeq 0.04$, based on the highest-redshift quasars
and the BBN baryon density.  As we will see later, CMB
polarization data have detected a signature consistent
with $\tau = 0.17 \pm 0.04$, implying reionization at
$z\simeq 20$.

\japfig{0}{0}{498}{252}{0.8}
{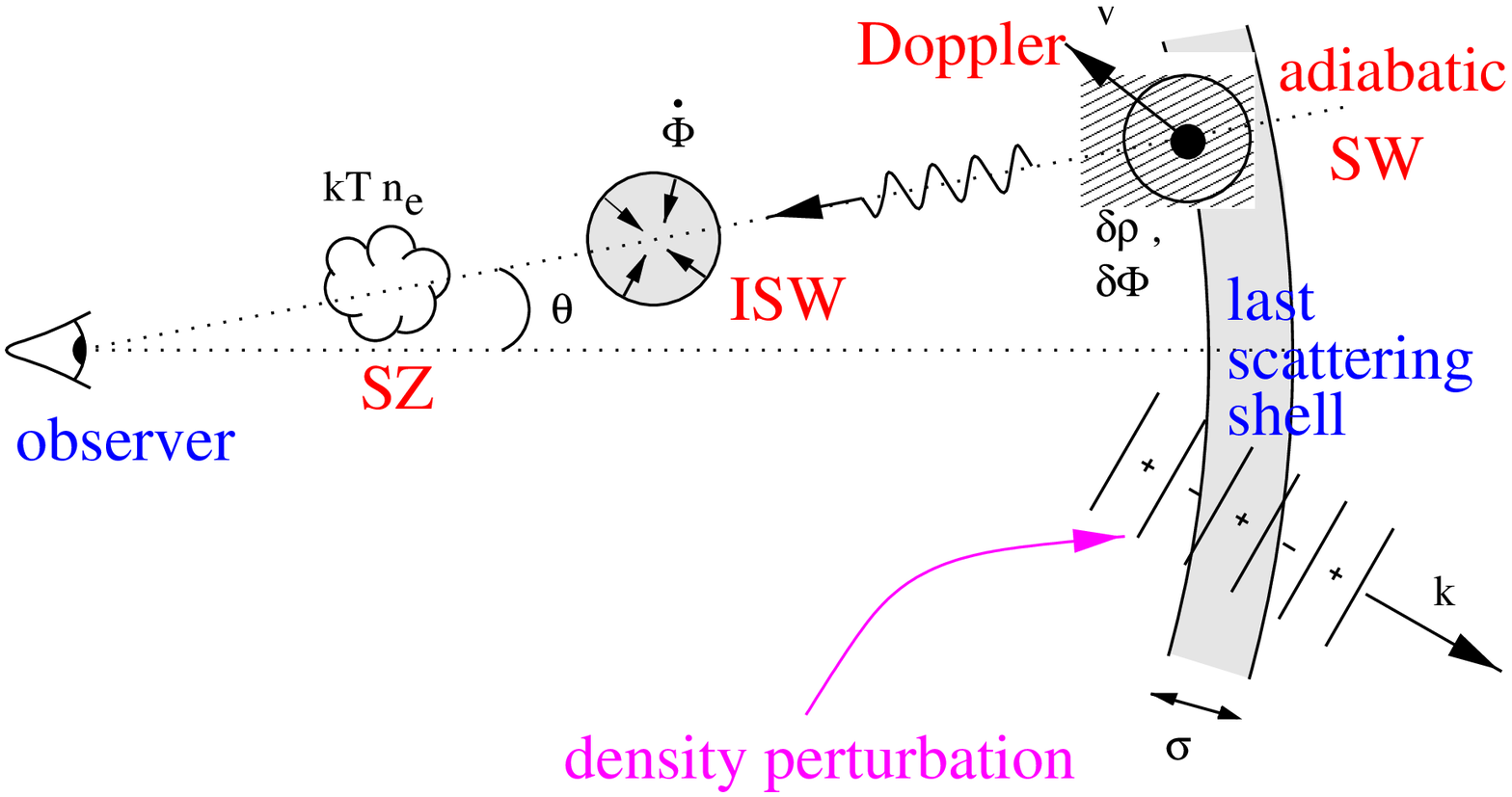}
{Illustrating the physical mechanisms that cause CMB anisotropies.
The shaded arc on the right represents the last-scattering shell;
an inhomogeneity on this shell affects the CMB through its
potential, adiabatic and Doppler perturbations.
Further perturbations are added along the
line of sight by time-varying potentials (Rees--Sciama effect)
and by electron scattering from hot gas (Sunyaev--Zeldovich effect).
The density field at last scattering can be Fourier analysed
into modes of wavevector ${\bf k}$. These spatial perturbation modes
have a contribution that is in general damped 
by averaging over the shell of last scattering.
Short-wavelength modes are more heavily affected (i) because
more of them fit inside the scattering shell, and (ii)
because their wavevectors point more nearly radially for
a given projected wavelength.}

\ssec{Inflationary predictions}

The most commonly-discussed mechanism for generating the inhomogeneities
that act as the source for $\delta T/T$ is inflation.
Of course, CMB anisotropies 
were calculated in largely the modern way well
before inflation was ever considered, by Peebles \& Yu (1970).
The standard approach involves
super-horizon fluctuations, which must
be generated by some acausal process. Inflation achieves this -- 
but we cannot claim that detection of super-horizon modes amounts
to a proof of inflation. Rather, we need some more
characteristic signature of the specific process used
by inflation: amplified quantum fluctuations
(see e.g. chapter 11 of Peacock 1999 or Liddle \& Lyth 2000
for details).

In the simplest models, inflation is driven by a scalar field $\phi$, with a potential
$V(\phi)$. As well as the characteristic energy density
of inflation, $V$, this can be characterized by two
parameters, $\epsilon$ \& $\eta$, which are dimensionless
versions of the first and second derivatives of $V$ with respect to $\phi$.
In these terms, the inflationary predictions for the perturbation 
index is
$$
n = 1-6\epsilon +2\eta.
$$
Since inflation continues while $\epsilon$ \& $\eta$ are small,
some tilt is expected ($|n-1| \sim 0.01$ to 0.05 in simple models).

The critical ingredient for testing inflation by making
further predictions is the possibility that, in addition to scalar modes,
the CMB could also be affected by gravitational waves 
(following the original insight of Starobinsky 1985). 
The relative amplitude of
tensor and scalar contributions depended on the inflationary
parameter $\epsilon$ alone:
$$
{C_\ell^{\japsub T}\over C_\ell^{\japsub S}} \simeq 12.4\epsilon \simeq 6(1-n_{\japsub S}).
$$
The second relation to the \japkey{tilt} is less general, as it assumes
a polynomial-like potential, so that $\eta$ is related to $\epsilon$.
For example,
$V=\lambda \phi^4$ implies $n_{\japsub S}\simeq 0.95$ and
$C_\ell^{\japsub T}/C_\ell^{\japsub S}\simeq 0.3$.
To be safe, we need one further observation, and this
is potentially provided by the spectrum of $C_\ell^{\japsub T}$.
Suppose we write separate power-law index definitions for the
scalar and tensor anisotropies:
$$
C_\ell^{\japsub S} \propto \ell^{n_{\japsub S}-3},\quad\quad
C_\ell^{\japsub T} \propto \ell^{n_{\japsub T}-3}.
$$
For the scalar spectrum, we had $n_{\japsub S}= n=1-6\epsilon +2\eta$;
for the tensors,
$n_{\japsub T}=1-2\epsilon$ [although different definitions of
$n_{\japsub T}$ exist; the convention
here is that $n=1$ always corresponds to a constant ${\mathcal T}^2(\ell)$].
Thus, a knowledge of $n_{\japsub S}$, $n_{\japsub T}$ and the
scalar-to-tensor ratio would overdetermine the model and allow
a genuine test of inflation.

\japfig{0}{0}{476}{485}{0.7}
{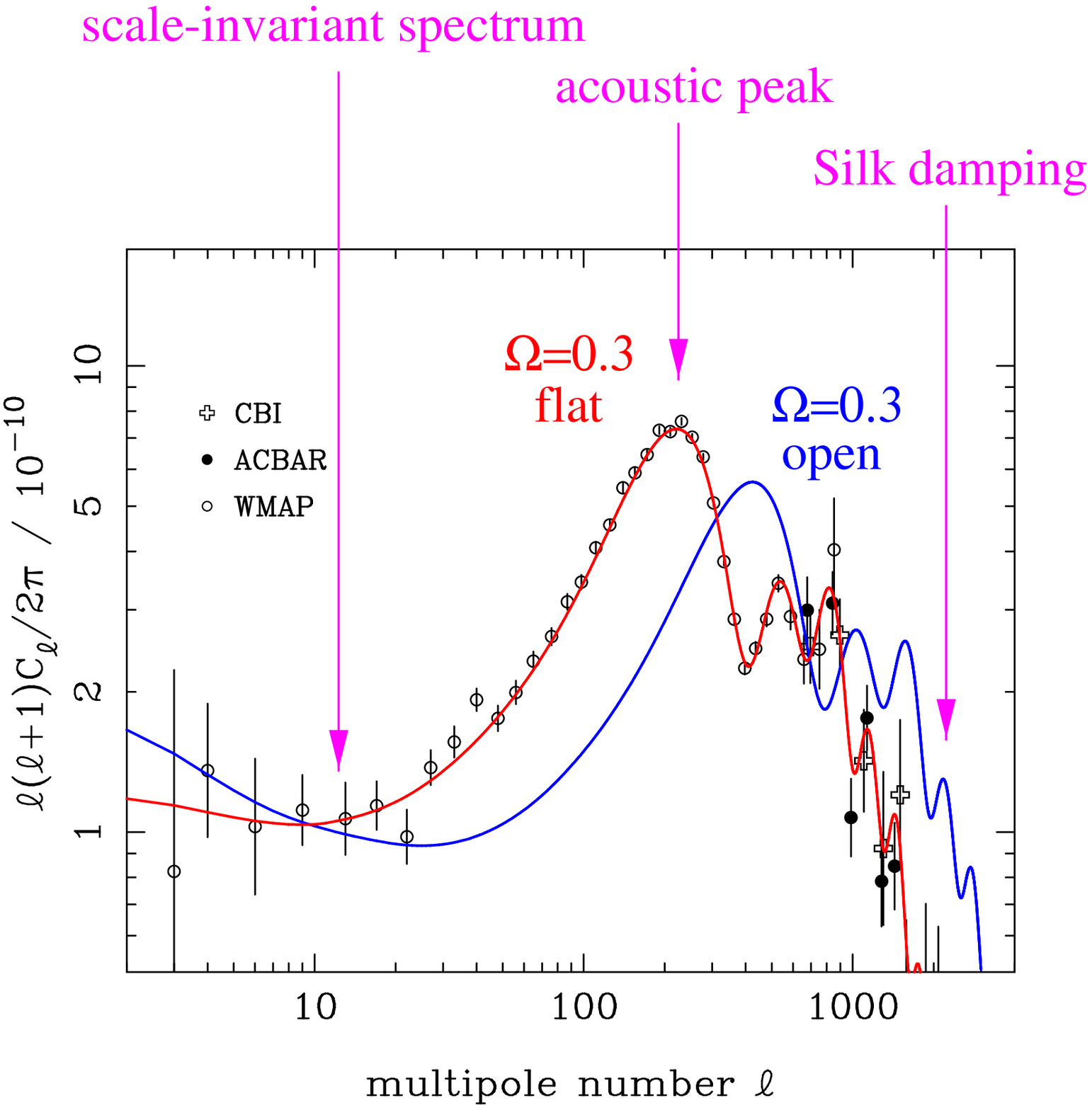}
{Angular power spectra ${\mathcal T}^2(\ell)=
\ell(\ell+1)C_\ell/2\pi$ for the CMB,
plotted against angular wavenumber $\ell$ in radians$^{-1}$.
For references to the experimental data,
see Spergel et al. (2003),  Kuo et al. (2002) and Pearson et al. (2002).
The two lines show
model predictions for adiabatic scale-invariant CDM fluctuations,
calculated using the ${\japsub CMBFAST}$ package
(Seljak \& Zaldarriaga 1996). These have
$(n,\Omega_m,\Omega_b,h)=(1,0.3,0.05,0.65)$ and have
respectively $\Omega_v=1-\Omega_m$ (`flat') and $\Omega_v=0$ (`open').
The main effect is that open models shift the peaks to the
right, as discussed in the text.
}

\ssec{Characteristic scales}

The current data are contrasted with some CDM models in
figure~\ref{fig:cmb_pow}. The key feature that is picked out is the peak
at $\ell\simeq 220$, together with harmonics of this
scale at higher $\ell$. Beyond $\ell\simeq 1000$, the spectrum
is clearly damped, in a manner consistent with the expected
effects of photons diffusing away from baryons (Silk damping),
plus smearing of modes with wavelength comparable to the
thickness of the last-scattering shell. This last effect
arises because recombination is not instantaneous, so the
redshift of last scattering shows a scatter around the mean,
with a thickness corresponding to approximately
$\sigma_r=7(\Omega_m h^2)^{-1/2}$~Mpc. On scales larger
than this, we see essentially an instantaneous imprint
of the pattern of potential perturbations and the
acoustic baryon/photon oscillations.

The significance of the main acoustic peak
scale is that it picks out the (sound) horizon at last scattering.
The redshift of last scattering is almost independent of
cosmological parameters at $z_{\japsub LS}\simeq 1100$, although a more
precise approximation is given in Appendix C of Hu \& Sugiyama (1995).
If we assume that the universe is matter dominated at
last scattering, the horizon size is
$$
D_{\japsub H}^{\japsub LS} = 184\, (\Omega_m h^2)^{-1/2} {\rm Mpc}.
$$
The angle this subtends is given by dividing by the current size
of the horizon (strictly, the comoving angular-diameter distance
to $z_{\japsub LS}$). Again, for a matter-dominated model, this is
$$
D_{\japsub H} = 6000\, \Omega_m^{-1}\, h^{-1} {\rm Mpc}
\quad\Rightarrow\quad \theta_{\rm \japsub H} = D_{\japsub H}^{\japsub LS} / D_{\japsub H}
\propto \Omega_m^{0.5}.
$$
This expression lies behind the common statement that
the CMB data require a flat universe. Figure~\ref{fig:cmb_pow} shows that
heavily open universes yield a main CMB peak at scales much smaller
than the observed $\ell\simeq 220$, and these can be ruled out.
Indeed, open models were disfavoured for this reason long
before any useful data existed near the peak, simply 
because of strict upper limits at $\ell\simeq 1500$
(Bond \& Efstathiou 1984). However, once a non-zero vacuum
energy is allowed, the story becomes more complicated,
and it turns out that large degrees of spatial curvature
cannot be excluded using the CMB alone.

\ssec{Evolution of CMB data}

The pace of progress in CMB experiments has maintained an
astonishing rate for a decade. Following the 1992 COBE
detection of fluctuations, 5 years of effort yielded the
unclear picture of the first panel in figure~\ref{fig:clplot_evol},
in which of order 10 experiments gave only vague evidence for a peak in
$\ell^2 C_\ell$. By the year 2000,
this had been transformed to a clear picture of a peak
at $\ell\simeq 200$, although there was no model-independent
evidence for higher harmonics.
The present situation is much more satisfactory,
with 3 peaks established in a way that does not require
any knowledge of the CDM model.

\japfig{80}{69}{486}{767}{0.75}{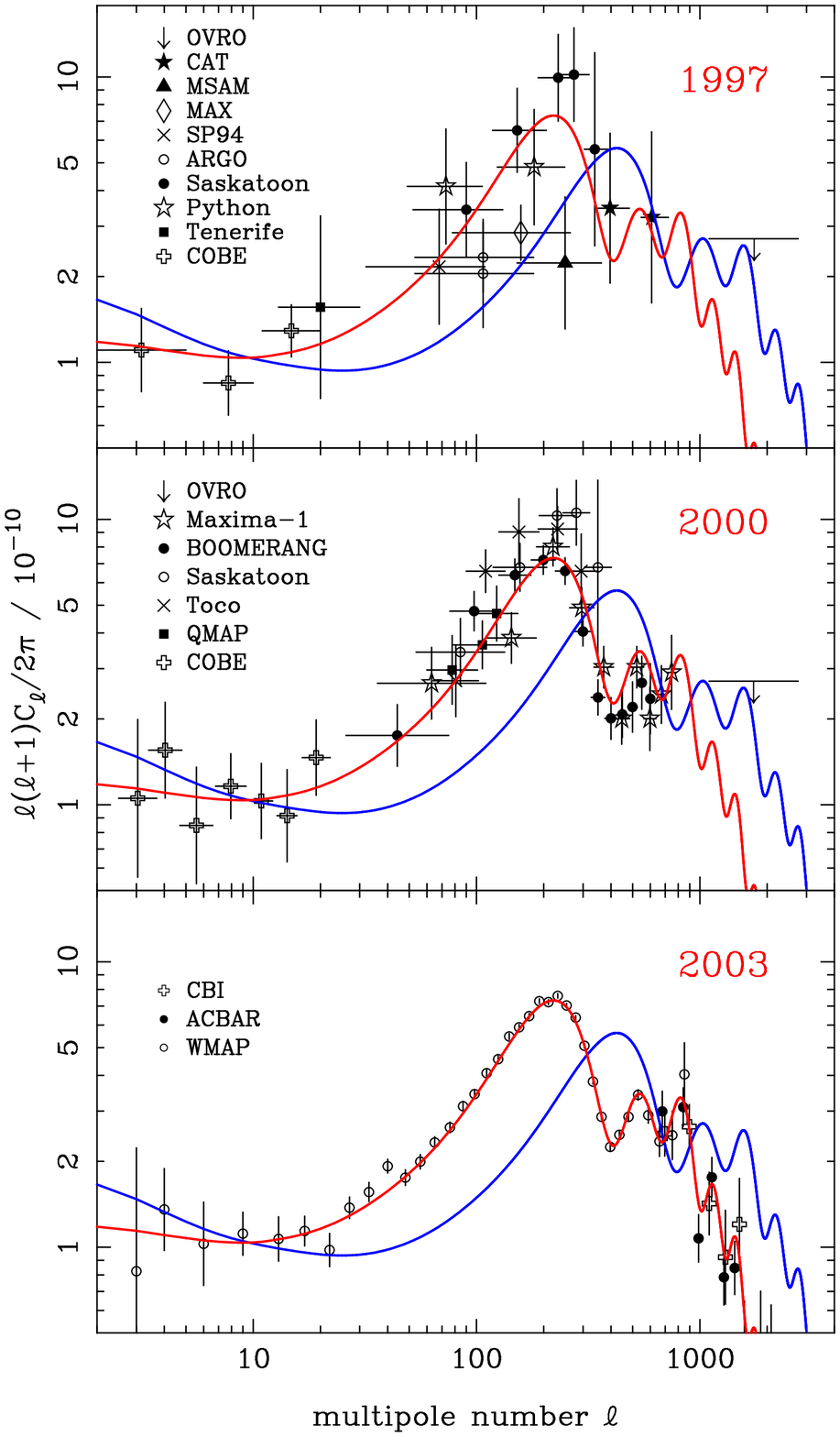}
{Dramatic change took place in CMB
power spectrum measurements around the turn of the 21st century.
Although some rise from the COBE level was arguably known even by 1997,
a clear peak around $\ell\simeq 200$ only became established in 2000,
whereas by 2003 definitive measurements of the spectrum at $\ell \ls 800$,
limited mainly by cosmic variance, had been made by the WMAP satellite.}

The WMAP results (Spergel et al. 2003) measure the power
spectrum about as well as possible (i.e. hitting the
limit of cosmic variance from a finite sky) up to
the second peak. At smaller scales, however, there is still
much scope for improvement, and the
rate of advance is unlikely to drop in the future. The
useful web page
{\tt http://background.uchicago.edu/{\tt\char'176}whu/cmbex.html}\ lists 
14 ongoing experiments, as well as 19 completed ones.

\sec{Model degeneracies inherent in CMB data}

\ssec{Geometrical degeneracy}

The normal argument for flatness from the CMB
starts with the comoving horizon size at last scattering
$$
D_{\japsub LS} = \frac{2c}{\Omega_m^{1/2} H_0}\,(1+z_{\japsub LS})^{-1/2} \simeq
184(\Omega_m h^2)^{-1/2}\; {\rm Mpc}
$$
and divides it by the present-day horizon size for a zero-$\Lambda$ universe,
$$
D_{\japsub H} = \frac{2c}{\Omega_m H_0},
$$
to yield a main characteristic angle that scales as $\Omega_m^{1/2}$.
Large curvature (i.e. low $\Omega_m$) is ruled out because the
main peak in the CMB power spectrum is not seen at very small
angles. However, introducing vacuum energy changes the conclusion.
If we take a
family of models with fixed initial perturbation spectra, fixed
physical densities $\omega_m \equiv \Omega_m h^2$,  
$\omega_b \equiv \Omega_b h^2$,
it is possible to vary both $\Omega_v$ and the curvature to keep a fixed value of
the angular size distance to last scattering, so that the
resulting CMB power spectra are identical.
This degeneracy was analyzed comprehensively by Efstathiou \& Bond (1999), and
we now summarize the main results.

\japfig{56}{53}{519}{764}{0.6}
{cmbdegen}
{The geometrical degeneracy in the CMB means
that models with fixed $\Omega_m h^2$ and
$\Omega_b h^2$ can be made to look identical by
varying the curvature against vacuum energy,
while also varying the Hubble parameter.
This degeneracy is illustrated here for the case
$\omega_m\equiv \Omega_m h^2=0.2$.
Models along a given line are equivalent from
a CMB point of view; corresponding lines in the
upper and lower panels have the same line style. 
Note that
supplying external information about $h$ breaks
the degeneracy.
This figure assumes scalar fluctuations only;
allowing tensor modes introduces additional
degeneracies -- mainly between the tensor
fraction and tilt.}

The usual expression for the comoving angular-diameter distance is
$$
\eqalign{
R_{0}S_{k}(r) = & {c \over H_{0}} |1-\Omega|^{-1/2} \quad \times \cr
& S_k \left[\int_{0}^{z}\!\!{|1-\Omega|^{1/2} \; dz' \over 
\sqrt{(1-\Omega)(1+z')^{2}+\Omega_v +
\Omega_m(1+z')^{3}}}\right],
}
$$
where $\Omega=\Omega_m+\Omega_v$. 
Defining $\omega_i \equiv \Omega_i h^2$, this can
be rewritten in a way that has no explicit $h$ dependence:
$$
R_{0}S_{k}(r) = {3000 \,{\rm Mpc} \over |\omega_k|^{1/2}}
S_k \left[\int_{0}^{z}\!\!{|\omega_k|^{1/2}\; dz' \over
\sqrt{\omega_k (1+z')^{2}+\omega_v +
\omega_m(1+z')^{3}}}\right],
$$
where $\omega_k \equiv (1-\Omega_m-\Omega_v)h^2$. This parameter
describes the curvature of the universe, treating it effectively as
a physical density that scales as $\rho \propto a^{-2}$.
This is convenient for the present formalism, but 
it is important to appreciate that curvature differs fundamentally
from a hypothetical fluid with such an equation of state.

The sound horizon distance at last scattering is 
governed by the relevant physical densities, $\omega_m$ and $\omega_b$;
if $\omega_m$ and $\omega_b$ are given, the shape of the
spatial power spectrum is determined. The translation of this
into an angular spectrum depends on the angular-diameter distance,
which is a function of these parameters, plus $\omega_k$ and
$\omega_v$. Models in which $\smash{\omega_m^{1/2}} R_0 S_k(r)$ is
a constant have the same angular horizon size.
For fixed $\omega_m$ and $\omega_b$, there is therefore a degeneracy between
curvature ($\omega_k$) and vacuum ($\omega_v$): these two parameters
can be varied simultaneously to keep the same apparent distance,
as illustrated in figure~\ref{fig:cmbdegen}.

In short, this degeneracy occurs because the physical densities
control the structure of the perturbations in physical Mpc at last
scattering, while curvature, $\Omega_v$ and $\Omega_m$
govern the proportionality
between length at last scattering and observed angle.
The degeneracy is not exact, and is
weakly broken by the \key{Integrated Sachs-Wolfe effect}
from evolving potentials at very low multipoles, 
and second-order effects at high $\ell$.
However, strong breaking of the degeneracy requires additional information.
This could be in the form of external data on the Hubble
constant, which obeys the relation 
$$
h^2 = \omega_m + \omega_v + \omega_k,
$$
so specifying $h$ in addition to the physical matter density
fixes $\omega_v + \omega_k$ and removes the degeneracy.
A more elegant approach is to add results from large-scale structure,
so that conclusions are based only on the shapes of power spectra.
Efstathiou et al. (2002) show that doing this
yields a total density ($|\Omega-1|<0.05$) at 95\% confidence.

\ssec{Horizon-angle degeneracy}

As we have seen, the geometrical degeneracy can be broken either by
additional information (such as a limit on $h$), or by invoking
a theoretical prejudice in favour of flatness. Even 
for flat models, however, there still
exists a version of the same degeneracy.
What determines the CMB peak locations for flat models?
The horizon size at last scattering is
$D_{\japsub H}^{\japsub LS} = 184\, (\Omega_m h^2)^{-1/2} {\rm Mpc}$.
The angular scale of these peaks
depends on the ratio between the horizon size at last scattering
and the present-day horizon size for flat models:
$$
D_{\japsub H} = 6000\, \Omega_m^{-0.4}\, h^{-1} {\rm Mpc}
\quad\Rightarrow\quad \theta_{\rm \japsub H} = D_{\japsub H}^{\japsub LS} / D_{\japsub H}
\propto \Omega_m^{-0.1}.
$$
(using the approximation of Vittorio \& Silk 1985).
This yields an angle scaling as $\Omega_m^{-0.1}$, so that the
scale of the acoustic peaks is apparently almost independent
of the main parameters.

However, this argument is incomplete because the earlier
expression for $D_{\japsub H}(z_{\japsub LS})$ assumes that
the universe is completely matter dominated at last scattering,
and this is not perfectly true.
The comoving sound horizon size at last scattering is defined by
(e.g. Hu \& Sugiyama 1995)
$$
 D_{\japsub S}(z_{\japsub LS}) \equiv \frac{1}{H_0 \Omega_m^{1/2}}
   \int_0^{a_{\japsub LS}} \frac{c_{\japsub S}}{(a + a_{\rm eq})^{1/2} } \, da
$$
where vacuum energy is neglected at these high redshifts;
the expansion factor $a \equiv (1+z)^{-1}$ and
$a_{\japsub LS}, a_{\rm eq}$ are the values at last scattering and
matter-radiation equality respectively.
In practice, $z_{\japsub LS}\simeq 1100$ independent of the matter
and baryon densities, and $c_{\japsub S}$ is fixed by 
$\Omega_b$. Thus the main effect is that $a_{\rm eq}$ depends
on $\Omega_m$.
Dividing by $D_{\japsub H}(z=0)$ therefore
gives the angle subtended today by the light horizon as
$$
  \theta_{\japsub H} \simeq  \frac{\Omega_m^{-0.1}}{\sqrt{1+z_{\japsub LS}}}
     \left[\sqrt{1 + \frac{a_{\rm eq}}{a_{\japsub LS}} } -
  \sqrt{\frac{a_{\rm eq}}{a_{\japsub LS}} }\, \right],
$$
where  $z_{\japsub LS} = 1100$ and $a_{\rm eq} = (23900 \,\omega_m)^{-1}$.
This remarkably simple result captures  most
of the parameter dependence of CMB peak locations within
flat $\Lambda$CDM  models.
Differentiating this equation near a fiducial $\omega_m = 0.147$ gives
$$
  \left.\frac{\partial\ln\theta_{\japsub
  H}}{\partial\ln\Omega_m}\right|_{\omega_m}
   = -0.1;
\quad
\quad
  \left.\frac{\partial\ln\theta_{\japsub
  H}}{\partial\ln\omega_m}\right|_{\Omega_m}=
  \frac{1}{2} \left( 1 + \frac{a_{\japsub LS}}{a_{\rm eq}}\right)^{-1/2} =  +0.24 ,
$$
in good agreement with the numerical derivatives 
in Eq.~(A15) of Hu et~al. (2001).  

Thus for moderate variations from a `fiducial' model, the CMB peak
multipole number scales approximately as $\ell_{\rm peak} \propto \Omega_m^{-0.14}
h^{-0.48}$, i.e.  the condition for constant CMB peak location is well
approximated as 
$$
\Omega_m h^{3.4} = {\rm constant}.
$$
However, information about the
peak heights does alter this degeneracy slightly; the relative peak
heights are preserved at constant $\Omega_m$, hence the actual likelihood
ridge is a `compromise' between constant peak location (constant
$\Omega_m h^{3.4}$) and constant relative heights (constant $\Omega_m
h^2$); the peak locations have more weight in this compromise, leading
to a likelihood ridge along approximately $\Omega_m h^{3.0} \simeq {\rm const}$
(Percival et al. 2002).
It is now clear how LSS data combines with the CMB: $\Omega_m h^{3.4}$
is measured to very high accuracy already, and
Percival et al. deduced $\Omega_m h^{3.4}= 0.078$ with an error of
about 6\% using pre-WMAP CMB data. The first-year WMAP results
in fact prefer $\Omega_m h^{3.4}= 0.084$ (Spergel et al. 2003); the slight
increase arises because WMAP indicates that previous datasets around
the peak were on average calibrated low.

\japfig{31}{183}{499}{579}{0.7}{tendegen}
{The tensor degeneracy. Adding a large tensor component to an
$n=1$ scalar model (solid line) greatly lowers the peak
(dashed line), once COBE normalization is imposed.
Tilting to $n=1.3$ cures this (dot-dashed line), but
the 2nd and subsequent harmonics are too high. Raising
the baryon density by a factor 1.5 (dotted line) leaves us
approximately back where we started.
}

\ssec{Tensor degeneracy}

All of the above applies to models in which scalar modes
dominate. The possibility of a large tensor component
yields additional degeneracies, as shown in figure~\ref{fig:tendegen}.
An $n=1$ model with a large
tensor component can be made to resemble a zero-tensor model
with large blue tilt ($n>1$) and high baryon content.
Efstathiou et al. (2002) show that adding LSS
data does not remove this degeneracy; this is reasonable,
since LSS data only constrain
the baryon content weakly.
A better way of limiting the possible tensor contribution is to
look at the amplitude of mass fluctuations today: this normalization
of the scalar component is naturally lower if the CMB signal is
dominated by tensors. These issues are discussed further below.

Another way in which the remaining degeneracy may be lifted is through
polarization of the CMB fluctuations.
A nonzero polarization is inevitable because the
electrons at last scattering experience an anisotropic
radiation field. Thomson scattering from an anisotropic
source will yield polarization, and the practical size of
the fractional polarization $P$ is of the order of the quadrupole radiation anisotropy
at last scattering: $P\gs 1$\%. 
This signal is expected to peak at $\ell \simeq 500$, and the
effect was first seen by the DASI experiment (Kovac et al. 2002).
Much more detailed polarization results were presented by
the WMAP satellite, including a critical detection of
large-scale polarization arising from secondary scattering at
low $z$, thus measuring the optical depth to last scattering
(Kogut et al. 2003).
On large scales, the polarization signature
of tensor perturbations differs from that of scalar perturbations
(e.g. Seljak 1997; Hu \& White 1997); the
different contributions to the total unpolarized $C_\ell$ can
in principle be disentangled, allowing the inflationary
test to be carried out.

\sec{Combination of the CMB and large-scale structure}

The 2dFGRS power spectrum contains important information about the
key parameters of the cosmological model, but we have seen that additional
assumptions are needed, in particular the values of $n$ and $h$.
Observations of CMB
anisotropies can in principle measure 
most of the cosmological parameters, and
combination with the 2dFGRS can lift most of the degeneracies inherent
in the CMB-only analysis. It is therefore of interest to see
what emerges from a joint analysis.

The clearest immediate result is that the geometrical
degeneracy becomes broken (Efstathiou et al. 2002).
A 95\% confidence upper limit on any curvature can be
set at $|\Omega-1|<0.05$. 
We can therefore be confident
that the universe is very nearly flat
so it is defensible to assume
hereafter that this is exactly true. The importance of tensors
will of course be one of the key questions for cosmology over the
next several years, but it is interesting to consider the limit
in which these are negligible. In this case, the standard model
for structure formation contains a vector of only 6 parameters:
$$
{\bf p} = (n_s, \Omega_m, \Omega_b, h, Q, \tau).
$$
Of these, the optical depth to last scattering, $\tau$, is almost entirely
degenerate with the normalization, $Q$ -- and indeed with
the bias parameter; we discuss this below.
The remaining four parameters are pinned down very precisely:
using a compilation of pre-WMAP CMB data plus the 2dFRGS power spectrum,
Percival et al. (2002) obtained
$$
(n_s, \omega_c, \omega_b, h) =
(0.963\pm 0.042, 0.115\pm 0.009, 0.021\pm0.002, 0.665 \pm 0.047),
$$
or an overall density parameter of $\Omega_m=0.313 \pm 0.055$.

It is remarkable how well these figures agree with completely
independent determinations: $h=0.72\pm 0.08$ from the HST key project
(Mould et al. 2000; Freedman et al. 2001);
$\Omega_b h^2 =0.020 \pm 0.001$ (Burles et al. 2001).
This gives confidence that the tensor component must
indeed be sub-dominant.

This analysis was published in Percival et al. (2002),
and is based on the preliminary version of the 2dFGRS power
spectrum, from Percival et al. (2001). We can make a
first estimate of how this is likely to change using the
$\Omega_m h = 0.18 \pm 0.02$ from the preliminary analysis
of $P(k)$ from the final dataset. In combination with
the WMAP $\Omega_m h^{3.4}= 0.084$ from the CMB peak degeneracy,
this yields
$$
\Omega_m = 0.25 \pm 15\% \quad
h = 0.73 \pm 5\%
$$
as the preferred current figures from an analysis of this type.
The matter density remains frustratingly imprecise, and
it is clear that it will be very hard to measure $h$ accurately
enough to cure this problem. However, complementary constraints
on $\Omega_m$ exist at similar precision (e.g. $\Omega_m = 0.28 \pm 18\%$
for a flat model from the SNe~Ia Hubble diagram;
Tonry et al. 2003).
With new results from gravitational lensing,
$\Omega_m$ should be measured to better than 10\%
precision within a year.

Perhaps the most striking conclusion from these results concerns
the nature of the primordial fluctuations, which remain
consistent with the $n=1$ scale-invariant form. The WMAP
analysis of Spergel et al. (2003) yields $0.97 \pm 0.03$
from CMB plus 2dFGRS (cf. $0.96 \pm 0.04$ from Percival et al. 2002).
The WMAP team also consider adding data from the Lyman-$\alpha$
forest, which pushes the solution away from a pure power-law:
$$
n=0.93\pm 0.03\quad\quad {d n\over d\ln k} = -0.031\pm0.016.
$$
This evidence for \key{running of $n$} is at best marginal, and
disappears completely when systematic uncertainties in the
Lyman-$\alpha$ data are considered (Seljak, McDonald \& Makarov 2003).
It would in any case be surprising if true, since simple inflation
models suggest that $dn/ d\ln k$ should be second order in $(n-1)$.
Although the tensor degeneracy prevents any very strong statements,
the data are best described by pure scalar fluctuations, and
Percival et al. (2002) set an upper limit of 0.7 to the 
tensor-to-scalar ratio.

The agreement with pure scalar $n=1$ is not yet a strong
embarrassment for inflation, but it is starting to bite on
some inflationary models. Leach \& Liddle (2003) show 
that CMB plus 2dFGRS are inconsistent with the $V=\lambda \phi^4$
model at just about 95\% confidence. It is possible
to set up inflation models in which tilt and tensors are
both negligible, but there has been a long-standing
hope for more substantial signs of inflationary
dynamics; if these are not seen soon, it will be
a major disappointment.

\ssec{Matter fluctuation amplitude and bias}

The above conclusions were obtained by considering the shapes
of the CMB and galaxy power spectra. However, it is also of
great interest to consider the amplitude of mass fluctuations,
since a comparison with the galaxy power spectrum
allows us to infer the degree of bias directly.
This analysis was performed by Lahav et al. (2002).
Given assumed values for the cosmological parameters, the
present-day linear normalization of the mass spectrum (e.g. $\sigma_8$)
can be inferred.
It is convenient to define a corresponding measure
for the galaxies, $\sigma_{8{\rm g}}$, such that we can express the bias parameter
as
$$
b = \frac{\sigma_{8{\rm g}}}{\sigma_{8{\rm m}} }.
$$
In practice, we define $\sigma_{8{\rm g}}$ to be the value  required
to fit a CDM model to the power-spectrum data on linear scales ($0.02<k<0.15 \hompc$).
The amplitude of 2dFGRS galaxies in real space estimated
by Lahav et al. (2002) is
$\smash{\sigma_{8{\rm g}}^R} (L^*) = 0.76$,
with a negligibly small random error.
This assumes no evolution in $\smash{\sigma_{8{\rm g}}}$,
plus the luminosity dependence of clustering measured by 
Norberg et al. (2001).

The value of $\sigma_8$ for the dark matter can be deduced from the
CMB fits. Percival et al. (2002) obtain 
$$
 \sigma_8 \exp(- \tau) = 0.72 \pm 0.04,
$$
where the error bar includes both data errors and theory
uncertainty. The WMAP number here is almost identical:
$\sigma_8 \exp(- \tau) = 0.71$, but no error is quoted
(Spergel et al. 2003).
The unsatisfactory feature is the degeneracy
with the optical depth to last scattering. For reionization
at redshift $8$, we would have $\tau\simeq 0.05$;
it is not expected theoretically that $\tau$ can be hugely larger,
and popular models would place reionization between $z=10$ and 
$z=15$, or $\tau\simeq 0.1$ (e.g. Loeb \& Barkana 2001). 
One of the many impressive aspects of the WMAP results is
that they are able to infer $\tau=0.17\pm 0.04$ from large-scale
polarization. Taken at face value, $\tau=0.17$ would  argue for reionization
at $z=20$, but the error means that more conventional figures are far from
being ruled out.
Taking all this together, it seems reasonable to assume that
the true value of $\sigma_8$ is within a few \% of 0.80.
Given the 2dFGRS figure of $\smash{\sigma_{8{\rm g}}^R} = 0.76$,
this implies that $L^*$ galaxies are very nearly exactly unbiased.
Since there are substantial variations
in the clustering amplitude with galaxy type, this outcome
must be something of a coincidence.
This conclusion of near-unity bias was reinforced
in a completely independent way, by using the
measurements of the bispectrum of galaxies in the 2dFGRS
(Verde et al. 2002). As it is based on three-point
correlations, this statistic is sensitive to the filamentary
nature of the galaxy distribution -- which is a signature of
nonlinear evolution. One can therefore split the degeneracy between
the amplitude of dark-matter fluctuations and the
amount of bias.
 
These conclusions point the way towards a possible limit on the
tensor contribution: a large contribution of tensors to the COBE
signal would lower the required scalar amplitude. As an extreme
example, a scalar-to-tensor ratio of 1 would reduce the
`apparent' $\sigma_8$ by roughly a factor of $\sqrt{2}$, to 0.5. Even
for an implausibly large value of $\tau$, this would be hard
to reconcile with the level of galaxy clustering plus the
requirement of a low degree of bias. Also, more direct limits
on $\sigma_8$ are now being derived from large-scale gravitational
lensing surveys, with $\sigma_8 \simeq 0.7$ to 0.8 being favoured
(e.g. Brown et al. 2003; Jarvis et al. 2003).

\sec{Less-standard ingredients}

\ssec{Limits to the neutrino mass}

Even though a CDM-dominated universe matches the
data very well, there are many plausible variations
to consider. Probably the most interesting is the
neutrino mass: experimental data on neutrino
oscillations mean that at least one
neutrino must have a mass of $\gs 0.05$~eV, so that
$\Omega_\nu \gs 10^{-3}$ -- the same order of magnitude
as stellar mass.

\japfigbasic{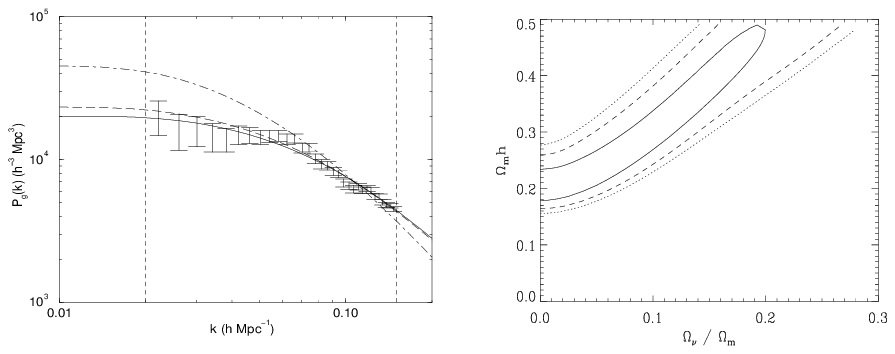}{1.0}
{Results from Elgaroy et al. (2002),
who considered constraints on the neutrino mass from 2dFGRS.
The first panel shows 
Power spectra for $\Omega_\nu = 0$ (solid line),
$\Omega_\nu=0.01$ (dashed line), and $\Omega_\nu=0.05$ (dot-dashed line)
with amplitudes fitted to the 2dFGRS power spectrum data (vertical bars).
Other parameters are fixed at
$\Omega_{\rm m}=0.3$, $\Omega_\Lambda=0.7$, $h=0.7$, $\Omega_{\rm b}h^2=0.02$.
The vertical dashed lines limit the range in $k$ used in the fits.
The second panel shows
68\% (solid line), 95\% (dashed line) and 99\% (dotted line)
confidence contours in the plane of
$f_\nu\equiv \Omega_{\nu}/\Omega_{\rm m}$
and $\Gamma\equiv\Omega_{\rm m} h$,
with marginalization over $h$ and $\Omega_{\rm b}h^2$ using Gaussian
priors.
}

As explained in earlier, a non-zero neutrino mass can
lead to relatively enhanced large-scale power, beyond the
neutrino free-streaming scale. This is illustrated in 
figure~\ref{fig:nu_fig}, taken from Elgaroy et al. (2002).
Broadly speaking, allowing a significant neutrino mass
changes the spectrum in a way that resembles lower density,
so there is a near-degeneracy between neutrino mass fraction
and $\Omega_m h$ (figure~\lastfig). A limit on the neutrino
fraction thus requires a prior on $\Omega_m h$. Based
on the cluster baryon fraction plus BBN, 
Elgaroy et al. adopt $\Omega_m<0.5$; together with the HST
Hubble constant, this yields a marginalized 95\% limit
of $f_\nu < 0.13$, or $m_\nu < 1.8$~eV.
Note that this is the sum of the eigenvalues of the mass matrix:
given neutrino oscillation results
(e.g. Ahmad et al. 2002; Eguchi et al. 2003), the only way a cosmologically
significant density can arise is via a nearly degenerate
hierarchy, so this allows us to deduce $m_\nu < 0.6$~eV
for any one species. Including the latest WMAP results in
order to set a more strict limit on
$\Omega_m h$, this limit falls to 0.23~eV (Spergel et al. 2003).

\ssec{The equation of state of the vacuum}

So far, we have assumed that the vacuum energy is exactly
a classical $\Lambda$, or at any rate indistinguishable from
one. This is a highly reasonable prior: there is no reason
for the asymptotic value of any potential to go exactly
to zero, so one always needs to solve the classical cosmological
constant problem -- for which probably the only reasonable
explanation is an anthropic one (e.g. Vilenkin 2001).
Therefore, dynamical provision of $w\equiv p_v/\rho_v \ne -1$
is not needed. Nevertheless, one can readily take an
empirical approach to $w$ (treated as a constant for 
a first approach).

\japfigbasic{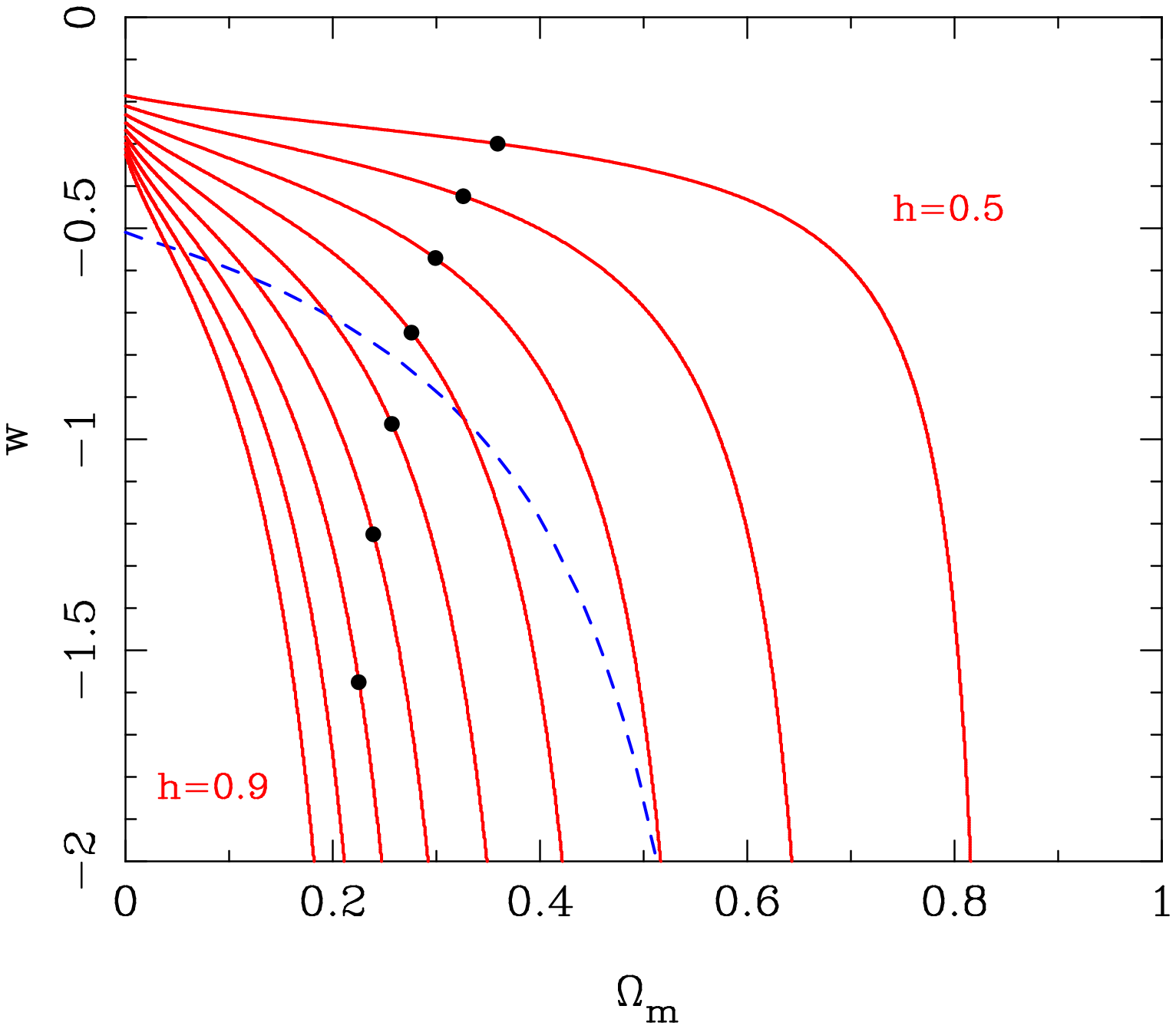}{0.7}
{The $\Omega_m h^{3.4}$ degeneracy for flat models
gives an almost exact value of $\Omega_m$ from the CMB
is $h$ is known, assuming the vacuum to be effectively
a classical $\Lambda$ ($w=-1$). If $w$ is allowed to vary,
this becomes a locus on the $(\Omega_m, w)$ plane (similar
to the locus for best-fitting  flat models from the SNe, showed dotted).
Solid circles show values of $\Omega_m h$ that satisfy the
updated 2dFGRS constraint of 0.18 (suppressing error bars).}

Figure~\lastfig\ shows a simplified approach to this, plotting the
locus on $(w,\Omega_m)$ space that is required for
a given value of $h$ if the location of the main CMB acoustic peak
is known exactly. For $h\simeq 0.7$, this is very similar
to the locus derived from the SN Hubble diagram
(Garnavich et al. 1998).
The solid circles show the updated 2dFGRS constraint of
$\Omega_m h =0.18$. In order to match the data with
$w$ closer to zero, $\Omega_m$ must increase and $h$ must
decrease. The latter trend means that the HST Hubble constant
sets an upper limit to $w$ of about $-0.54$ (Percival et al. 2002).
This is very similar to the SNe constraint of Garnavich et al. (1998),
so the combined limit is already close to $w<-0.8$. The vacuum
energy is indeed looking rather similar to $\Lambda$.

\ssec{The total relativistic density}

Finally, an interesting aspect of figure~\lastfig\ is that it reminds
us of history. When the COBE detection was announced in 1992,
a popular model was `standard' CDM with $\Omega_m=1$, $h=0.5$.
As we see, this comes close to fitting the CMB data, and such
a model is not unattractive in some ways. Can we be sure it
is ruled out? Leaving aside the SNe data, one might think to
evade the 2dFGRS constraint by altering the total relativistic
content of the universe (for example, by the decay of a
heavy neutrino after nucleosynthesis). Since 2dFGRS measures
the horizon at matter-radiation equality, this will be changed.
If the radiation density is arbitrarily boosted by a factor
$X$, the constraint from LSS becomes
$$
(\Omega_m h)_{\rm apparent} = X^{-1/2} (\Omega_m h)_{\rm true}.
$$
Therefore $X\simeq 8$ is required to allow an Einstein--de Sitter
universe.

However, this argument fails, because it does not take into account
the effect of the extra radiation on the CMB. As argued above, the
location of the acoustic peaks depends on $a_{\rm eq}$, which depends
on $\omega_m$. However, if we change the radiation content, then
what matters is $\omega_m/X$. Thus, the CMB peak constraint now reads
$$
\Omega_m^{-0.1} (\omega_m/X)^{0.24} = {\rm constant};
$$
when combining LSS and CMB, everything is as before except that the
effective Hubble parameter is $h/X^{1/2}$. Thus, a model
with $\Omega_m=1$ but boosted radiation would only fit the CMB with
$h\simeq 0.5\sqrt{8} \simeq 1.4$, and the attractiveness
of a low age is lost. In any case, combining LSS and CMB would
give the same $\Omega_m \simeq 0.3$ independent of $X$, so it is
impossible to save models with $\Omega_m=1$ by this route.

Finally, it is interesting to invert this argument.
Since Percival et al. (2002) obtain an effective
$h$ of $0.665\pm 0.047$ and Freedman et al. (2001) measure
$h=0.72\pm 0.08$, we deduce
$$
1.68 X = 1.82 \pm 0.24.
$$
This convincingly rules out the $1.68 X=1$ that would apply
if the universe contained only photons, and amounts to
a detection of the neutrino background. In
terms of the number of neutrino species, this is
$N_\nu = 3.6 \pm 1.1$. A more precise result is of
course obtained from primordial nucleosynthesis, but this
applies at a much later epoch, thus constraining models
with decaying particles.

\sec{Conclusions}

The beautiful data on the large-scale structure of the universe
revealed in particular by the 2dF Galaxy Redshift Survey
combine with the incredible recent progress in CMB
data to show spectacularly good agreement with a
`standard model' for structure formation. This consists
of a scalar-mode adiabatic CDM universe with scale-invariant
fluctuations. Measuring the exact parameters of this
model is rendered difficult by the intrinsic degeneracies
of the structure-formation process, but  progress is
being made. The most recent data yield
$\Omega_m = 0.25 \pm 15\%$ and $h=0.73 \pm 5\%$; these
figures accord well with independent constraints, and it
is very hard to believe that they are incorrect.

Allowing extra degrees of freedom, such as massive neutrinos,
vacuum equation of state $w\ne 1$, or extra relativistic content
worsens the agreement with independent constraints on
$h$ and $\Omega_m$. This both supports the simplest picture
and allows us to set interesting limits on these 
non-standard ingredients.

For the future, we can look with anticipation to 
meaningful tests of inflation: the current data are consistent
with $n=1$ to an error of $\pm 0.03$, and the errors may be
expected to halve over the next couple of years, bringing
plausible levels of tilt well within the reach of
experimental detection. 
A demonstration that $n\ne 1$ would be a large step in the direction
of proving inflation to be true, so the cosmological stakes
over the next few years will be high.
The tensor fraction is a less clear target, but the motivation
to improve on the current weak upper limits will remain strong.

It should of course not be forgotten that the large-scale structure
we measure locally consists of galaxies. In these lectures,
the physics of galaxy formation has been largely ignored,
but this will be the increasing focus of LSS studies:
not just the global parameters of the universe, but the
detailed understanding of how the complex structures
around us formed.

\section*{Acknowledgements}

I have drawn on the body of results achieved by my colleagues
in the 2dF Galaxy Redshift Survey team: Matthew Colless (ANU), 
Ivan Baldry (JHU), Carlton  Baugh (Durham), 
Joss Bland-Hawthorn (AAO), Terry Bridges (AAO),
Russell Cannon (AAO), Shaun Cole (Durham), 
Chris Collins (LJMU), Warrick Couch (UNSW), 
Gavin Dalton (Oxford), Roberto De Propris (UNSW), 
Simon Driver (St Andrews), George Efstathiou (IoA), 
Richard  Ellis (Caltech), Carlos Frenk (Durham), 
Karl Glazebrook (JHU), Carole Jackson (ANU), Ofer Lahav (IoA), Ian Lewis (AAO), 
Stuart Lumsden (Leeds), Steve Maddox (Nottingham), Darren Madgwick (IoA), 
Peder Norberg (Durham), Will Percival (ROE), 
Bruce Peterson (ANU), Will Sutherland (ROE), Keith Taylor (Caltech).
The 2dF Galaxy Redshift Survey
was made possible by the dedicated efforts of the staff
of the Anglo-Australian Observatory, both in creating the 2dF
instrument, and in supporting it on the telescope.

\section*{References}

{

\def\mnras{Mon. Not. R. Astr. Soc.}
\def\mn{Mon. Not. R. Astr. Soc.}
\def\qjras{Q. J. R. Astr. Soc.}
\def\apj{Astrophys. J.}
\def\apjs{Astrophys. J. Suppl.}
\def\aj{Astr. J.}
\def\pr{Phys. Rev.}
\def\pl{Phys. Lett.}
\def\pasp{Proc. Astr. Soc. Pacif.}
\def\araa{Ann. Rev. Astr. Astrophys.}
\def\nature{Nature}
\def\aa{Astr. Astrophys.}
\def\aas{Astr. Astrophys. Suppl.}
\def\prl{Phys. Rev. Lett.}
\def\prd{Phys. Rev. D}
\def\rmp{Rev. Mod. Phys.}
\def\science{Science}

\japref Adelberger K. et al., 1998, ApJ, 505, 18
\japref Adler R.J., 1981, {\it The Geometry of Random Fields}, Wiley
\japref Ahmad Q.R. et al. (the SNO Consortium), 2002, \prl, 89, 011301
\japref Bardeen J.M., 1980, Phys. Rev. D, 22, 1882
\japref Bardeen J.M., Bond J.R., Kaiser N., Szalay A.S., 1986, \apj, 304, 15 
\japref Baugh C.M., Efstathiou G., 1993, MNRAS, 265, 145
\japref Baugh C.M., Efstathiou G., 1994, MNRAS, 267, 323
\japref Bahcall N.A., Soneira R.M., 1983, ApJ, 270, 20
\japref Benoist C., Maurogordato S., da Costa L.N., Cappi A., Schaeffer R., 1996, ApJ, 472, 452
\japref Benson A.J., Cole S., Frenk C.S., Baugh C.M., Lacey C.G., 2000a, \mn, 311, 793
\japref Benson A.J., Baugh C.M., Cole S., Frenk C.S., Lacey C.G., 2000b, \mn, 316, 107
\japref Benson, A.J., Frenk, C.S., Baugh, C.M., Cole, S., Lacey, C.G., 2001, MNRAS, 327, 1041
\japref Bond J.R., Cole S., Efstathiou G., Kaiser N., 1991, \apj, 379, 440
\japref Bond J.R., Efstathiou G., 1984, \apj, 285, L45
\japref Bond J.R., Szalay A., 1983, ApJ, 274, 443
\japref Bouchet F.R., Colombi S., Hivon E., Juszkiewicz R., 1995, \aa, 296, 575
\japref Brown M.L. et al., 2003, MNRAS, 341, 100
\japref Bucher M., Moodley K., Turok, N., 2002, Phys. Rev. D, 66, 023528
\japref Burles S., Nollett K.M., Turner M.S., 2001, ApJ, 552, L1
\japref Carlberg R.G., Couchman H.M.P., Thomas P.A., 1990, \apj, 352, L29
\japref Carroll S.M., Press W.H., Turner E.L., \araa, 30, 499
\japref Colberg J. et al., 2000, MNRAS, 319, 209
\japref Cole S., Arag\'on-Salamanca A., Frenk C.S., Navarro J.F., Zepf S.E., 1994, \mn, 271, 781
\japref Cole S., Hatton S., Weinberg D.H., Frenk C.S., 1998, MNRAS, 300, 945
\japref Cole S., Lacey C.G., Baugh C.M., Frenk C.S., 2000, \mn, 319, 168
\japref Cole S., Kaiser N., 1989, \mn, 237, 1127
\japref Coles P., 1993, \mn, 262 ,1065
\japref Coles P., Jones B.J.T., 1991, \mnras, 248, 1
\japref Colless M. et al., 2001, MNRAS, 328, 1039
\japref Croft R.A.C., Dalton G.B., Efstathiou G., Sutherland W.J., Maddox S.J., 1997, MNRAS, 291, 305
\japref Davis M., Peebles P.J.E., 1983, ApJ, 267, 465
\japref Davis M., Geller M.J., 1976, ApJ, 208, 13
\japref Davis M., Efstathiou G., Frenk C.S., White S.D.M., 1985, \apj, 292, 371
\japref Dekel A., Rees M.J., 1987, Nat, 326, 455
\japref Dekel A., Lahav O., 1999, ApJ, 520, 24
\japref de Lapparant V., Geller M.J., Huchra J.P., 1986, ApJ, 302, L1
\japref Dodelson S., 2003, {\it Modern Cosmology}, Academic Press
\japref Efstathiou G. et al., 2002, MNRAS, 330, L29
\japref Efstathiou G., Bond J.R., 1986, MNRAS, 218, 103
\japref Efstathiou G., Bond J.R., 1999, MNRAS, 304, 75
\japref Efstathiou G., Davis M., White S.D.M., Frenk C.S., 1985, \apjs, 57, 241
\japref Eguchi K. et al., 2003, \prl, 90, 1802
\japref Eisenstein D.J., Hu W., 1998, ApJ, 496, 605
\japref Eisenstein D.J., Hu W., 1999, ApJ, 511, 5
\japref Eke V.R., Cole S., Frenk C.S., 1996, \mn, 282, 263
\japref Elgaroy O. et al., 2002, Phys. Rev. Lett., 89, 061301
\japref Evrard A. et al., 2002, ApJ, 573, 7
\japref Evrard A., 1997, MNRAS, 292, 289
\japref Feldman H.A., Kaiser N., Peacock J.A., 1994, ApJ, 426, 23
\japref Freedman W.L. et al., 2001, ApJ, 553, 47 
\japref Fry J.N., 1986, ApJ, 461, L65
\japref Garnavich P.M. et al., 1998, ApJ, 509, 74
\japref Ghigna S., Moore B., Governato F., Lake G., Quinn T., Stadel J., 1998, \mn, 300, 146
\japref Goldberg D.M., Strauss M., 1998, ApJ, 495, 29
\japref Gordon C., Lewis A., 2002, astro-ph/0212248 
\japref Hambly N.C., Irwin M.J., MacGillivray H.T., 2001, MNRAS, 326 1295
\japref Hawking S., 1982, Phys. Lett., B115, 29 
\japref Hawkins E. et al., 2002, astro-ph/0212375
\japref Heath D., 1977, \mnras, 179, 351
\japref Hernquist L., Bouchet F.R., Suto Y., 1991, \apjs, 75, 231
\japref Hockney R. W., Eastwood J. W., 1988, {\it Computer Simulations Using Particles}, IOP Publishing
\japref Hu W., Dodelson S., 2002, ARAA, 40, 171
\japref Hu W., Fukugita M., Zaldarriaga M., Tegmark M., 2001, ApJ, 549, 669
\japref Hu W., Sugiyama N., 1995, ApJ, 444, 489
\japref Hu W., White M., 1997, New Astronomy, 2, 323
\japref Hubble E.P., 1934, ApJ, 79, 8 
\japref Jarvis M. et al., 2003, AJ, 125, 1014
\japref Jenkins A., Frenk C.S., Pearce F.R., Thomas P.A., Colberg J.M., White S.D.M., Couchman H.M.P., Peacock J.A., Efstathiou G., Nelson A.H., 1998, ApJ, 499, 20
\japref Jing Y.P., Mo H.J., B\"orner G., 1998, ApJ, 494, 1
\japref Kaiser N., 1984, ApJ, {284}, L9
\japref Katz N., Weinberg D.H., Hernquist L., 1996, ApJ Suppl., 105, 19 
\japref Kauffmann G., Colberg J.M., Diaferio A., White S.D.M., 1999, \mn, 303, 188
\japref Kauffmann G., Nusser A., Steinmetz M., 1997, \mn, 286, 795
\japref Kauffmann G., White S.D.M., Guiderdoni B., 1993, \mn, 264, 201
\japref Kirshner R.P., Oemler A., Schechter P.L., Shectman S.A., 1981, \apj, 248, L57
\japref Klypin A., Kopylov A.I., 1983, Soviet Astron. Lett., 9, 41
\japref Klypin A.,  Primack J., Holtzman J., 1996, \apj, 466, 13
\japref Klypin A., Gottl\"ober S., Kravtsov A.V., Khokhlov A.M., 1999, ApJ, 516, 530
\japref Kodama H., Sasaki M., 1984, Prog. Theor. Phys. Suppl., 78, 1
\japref Kovac J.M. et al., 2002, Nature, 420, 772
\japref Kogut A. et al., 2003, astro-ph/0302213
\japref Kravtsov A.V., Klypin A.A., Khokhlov A.M., 1997, ApJ Suppl., 111, 73
\japref Kuo C.L. et al., 2002, astro-ph/0212289 
\japref Lahav O., Lilje P.B., Primack J.R., Rees M.J., 1991, \mnras, 251, 128
\japref Lahav O. et al., 2002, MNRAS, 333, 961
\japref Leach S.M., Liddle A.R., 2003, astro-ph/0306305
\japref Lewis I.J. et al., 2002, MNRAS, 333, 279
\japref Liddle A.R., Lyth D., 2000, {\it Cosmological inflation \& large-scale structure}, CUP
\japref Loeb A., Barkana R., 2001, ARAA, 39, 19
\japref Loveday J., Maddox S.J., Efstathiou G., Peterson B.A., 1995, \apj, 442, 457
\japref Lyth D., Wands D., 2002, Phys. Lett., B524, 5
\japref Ma C.-P., 1999, \apj, 510, 32
\japref Ma C.-P., 1996, \apj, 471, 13
\japref Maddox S.J., Efstathiou G., Sutherland W.J., Loveday J., 1990a, MNRAS, 242, 43{\sc p}
\japref Maddox S.J., Sutherland W.J., Efstathiou G., Loveday J., 1990b, MNRAS, 243, 692
\japref Maddox S.J., Efstathiou G., Sutherland W.J., 1990c, MNRAS, 246, 433
\japref Mann R.G., Peacock J.A., Heavens A.F., 1998, MNRAS, 293, 209
\japref Meiksin A.A., White M., Peacock J.A., 1999, MNRAS, 304, 851
\japref M\'esz\'aros P., 1974, \aa, 37, 225
\japref Mo H.J., White S.D.M., 1996, MNRAS, 282, 1096
\japref Moore B., Frenk C.S., White S.D.M., 1993, \mn, 261, 827
\japref Moore B., Quinn T., Governato F., Stadel J., Lake G., 1999, \mn, 310, 1147 [M99]
\japref Mould J.R. et al., 2000, ApJ, 529, 786
\japref Mukhanov V.F., Feldman H.A., Brandenberger R.H., 1992, Phys. Reports, 215, 203
\japref Narayanan V.K., Spergel D.N., Dav\'e R., Ma C.-P., 2000, ApJ, 543, L103
\japref Navarro J.F., Frenk C.S., White S.D.M., 1996, ApJ, 462, 563
\japref Neyman J., Scott E.L. \& Shane C.D., 1953, ApJ, 117, 92
\japref Norberg P. et al., 2001, MNRAS, 328, 64
\japref Nusser A., 2000, MNRAS, 317, 902
\japref Papovich C., Dickinson M., Ferguson H.C., 2001, ApJ, 559, 620
\japref Peacock J.A., 1997, \mn, 284, 885
\japref Peacock J.A., 1999, {\it Cosmological Physics}, CUP
\japref Peacock J.A., Smith R.E., 2000, \mnras, 318, 1144
\japref Pearce F.R. et al., 1999, \apj, 521, L99
\japref Pearce F.R. et al., 2001, MNRAS, 326, 649
\japref Pearson T.J. et al., 2002, astro-ph/0205388
\japref Peebles P.J.E., 1974, A\&A, 32, 197
\japref Peebles P.J.E., 1980, {\it The Large-Scale Structure of the  Universe}, Princeton Univ. Press, Princeton, NJ
\japref Peebles P.J.E., Yu J.T., 1970, \apj, 162, 815
\japref Peebles P.J.E., 1987, \nature, 327, 210
\japref Pen U.-L., 1998, ApJ, 504, 601
\japref Percival W.J. et al., 2001, MNRAS, 327, 1297
\japref Percival W.J. et al., 2002, MNRAS, 337, 1068
\japref Pogosyan D., Starobinsky A.A., 1995, \apj, 447, 465
\japref Press W.H., Schechter P., 1974, {\apj}, {187}, 425
\japref Press W.H., Teukolsky S.A., Vetterling W.T., Flannery B.P.,  {\it Numerical Recipes\/} (2nd edition), Cambridge University Press
\japref Rees M.J., 1985, \mnras, 213, 75
\japref Saunders W., Frenk C., Rowan-Robinson M., Efstathiou G.,  Lawrence A., Kaiser N., Ellis R., Crawford J., Xia X.-Y., Parry I., 1991, Nature, 349, 32
\japref Schlegel D.J., Finkbeiner D.P., Davis M., 1998, ApJ, 500, 525
\japref Seljak U., 1997, \apj, 482, 6
\japref Seljak U., 2000, MNRAS, 318, 203
\japref Seljak U., Zaldarriaga M., 1996, \apj, 469, 437
\japref Seljak U., McDonald P. Makarov A., 2003, astro-ph/0302571
\japref Sheth R.K., Tormen G., 1999, \mn, 308, 119
\japref Sheth R.K., Mo H.J., Tormen G., 2001, MNRAS, 323, 1
\japref Smoot G.F. et al., 1992, \apj, 396, L1
\japref Somerville R.S., Primack J.R., 1999, \mn, 310, 1087
\japref Spergel D.N. et al., 2003, astro-ph/0302209
\japref Starobinsky A.A., 1985, Sov. Astr. Lett., 11, 133
\japref Stoughton C.L. et al., 2002, AJ, 123, 485
\japref Sugiyama N., 1995, \apjs, 100, 281
\japref Tegmark M., Peebles P.J.E., 1998, ApJ, 500, L79
\japref Tonry J.L. et al., 2003, astro-ph/0305008
\japref Valageas P., 1999, A\&A, 347, 757
\japref van Kampen E., 2000, astro-ph/0002027
\japref van Kampen E., Jimenez R., Peacock J.A.,  1999, \mn, 310, 43
\japref Verde L. et al., 2002, MNRAS, 335, 432
\japref Vilenkin A., 2001, hep-th/0106083
\japref Vittorio N., Silk J., 1985, ApJ, 297, L1
\japref White S.D.M., Rees M., 1978, \mn, 183, 341
\japref White S.D.M., Davis M., Efstathiou G., Frenk C.S., 1987, \nature, 330, 451
\japref Zehavi I. et al., 2003, astro-ph/0301280
\japref Zeldovich Y.B., 1970, \aa, 5, 84
\japref Zwicky F., 1933, Helv. Phys. Acta, 6, 110

}

\end{document}